%% file: MGFAfinal3.tex
\documentclass[10pt]{iopart}
\usepackage{epsfig}
\usepackage{graphicx}
\input Commands.tex

\begin{document}\small

\title{Theory of agent-based market models with controlled levels of greed and anxiety}
\author{P Papadopoulos and ACC Coolen}
\address{Department of Mathematics,
King's College London\\
The Strand, London WC2R 2LS, United Kingdom}
\ead{panagiotis.2.papadopoulos@kcl.ac.uk, ton.coolen@kcl.ac.uk}

\begin{abstract}
We use generating functional analysis to study minority-game type market models with generalized strategy valuation updates that control the psychology of agents' actions. The agents' choice between trend following and contrarian trading, and their vigor in each, depends on the overall state of the market. Even in `fake history' models, the theory now involves an effective overall bid process (coupled to the effective agent process) which can exhibit profound remanence effects and new phase transitions. For some models the bid process can be solved directly,
others require Maxwell-construction type approximations.
\end{abstract}

\section{Introduction}

Minority games (MG)  \cite{1,2} are simple
models proposed about a decade ago to understand the origin of the cooperative phenomena and nontrivial
fluctuations observed in financial markets, on the basis of so-called inductive decision making by imperfect
interacting agents \cite{Brian Arthur}. With the recent exposure of the inadequacy of classical economic modeling that assumes full rationality, intelligence and honesty of market players and evolution to an efficient stable market, the case for the alternative MG approach to market models is probably beyond discussion. The further strength of MG-type models is that they can be solved analytically; especially the generating functional analysis (GFA) method has
 proven to be an effective tool in this field \cite{DeDom,Ton1,Ton2}.
We refer to the recent textbooks \cite{Book1, Book2} for historical backgrounds, the connection between MGs and markets,
 details on mathematical methods, and full references.

Standard MG models are now understood reasonably well, so the next task is to generalize the mathematical technology
developed so as to apply to models that are more realistic economically. In this paper we try to contribute to this aim, by studying analytically
 MG-type models in which we generalize the agents' strategy valuation rules: the strategy type to be rewarded (and by how much) depends on the overall market bid $A$. Agents can switch from a contrarian to a trend-following strategy, and adjust their level of caution, on the basis of how anomalous they perceive the market to be. Such ingredients are incorporated easily into the fabric of MG-type models,
by enhancing the valuation update with a function $F[A]$ of the excess demand; the question is how to decide which agent
behaviour should be encoded in this function, and how to solve
mathematically the resulting GFA equations.

By changing the magnitude of the valuation updates in the MG in an $A$-dependent way, we can model a preference of agents  (who remain contrarians) for profitable decisions with large excess demand/supply (high-risk high-gain behaviour), or model the opposite situation where they prefer small excess demand/supply (low-risk low-gain behaviour). If we also allow agents to switch to trend-following, there are two possible routes.
One could argue that when markets are booming, equity prices tend to rise slowly and the trading volume
increases (so agents are trend-followers, as in the majority game \cite{Kozlowski}), until the magnitude $|A|$ of the overall market bid becomes too large, leading to turmoil and contrarian behaviour. This was the view of \cite{demartino2,demartino1}.
Alternatively, one could argue that for small $|A|$ the market would be regarded as normal, and agents would seek advantage by contrarian trading, whereas large $|A|$ would be regarded as anomalous, prompting panic-driven trend-following.

In this paper we first develop the theory for MG-type models with arbitrary valuation update functions $F[A]$. Then we focus on the above three agent behaviour scenarios: a model of strict contrarians  with controlled degrees of caution versus greed, one where non-volatile markets are interpreted as risk free and prompt trend-following (agents only become contrarians under volatile conditions),
and one where in non-volatile markets agents are contrarians (switching to panic-driven trend-following under volatile conditions).
Although the latter two cases are mathematically similar, they
lead to qualitatively different solutions. We will find that certain regions
of the phase diagram are characterized by prominent remanence effects, with multiple locally stable states, prompting us to
rely on approximations of the Maxwell-construction type. As a result, whereas our predictions are confirmed satisfactorily by numerical simulations in non-remanent cases, when there is remanence, there remain deviations and gaps in our understanding which will require further investigation.

\section{Definitions}

Let there to be $N$ agents in the game, labeled by
$i\in\{1,\ldots,N\}$, where $N$ is odd. The game proceeds at discrete steps
 $t=0,1,2,\ldots$.
At every step $t$ each agent takes a binary decision
$b_i(t)\in\{-1,1\}$, the `bid',
representing e.g. buying $(-1)$ versus selling $(+1)$.
 A re-scaled aggregate overall bid at step $t$ is defined as
$A(t)=N^{-1/2}\sum_{i=1}^N b_i(t)+A_e(t)$, where $A_e(t)$ is an external contribution,
representing perturbations, the impact of large market operators, or the action of regulators.
At each step, the agents are provided with external information, which in fake history MGs is a random number
 $\mu(t)\in\{1,\ldots,p\}$.
Each agent $i$ has two strategies $\bR_{ia}$, with $a=1,2$, with which to map each
 $\mu$ to a trading action:
\bd
\bR^{ia}:\{1,\ldots,p\}\to \{-1,1\}
\ed
A strategy functions as a
lookup-table with $p$ entries, all of which  are drawn independently at random from $\{-1,1\}$
(with equal probabilities) before the start of the game.
If agent $i$ uses
strategy $a$ at step $t$, then he will act (deterministically)
according to $b_i(t)=R_{ia}^{\mu(t)}$.

In the standard MG an agent makes profit if he finds himself in the minority, i.e.
if $A(t)b_i(t)<0$. Each agent monitors the performance of his two strategies
(irrespective of whether they were used), by measuring the so-called valuations $p_{ia}(t)$, defined via
$p_{ia}(t+1)=p_{ia}(t)- \rate A(t)~
R_{ia}^{\mu(t)}$. Here $\rate>0$ denotes a `learning rate'.
At each step $t$ of the game, each agent $i$ will select his
best strategy $a_i(t)$ at that stage of the process, defined as
$a_i(t)={\rm arg}~ {\rm max}_{a\in\{1,2\}}
\{p_{ia}(t)\}$.
Agents thus behave as `contrarians'. Here we generalize this rule in the spirit of \cite{demartino2}, to incorporate
the possibility that whether agents behave as contrarians or trend-followers may depend dynamically on whether they perceive the market
state $A(t)$ to be close to or far from its `natural' value:
$p_{ia}(t+1)=p_{ia}(t)- \rate
R_{ia}^{\mu(t)} F[A(t)]$,
with $F[-A]=-F[A]$ (to retain $A=0$ as the `natural' value for the overall bid).
In the `batch' version of the above dynamics one replaces in the evolution equations for the valuations
the randomly drawn information $\mu(t)$ by an
average over $\{1,\ldots,p\}$. The result is
\begin{eqnarray}
p_{ia}(t\!+\!1)&=& p_{ia}(t)- \frac{\rate}{\sqrt{N}}\sum_{\mu=1}^p F[A_\mu(t)] R_{ia}^\mu \label{eq:basicMG1}
\\
A_\mu(t)&=&A_e(t)+\frac{1}{\sqrt{N}}\sum_{i=1}^N R_{i
a_i(t)}^\mu
\end{eqnarray}
We switch in the usual manner to new variables $q_i(t)=\frac{1}{2}[p_{i1}(t)-p_{i2}(t)]$, $\xi_\mu^i=\frac{1}{2}(R_{i1}^\mu-R_{i2}^\mu)$,
$\omega^i_\mu=\frac{1}{2}(R_{i1}^\mu+R_{i2}^\mu)$, and $\Omega_\mu=N^{-1/2}\sum_i \omega^\mu_i$. We choose  $\rate=2$, and add perturbations $\theta_i(t)$ in order to define response functions. We introduce decision noise
 by replacing
$\sgn[q_i(t)]\to \sigma[q_i(t),z_i(t)]$,
in which the $\{z_{j}(t)\}$ are
independent and zero average Gaussian random numbers (the standard
 examples are
 additive and multiplicative noise definitions, corresponding to $\sigma[q,z]=\sgn[q+Tz]$ and
 $\sigma[q,z]=\sgn[q]~\sgn[1+Tz]$).
 This leaves us with
\begin{eqnarray}
q_{i}(t\!+\!1)&=& q_{i}(t)+\theta_i(t)- \frac{2}{\sqrt{N}}\sum_{\mu=1}^p \xi^\mu_i F[A_\mu(t)] \label{eq:basicMG3}
\\
A_\mu(t)&=&A_e(t)+\Omega_\mu+\frac{1}{\sqrt{N}}\sum_{j=1}^N\xi^\mu_j~\sigma[q_j(t),z_j(t)]
\label{eq:basicMG4}
\end{eqnarray}
One obtains the standard batch MG for $F[A]=A$, and a batch {\em majority} game for $F[A]=-A$.
We will develop the theory initially for arbitrary $F[A]$, but focus ultimately on the choices
\begin{eqnarray}
&&
F[A]=\sgn(A)|A|^\gamma,~~~~~~~~
F[A]= \tau A\big[1-A^2/A_0^2\big]
\end{eqnarray}
with $\tau=\pm 1$ and $A_0>0$. In the second formula,
for $\tau=-1$ the agents behave as a trend-followers (majority game play) when $|A|<A_0$, but switch to contrarians
(minority game play) for $|A|>A_0$; this is the model of \cite{demartino2,demartino1}. For $\tau=1$ the situation is reversed.

 We
will write averaging over the
stochastic process (\ref{eq:basicMG3}) as $\bra \ldots
\ket$.
The global bid
fluctuations in the system are characterized by the volatility
matrix $\Xi_{tt^\prime}$:
\begin{eqnarray}
\Xi_{tt^\prime}&=&\frac{1}{p }\sum_{\mu=1}^p \Big\bra\Big[
 A_\mu(t)-
\frac{1}{p}\sum_{\nu=1}^p \bra A_\nu(t)\ket\Big]
 \Big[
 A_\mu(t^\prime)-
\frac{1}{p}\sum_{\nu=1}^p \bra
A_\nu(t^\prime)\ket\Big]\Big\ket
 \label{eq:volatility_matrix}
\end{eqnarray}
From this follows the conventional market volatility $\sigma$ via
$\sigma^2=\lim_{\tau\to\infty}\tau^{-1}\sum_{t=1}^{\tau}\Xi_{tt}$.

\section{Generating functional analysis}

 The appropriate
moment generating functional for a
stochastic process of the type
(\ref{eq:basicMG3}), given that now we will be interested also in overall bid statistics, is
\begin{eqnarray}
  Z[\bpsi,\bphi]
  &=&
  \Big\bra \rme^{\rmi\sum_{t\geq 0} \sum_i  \psi_i(t) \sigma[q_i(t),z_i(t)]+\rmi\sqrt{2}\sum_{t\geq 0}\sum_\mu
\phi_\mu(t)A_\mu(t)}  \Big\ket
  \label{eq:defZ}
\end{eqnarray}
with $A_\mu(t)$ as defined in (\ref{eq:basicMG4}).
  For (\ref{eq:defZ}) to be
well defined, we specify an upper time limit $t_{\rm max}$. Taking suitable derivatives of the
 (\ref{eq:defZ}) with respect to the
variables $\{\psi_i(t),\phi_\mu(t)\}$  generates all moments of the
random variables $\{\sigma[q_i(t),z_i(t)]\}$ and $\{A_\mu(t)\}$,  at arbitrary times, e.g.
\begin{eqnarray}
\bra \sigma[q_i(t),z_i(t)]\ket&=& -\rmi\lim_{\bpsi,\bphi\to
\bnull}\frac{\partial Z[\bpsi,\bphi]}{\partial\psi_i(t)}
\label{eq:generate1}
\\
\bra
\sigma[q_i(t),z_i(t)]\sigma[q_j(t^\prime),z_j(t^\prime)]\ket&=&
-\lim_{\bpsi,\bphi\to
\bnull}\frac{\partial^2 Z[\bpsi,\bphi]}{\partial\psi_i(t)\partial\psi_j(t^\prime)}
\label{eq:generate2}
\\
\bra A^\mu[\bq(t),\bz(t)]\ket &=&
-\frac{\rmi}{\sqrt{2}}\lim_{\bpsi,\bphi\to\bnull}\frac{\partial Z[\bpsi,\bphi]
}{\partial\phi_\mu(t)} \label{eq:generate4}
\\
\bra A^\mu[\bq(t),\bz(t)] A^\nu[\bq(t^\prime),\bz(t^\prime)]\ket
&=& -\frac{1}{2}\lim_{\bpsi,\bphi\to\bnull}\frac{\partial^2 Z[\bpsi,\bphi]
}{\partial\phi_\mu(t)\partial\phi_\nu(t^\prime)}
\label{eq:generate5}
\end{eqnarray}
 Averaging (\ref{eq:defZ})
over the disorder (the
strategies) allows us to obtain from
(\ref{eq:generate1},\ref{eq:generate2},\ref{eq:generate4},\ref{eq:generate5})
expressions for disorder-averaged
observables, such as correlation and response
functions:
\begin{eqnarray}
\hspace*{-15mm}
C_{tt^\prime}&= \frac{1}{N}\!\sum_i \overline{\bra
\sigma[q_i(t),z_i(t)]\sigma[q_i(t^\prime),z_i(t^\prime)]\ket}
~&= -\lim_{\bpsi\to
\bnull}\frac{1}{N}\!\sum_i\frac{\partial^2 \overline{Z[\bpsi]}}{\partial\psi_i(t)\partial\psi_i(t^\prime)}
\label{eq:defineC}
 \\
 \hspace*{-15mm}
  G_{tt^\prime}&=
 \frac{1}{N}\!\sum_i
\frac{\partial}{\partial\theta_i(t^\prime)}\overline{\bra
\sigma[q_i(t),z_i(t)]\ket}
~&= - \lim_{\bpsi\to \bnull}\frac{\rmi}{N}\!\sum_i
\frac{\partial^2 \overline{Z[\bpsi]}}{\partial\psi_i(t)\partial\theta_i(t^\prime)}
\label{eq:defineG}
 \end{eqnarray}

\subsection{Evaluation of the disorder average}

 We write
(\ref{eq:defZ}) in the usual way as an integral over all possible values of the $q_i(t)$ and the $A_\mu(t)$ at all times,
and insert $\delta$-functions to select the solution of (\ref{eq:basicMG3}), followed by
averaging over the noise $\bz=\{z_i(t)\}$.
 With the short-hand
$s_i(t)=\sigma[q_i(t),z_i(t)]$ we obtain:
\begin{eqnarray}
  Z[\bpsi,\bphi]
  &=&
  \int\!\Big[\prod_{\mu t}\frac{\rmd A_\mu(t)\rm d\hat{A}_\mu(t)}{2\pi} \rme^{\rmi\hat{A}_\mu(t)[A_\mu(t)-A_e(t)]+\rmi\sqrt{2}
\phi_\mu(t)A_\mu(t)}\Big]
\nonumber
\\
&&\hspace*{-5mm}\times
\Big\bra
 \int\!p_0(\bq(0))\Big[\prod_{i t}\frac{\rm dq_i(t)\rm d\hat{q}_i(t)}{2\pi} \rme^{\rmi\hat{q}_i(t)\left[
     q_i(t+1)-q_i(t)-\theta_i(t)
   \right]+\rmi\psi_i(t)s_i(t)}\Big]
 \nonumber \\
 &&
 \hspace*{10mm}\times
\prod_{\rmi\mu}\rme^{\frac{i}{\sqrt{N}}\sum_t\{2\xi_i^{\mu}\hat{q}_i(t)
      F[A_\mu(t)]-\hat{A}_\mu(t)[\omega_i^\mu+\xi^\mu_i s_i(t)]\}}
\Big\ket_{\!\bz}
\label{eq:Zbatch_early}
\end{eqnarray}
The disorder variables appear only in the last line, so
\begin{eqnarray}
 \overline{Z[\bpsi,\bphi]}
  &=&
  \int\!\Big[\prod_{\mu t}\frac{\rm dA_\mu(t)\rm d\hat{A}_\mu(t)}{2\pi} \rme^{\rmi\hat{A}_\mu(t)[A_\mu(t)-A_e(t)]+\rmi\sqrt{2}
\phi_\mu(t)A_\mu(t)}\Big]
\\
&&\hspace*{-10mm}\times
\Big\bra
 \int\!p_0(\bq(0))\Big[\prod_{i t}\frac{\rm dq_i(t)\rm d\hat{q}_i(t)}{2\pi} \rme^{\rmi\hat{q}_i(t)\left[
     q_i(t+1)-q_i(t)-\theta_i(t)
   \right]+\rmi\psi_i(t)s_i(t)}\Big] ~\Xi\Big\ket_{\!\bz}
   \nonumber
\end{eqnarray}
with
\begin{eqnarray}
\Xi
&=& \prod_{i\mu}\overline{\rme^{\frac{\rmi}{\sqrt{N}}\sum_t\{(R_1-R_2)\hat{q}_i(t)
      F[A_\mu(t)]-\frac{1}{2}\hat{A}_\mu(t)[R_1+R_2+(R_1-R_2) s_i(t)]\}}}
\nonumber
\\
&=& \prod_{\mu}
\rme^{ -\frac{1}{4}\!\sum_{tt^\prime}\hat{A}_\mu(t)\hat{A}_\mu(t^\prime)
-\sum_{tt^\prime}L_{tt^\prime}
      F[A_\mu(t)]  F[A_\mu(t^\prime)]
      -\frac{1}{4}\!\sum_{tt^\prime}C_{tt^\prime}\hat{A}_\mu(t)\hat{A}_\mu(t^\prime)}
      \nonumber
      \\
&&\hspace*{40mm} \times
      \rme^{\sum_{tt^\prime}K_{tt^\prime}\hat{A}_\mu(t)
      F[A_\mu(t^\prime)]
      +\order(N^{-1})}
\end{eqnarray}
where we
 introduced the temporary abbreviations
 \begin{eqnarray*}
 \hspace*{-10mm}
C_{t t^\prime}&=&\frac{1}{N}\sum_i s_i(t) s_i(t^\prime)~~~~~~ K_{t
t^\prime}=\frac{1}{N}\sum_i s_i(t) \hq_i(t^\prime)~~~~~~
 L_{t
t^\prime}=\frac{1}{N}\sum_i \hq_i(t) \hq_i(t^\prime)
\end{eqnarray*}
We isolate these as usual by inserting appropriate $\delta$-functions (in integral
representation, which generates conjugate integration
variables), and define the short-hands $\D
C=\prod_{t t^\prime}[\rmd C_{t t^\prime}/\sqrt{2\pi}]$ and $\D A=\prod_{t
}[\rmd A(t)/\sqrt{2\pi}]$ (with similar definitions
for the other kernels and functions). Substitution
into $\overline{Z[\bpsi,\bphi]}$, followed by
re-arranging terms, then leads us to
\begin{eqnarray}
\hspace*{-15mm}
 \overline{Z[\bpsi,\bphi]}
  &=&
  \int\![\D C \D\hat{C}][\D K \D\hat{K}][ \D L \D\hat{L}]~
 \rme^{\rmi N\sum_{tt^\prime}[\hat{C}_{tt^\prime}C_{t t^\prime}+\hat{K}_{t
t^\prime}K_{tt^\prime}+\hat{L}_{tt^\prime}L_{t t^\prime}]
+\order(\log(N))}
  \nonumber
  \\
\hspace*{-15mm}  && \times
\Big\bra
 \int\! \D\bq \D\hat{\bq}~p_0(\bq(0))~
  \rme^{\rmi\sum_{t i}\hq_i(t) [
     q_i(t+1) - q_i(t)-\theta_i(t)]+\rmi\sum_{i t}\psi_i(t) s_i(t)}
 \nonumber
 \\
 \hspace*{-15mm}
 &&\times~
\rme^{-\rmi\sum_i\sum_{tt^\prime}[\hat{C}_{t t^\prime}s_i(t)
s_i(t^\prime)+\hat{K}_{t t^\prime} s_i(t)
\hq_i(t^\prime)+\hat{L}_{t t^\prime} \hq_i(t) \hq_i(t^\prime)]}
\Big\ket_{\bz}
  \nonumber
  \\
  \hspace*{-15mm}
  &&\times
 \prod_\mu \int\!\D\bA \D\hat{\bA}~ \rme^{\rmi\sum_t\big[\hat{A}(t)[A(t)-A_e(t)]+\rmi\sqrt{2}
\phi_\mu(t)A(t)\big]-\frac{1}{4}\sum_{tt^\prime}\hat{A}(t)[1+C_{tt^\prime}]\hat{A}(t^\prime)}
   \nonumber
   \\
   \hspace*{-15mm}
   &&\hspace*{20mm} \times
\rme^{
-\sum_{tt^\prime}L_{tt^\prime}
      F[A(t)]  F[A(t^\prime)]
      +\sum_{tt^\prime}K_{tt^\prime}\hat{A}(t)
      F[A(t^\prime)]
      }
      \label{eq:Zbatchintermediate}
\end{eqnarray}
Upon
assuming simple initial conditions of the form $p_0(\bq)=\prod_i
p_{0}(q_i)$ we then arrive at
\begin{equation}\label{eq:Zafteraverage}
  \overline{Z[\bpsi,\bphi]}=
  \int[\D C \D\hat{C}][\D K \D\hat{K}][ \D L \D\hat{L}]~
    \rme^{N\left[\Psi+\Phi+\Omega \right]+\order(\log(N))}
\end{equation}
with
\begin{eqnarray}
\Psi&=&\rmi\sum_{tt^\prime}[ \hC_{tt^\prime} C_{tt^\prime} +\hK_{t
t^\prime} K_{tt^\prime} + \hL_{tt^\prime} L_{tt^\prime}]
\label{eq:Psi}
\\
\Phi&=& \frac{1}{N}\sum_\mu\log
\int\!\D\bA \D\hat{\bA}~ \rme^{\rmi\sum_t\big[\hat{A}(t)[A(t)-A_e(t)]+\rmi\sqrt{2}
\phi_\mu(t)A(t)\big]}
   \label{eq:Phi}
   \\
   &&\times
\rme^{-\frac{1}{4}\sum_{tt^\prime}\hat{A}(t)[1+C_{tt^\prime}]\hat{A}(t^\prime)
-\sum_{tt^\prime}L_{tt^\prime}
      F[A(t)]  F[A(t^\prime)]
      +\sum_{tt^\prime}K_{tt^\prime}\hat{A}(t)
      F[A(t^\prime)]
      }
\nonumber
\\
\Omega &=& \frac{1}{N}\sum_i\log \Big\langle \int\! \D q
\D\hat{q}~p_0(q(0)) \nonumber\\ &&  \times~
\rme^{\rmi\sum_{t}\left[\hq(t)[
     q(t+1) - q(t)-\theta_i(t)]+\psi_i(t) \sigma[q(t),z(t)] \right]-\rmi\sum_{t t^\prime}\hq(t)\hat{L}_{t t^\prime}  \hq(t^\prime)}
\nonumber \\[1mm] &&
 \times~ \rme^{-\rmi\sum_{t
t^\prime}[\hat{C}_{t t^\prime} \sigma[q(t),z(t)]
\sigma[q(t^\prime),z(t^\prime)] +\hat{K}_{t
t^\prime}\sigma[q(t),z(t)] \hq(t^\prime)]}\Big\rangle_{\!\bz}
\label{eq:Omega}
\end{eqnarray}
The $\order(\log(N))$ term in the exponent of
(\ref{eq:Zafteraverage}) is independent of
  $\{\psi_i(t),\phi_\mu(t),\theta_i(t)\}$, and
 the external bid $A_e(\ell)$ and the introduced valuation
update function $F[A]$ appear only in $\Phi$.

\subsection{Saddle-point equations and interpretation of order parameters}

The disorder-averaged functional
(\ref{eq:Zafteraverage}) is evaluated by steepest descent
integration, leading to coupled equations from which to
solve the dynamic order parameters
$\{C,\hat{C},K,\hat{K},L,\hat{L}\}$:
\begin{eqnarray}
 C_{t t^\prime}&=&\bra ~\sigma[q(t),z(t)]~
\sigma[q(t^\prime),z(t^\prime)]~ \ket_\star \label{eq:C1}\\[2mm]
 K_{t t^\prime}&=& \bra~ \sigma[q(t),z(t)]~ \hat{q}(t^\prime)~\ket_\star
\label{eq:K1}
\\[2mm]
 L_{tt^\prime}&=& \bra ~\hat{q}(t)\hat{q}(t^\prime)~\ket_\star
\label{eq:L1} \\[1mm] \hC_{tt^\prime}&=&\frac{\rmi\partial
\Phi}{\partial C_{t t^\prime}} ~~~~~~
\hK_{tt^\prime}=\frac{\rmi\partial \Phi}{\partial K_{t t^\prime}}
~~~~~~ \hL_{tt^\prime}=\frac{\rmi\partial \Phi}{\partial L_{t
t^\prime}}
 \label{eq:conjugates}
\end{eqnarray}
The notation $\bra\ldots\ket_\star$ in the above expressions is a short-hand for
\begin{eqnarray}
\bra g[\{q,\hat{q},z\}]\ket_\star &=& \lim_{N\to\infty}
\frac{1}{N}\sum_i \frac{\int\! \D q \D\hat{q}~\bra M_i[\{q,\hat{q},z\}]
g[\{q,\hat{q},z\}]\ket_{\bz}} {\int\! \D q \D\hat{q}~\bra
M_i[\{q,\hat{q},z\}]\ket_{\bz}}
\\[1mm]
M_i[\{q,\hat{q},z\}]&=& p_0(q(0))~
 \rme^{\rmi\sum_{t}\hq(t)[
     q(t+1) - q(t)-\theta_i(t)]+\rmi\sum_t\psi_i(t) \sigma[q(t),z(t)]}
 \nonumber \\ &&
 \times~\rme^{-\rmi\sum_{t
t^\prime}[\hat{C}_{t t^\prime} \sigma[q(t),z(t)]
\sigma[q(t^\prime),z(t^\prime)] +\hat{K}_{t
t^\prime}\sigma[q(t),z(t)] \hq(t^\prime)]}
 \nonumber \\ &&
 \times~\rme^{-\rmi\sum_{t
t^\prime}\hat{L}_{t t^\prime} \hq(t)\hq(t^\prime)}
 \label{eq:Mi}
\end{eqnarray}
Since the bid contribution $A_e(t)$ appears only in $\Phi$, one can take over
standard results from the theory of batch MGs with regard to interpretation and simplification of order parameters (see e.g.
\cite{Book2}).
Using the  normalization identity
$\lim_{\bpsi,\bphi\to\bnull}\overline{Z[\bpsi,\bphi]}=1$ we
find
\begin{eqnarray}
C_{tt^\prime}&=&
    \lim_{\bpsi,\bphi\to
\bnull} \bra
    \sigma[q(t),z(t)]\sigma[q(t^\prime),z(t^\prime)]\ket_\star
    \label{eq:CinM}
 \\
  G_{tt^\prime}
&=&
 -\rmi\lim_{\bpsi,\bphi\to\bnull} \bra
    \sigma[q(t),z(t)]\hq(t^\prime)\ket_\star
    \label{eq:GinM}
    \\
    0
    &=&-\lim_{\bpsi,\bphi\to
\bnull} \bra
    \hq(t)\hq(t^\prime)\ket_\star
    \label{eq:0inM}
 \end{eqnarray}
These expressions (\ref{eq:CinM},\ref{eq:GinM},\ref{eq:0inM})
are to be evaluated {\em at} the physical saddle-point of
$\Psi+\Omega+\Phi$. We conclude from (\ref{eq:GinM}) and
(\ref{eq:0inM}), in combination with
(\ref{eq:K1},\ref{eq:L1}) that for all $(t,t^\prime)$
\be
K_{tt^\prime}=\rmi G_{tt^\prime},~~~~~~~~ L_{tt^\prime}=0
\label{eq:KandL}
 \ee
We now send the fields $\{\psi_i,\phi_\mu\}$ to zero, and
choose the $\{\theta_i\}$ to be independent of $i$.
 The measure $M_i[\{q,\hq,z\}]$ then loses its
dependence on $i$, and equations
(\ref{eq:C1},\ref{eq:K1},\ref{eq:L1},\ref{eq:conjugates}) simplify to
\begin{eqnarray}
 C_{t t^\prime}&=&\bra ~\sigma[q(t),z(t)]~
\sigma[q(t^\prime),z(t^\prime)]~ \ket_\star \label{eq:C2}\\[2mm]
 G_{t t^\prime}&=& -\rmi\bra~ \sigma[q(t),z(t)]~ \hat{q}(t^\prime)~\ket_\star
\label{eq:G2}
\\[1mm]
\hC_{tt^\prime}&=&\lim_{L\to 0}\frac{\rmi\partial \Phi}{\partial C_{t
t^\prime}}, ~~~~~~ \hK_{tt^\prime}=\lim_{L\to 0}\frac{\rmi\partial
\Phi}{\partial K_{t t^\prime}}, ~~~~~~ \hL_{tt^\prime}=\lim_{L\to
0}\frac{\rmi\partial \Phi}{\partial L_{t t^\prime}}
 \label{eq:conjugates2}
\end{eqnarray}
with
\begin{eqnarray}
\bra g[\{q,\hat{q},z\}]\ket_\star &=&  \frac{\int\! \D q
\D\hat{q}~\bra M[\{q,\hat{q},z\}] g[\{q,\hat{q},z\}]\ket_{\bz}}
{\int\! \D q \D\hat{q}~\bra M[\{q,\hat{q},z\}]\ket_{\bz}}
\\[1mm]
M[\{q,\hat{q},z\}]&=& p_0(q(0))~
 \rme^{\rmi\sum_{t}\hq(t)[
     q(t+1) - q(t)-\theta(t)]-\rmi\sum_{t
t^\prime}\hat{L}_{t t^\prime} \hq(t)\hq(t^\prime)}
 \nonumber \\ &&
 \times~\rme^{-\rmi\sum_{t
t^\prime}[\hat{C}_{t t^\prime} \sigma[q(t),z(t)]
\sigma[q(t^\prime),z(t^\prime)] +\hat{K}_{t
t^\prime}\sigma[q(t),z(t)] \hq(t^\prime)]}
 \label{eq:M}
\end{eqnarray}

\subsection{Evaluation of $\Phi$}

We turn our attention to the function $\Phi$ (\ref{eq:Phi}),
which, according to (\ref{eq:conjugates2}), we only need to know
for small $L$.
We define the matrices $\one$ and $D$, with
 entries
$\one_{tt^\prime}=\delta_{tt^\prime}$ and $D_{tt^\prime}=
1+C_{tt^\prime}$, and
 the short-hands $\bA=\{A(t)\}$, $\bA_e=\{A_e(t)\}$ and $F[\bA]=\{F[A(t)]\}$. This allows us to write
\begin{eqnarray}
\Phi&=&
\frac{1}{N}\sum_\mu\log\int\!\rm d\bA \rm d\bxi~P(\bxi)\delta[\bA-\bA_e+G F[\bA]-\bxi] \nonumber
\\[-1mm]
  &&\hspace*{20mm}
  \times
\rme^{
\rmi\sqrt{2}\sum_t
\phi_\mu(t)A(t)-\sum_{tt^\prime}L_{tt^\prime}
      F[A(t)]  F[A(t^\prime)]}
\end{eqnarray}
with
\begin{eqnarray}
P(\bxi)&=&\frac{\prod_t(1/\sqrt{2\pi})}{\sqrt{{\rm det}[\frac{1}{2}D]}}\rme^{-\bxi\cdot \bD^{-1}\bxi}
\end{eqnarray}
At the saddle-point, $G$ must obey causality,
so we can interpret the above expression in terms of an effective stochastic process
for the bid, with a zero average Gaussian noise field $\xi(t)$:
\begin{eqnarray}
&&
A(t)= A_e(t)-\sum_{t^\prime<t}G_{tt^\prime}F[A(t^\prime)]+\xi(t)
\label{eq:bidproces1}
\\
&&
\bra \xi(t)\ket=0,~~~~~~\bra \xi(t)\xi(t^\prime)\ket=\frac{1}{2}D_{tt^\prime}
\label{eq:bidproces2}
\end{eqnarray}
The situation is clearly reminiscent of real history MGs \cite{RealHistory}, although the equations are simpler.
From now on we write averages over (\ref{eq:bidproces1},\ref{eq:bidproces2}) simply as $\bra \ldots\ket$.
We also define
\begin{eqnarray}
\Phi_0&=& \alpha\log\int\!\rm d\bA \rm d\bxi~P(\bxi)\delta[\bA-\bA_e+G F[\bA]-\bxi]
\end{eqnarray}
At the saddle-point we have $\Phi_0=0$.
We expand $\Phi$
for small $L$ and small $\bphi$:
\begin{eqnarray}
\Phi&=&
\Phi_0+
\frac{\rmi\sqrt{2}}{N}\sum_{\mu t}
\phi_\mu(t)\bra A(t)\ket -\alpha\sum_{tt^\prime}L_{tt^\prime}
     \bra F[A(t)]  F[A(t^\prime)]\ket
\nonumber
\\
&&
-\frac{1}{N}\sum_{\mu tt^\prime}
\phi_\mu(t)\phi_\mu(t^\prime)
[\bra A(t)A(t^\prime)\ket-\bra A(t)\ket\bra A(t^\prime)\ket]
+\order(\phi^3\!,L^2)
\end{eqnarray}
Clearly, $\lim_{\bphi\to\bnull,L\to\bnull} \Phi=\Phi_0$, and $\Phi_0$ depends on $G$ only, not on $C$ or $L$.
From our expansion of $\Phi$ we can extract all the quantities we are interested in.
We only have to be careful that causality can only be assumed {\em after} the differentiations:
\begin{eqnarray}
\hC_{tt^\prime}&=&\lim_{L\to 0,\bphi\to\bnull}\frac{\rmi\partial \Phi}{\partial C_{t
t^\prime}}=0
\label{eq:conj_final1}
\\
\hK_{tt^\prime}&=&\lim_{L\to 0,\bphi\to\bnull}\frac{\partial
\Phi}{\partial G_{t t^\prime}}=\frac{\partial
\Phi_0}{\partial G_{t t^\prime}}
= -\alpha \frac{\partial}{\partial A_e(t)}\bra F[A(t^\prime)]\ket
\label{eq:conj_final2}
\\
\hL_{tt^\prime}&=&\lim_{L\to
0}\frac{\rmi\partial \Phi}{\partial L_{t t^\prime}}=
-\rmi\alpha
     \bra F[A(t)]  F[A(t^\prime)]\ket
\label{eq:conj_final3}
\end{eqnarray}
We define the bid response function $R_{tt^\prime}=\partial \bra F[A(t)]\ket/\partial A_e(t^\prime)$
and the bid covariance function $\Sigma_{tt^\prime}=2\bra F[A(t)]  F[A(t^\prime)]\ket$,
so $\hK_{tt^\prime}=-\alpha R_{t^\prime t}$ and $\hL_{tt^\prime}=-\frac{1}{2}\rmi\alpha \Sigma_{tt^\prime}$.
Causality ensures that $R_{tt^\prime}=0$ for $t<t^\prime$, and $R_{tt}=1$.
We can also calculate bid statistics:
\begin{eqnarray}
\hspace*{-10mm}
\overline{\bra A^\mu[\bq(t),\bz(t)]\ket} &=
-\frac{\rmi}{\sqrt{2}}\lim_{\bpsi,\bphi\to\bnull}\frac{\partial \overline{Z[\bpsi,\bphi]}
}{\partial\phi_\mu(t)}
&= \bra A(t)\ket
\label{eq:bidaverage}
  \\
  \hspace*{-10mm}
\bra A^\mu[\bq(t),\bz(t)] A^\nu[\bq(t^\prime),\bz(t^\prime)]\ket
&= -\frac{1}{2}\lim_{\bpsi,\bphi\to\bnull}\frac{\partial^2 Z[\bpsi,\bphi]
}{\partial\phi_\mu(t)\partial\phi_\nu(t^\prime)}
&=\bra A(t)A(t^\prime)\ket
\end{eqnarray}

\subsection{The effective single agent equation}

The above results lead to a further simplification of our saddle-point equations
(\ref{eq:C2},\ref{eq:G2},\ref{eq:conjugates2}):
\begin{eqnarray}
 C_{t t^\prime}&=&\bra ~\sigma[q(t),z(t)]~
\sigma[q(t^\prime),z(t^\prime)]~ \ket_\star \label{eq:C3}\\[2mm]
 G_{t t^\prime}&=& -\rmi\bra~ \sigma[q(t),z(t)]~ \hat{q}(t^\prime)~\ket_\star
\label{eq:G3}
\end{eqnarray}
with
\begin{eqnarray}
\hspace*{-22mm}
\bra g[\{q,\hat{q},z\}]\ket_\star &=&  \frac{\int\! \D q
\D\hat{q}~\bra M[\{q,\hat{q},z\}] g[\{q,\hat{q},z\}]\ket_{\bz}}
{\int\! \D q \D\hat{q}~\bra M[\{q,\hat{q},z\}]\ket_{\bz}}
\label{eq:star_average2}
\\[1mm]
\hspace*{-22mm}
M[\{q,\hat{q},z\}]&=& p_0(q(0))
 \rme^{\rmi\sum_{t}\hq(t)[
     q(t+1) - q(t)-\theta(t)+\alpha\sum_{t^\prime}R_{tt^\prime}\sigma[q(t^\prime),z(t^\prime)]]-\frac{1}{2}\alpha\sum_{t
t^\prime}\hq(t)\Sigma_{tt^\prime}\hq(t^\prime)
}\nonumber
\\[-1mm]
\hspace*{-22mm}&&
 \label{eq:M2}
\end{eqnarray}
We eliminate the $\{\hq(t)\}$ by exploiting causality: the term
$\sum_{t^\prime}R_{tt^\prime}\sigma[q(t^\prime),z(t^\prime)]$
in (\ref{eq:M2})
involves only values of $q(t^\prime)$ with $t^\prime\leq t$. This
allows us to calculate the denominator of the
fraction (\ref{eq:star_average2}) by integrating out the variables
$\{q(t)\}$  iteratively, first over $q(t_{\rm max})$ (which gives
us $\delta[\hat{q}(t_{\rm max}-1)]$), followed by
integration over $q(t_{\rm max}-1)$, etc. The
result is simply
\begin{eqnarray}
\int\!\D q \D\hat{q}~M[\{q,\hat{q},z\}]&=&
\int\!\rmd q(0)~p_0(q(0))=1
\end{eqnarray}
This, in turn, implies that \bd \bra
\sigma[q(t),z(t)]\hq(t^\prime)\ket_\star=\rmi\frac{\partial}{\partial\theta(t^\prime)}\bra
\sigma[q(t),z(t)] \ket_\star \ed
 We do the remaining integrals over $\{\hq\}$,  and write
our equations in the simpler form
\begin{eqnarray}
 C_{t t^\prime}&=&\bra  \sigma[q(t),z(t)]
\sigma[q(t^\prime),z(t^\prime)]\ket_\star ~~~~~~~~
 G_{t t^\prime}= \frac{\partial\bra\sigma[q(t),z(t)] \ket_\star}{\partial\theta(t^\prime)}
\label{eq:CG}
\end{eqnarray}
with $\bra g[\{q,z\}]\ket_\star = \int\![\prod_t \rmd q(t)]~\bra
M[\{q,z\}] g[\{q,z\}]\ket_{\bz}$ and
\begin{eqnarray}
\hspace*{-15mm} M[\{q,z\}]
&=&p_0(q(0))\int\!\prod_t\Big[\frac{\rm d\eta(t)}{\sqrt{2\pi}}\Big]
\frac{\rme^{-\frac{1}{2}\sum_{tt^\prime}\eta(t)(\Sigma^{-1})_{tt^\prime}\eta(t^\prime)}}{\sqrt{\det
\Sigma}} \nonumber
\\
\hspace*{-15mm}
&& \times \prod_{t\geq 0}
\delta\Big[q(t+1)-q(t)-\theta(t)+\alpha\sum_{t^\prime}R_{tt^\prime}\sigma[q(t^\prime),z(t^\prime)]-\sqrt{\alpha}\eta(t)
\Big]
\label{eq:single_trader_measure}
\end{eqnarray}
We recognize that
(\ref{eq:single_trader_measure}) is the measure corresponding to a single-agent
process of the form
\begin{eqnarray}
\label{eq:singlebatchagent}
  q(t+1) &=& q(t)+\theta(t) - \alpha \sum_{t^\prime\leq t} R_{t t^\prime} \sigma[q(t^\prime),z(t^\prime)]
  +\sqrt{\alpha}~\eta(t)
\end{eqnarray}
in which
 $\eta(t)$ is a zero-mean Gaussian noise,  with
  temporal correlations
  $\bra\eta(t)\eta(t^\prime)\ket=\Sigma_{tt^\prime}$.
  The two kernels $\Sigma$ and $R$ are to be calculated from
  \begin{eqnarray}
  R_{tt^\prime}&=& \frac{\partial}{\partial A_e(t^\prime)} \bra F[A(t)]\ket,~~~~~~
\Sigma_{tt^\prime}=2\bra F[A(t)]  F[A(t^\prime)]\ket
\label{eq:RS}
\end{eqnarray}
where the averages refer to the process (\ref{eq:bidproces1},\ref{eq:bidproces2}).
 The
correlation and response functions
(\ref{eq:defineC},\ref{eq:defineG}), the order
parameters of our problem, are to be solved from (\ref{eq:CG}), in which $\bra \ldots\ket_\star$ now
denotes averaging over
(\ref{eq:singlebatchagent}) and the zero-average Gaussian
noise $\{z(t)\}$, with $\bra
z(t)z(t^\prime)\ket=\delta_{tt^\prime}$. These results represent a
 fully exact and closed theory for $N\to\infty$.
\vsp

As a simple test we could go back to the standard MG.
If we choose $F[A]=A$, the effective bid process (\ref{eq:bidproces1},\ref{eq:bidproces2}) becomes linear, and  is solved easily:
\begin{eqnarray}
A(t)&=& \sum_{t^\prime}(\one+G)^{-1}_{tt^\prime}[ A_e(t^\prime)+\xi(t^\prime)]
\label{eq:simple_bid_process}
\end{eqnarray}
so that we can calculate the kernels $\Sigma$ and $R$ explicitly:
  \begin{eqnarray}
  \hspace*{-10mm}
  R_{tt^\prime}&=& \frac{\partial}{\partial A_e(t^\prime)} \bra A(t)\ket
 =
  \frac{\partial}{\partial A_e(t^\prime)} \sum_{s}(\one+G)^{-1}_{ts} A_e(s)
 =(\one+G)^{-1}_{tt^\prime}
  \\
  \hspace*{-10mm}
\Sigma_{tt^\prime}&=& 2\bra A(t) A(t^\prime)\ket
=2\sum_{ss^\prime}(\one+G)^{-1}_{ts}(\one+G)^{-1}_{t^\prime s^\prime}\bra [A_e(s)+\xi(s)][A_e(s^\prime)+\xi(s^\prime)]\ket
\nonumber
\\
\hspace*{-10mm}
&=&
[(\one+G)^{-1}D[A_e](\one+G^\dag)^{-1}]_{tt^\prime}
\end{eqnarray}
with $D[A_e]_{tt^\prime}=2A_e(t) A_e(t^\prime)+1+C_{tt^\prime}$.
One confirms readily that this is the correct solution.

\section{Ergodic stationary states for $A_e(t)=A_e$}

We now take $A_e(t)=A_e$ for all $t$.
In time-translation
invariant stationary states
without long-term memory one has $G_{tt^\prime}=G(t-t^\prime)$,
$C_{tt^\prime}=C(t-t^\prime)$, $\Sigma_{t
t^\prime}=\Sigma(t-t^\prime)$, and $R_{tt^\prime}=R(t-t^\prime)$; all three
operators $\{C,G,\Sigma\}$ and their powers commute.
We try to calculate the four persistent order parameters,
 $\chi=\sum_{t>0}
G(t)$, $\chiR=\sum_{t\geq 0}R(t)$,  $c=\lim_{t\to\infty}C(t)$, and $S^2_0=\lim_{t\to\infty}\Sigma(t)$, from the
closed equations
(\ref{eq:CG},\ref{eq:RS}).
We assume a stationary state without
anomalous response, i.e. both $\chi$ and $\chiR$ are finite numbers.
From now on we will use the the
following notation for time averages:
$\overline{x}=\lim_{\tau\to\infty}\tau^{-1}\sum_{t=1}^{\tau}x(t)$.
Much of the analysis is standard, and we will where appropriate skip those details
that are familiar.

\subsection{Equations for persistent
order parameters - the effective agent process}

 We define
$\tilde{q}=\lim_{t\to\infty}q(t)/t$, assuming that this
limit exists, and send $t\to\infty$ in the integrated version of (\ref{eq:singlebatchagent}).
The result is
\begin{eqnarray}
  \tilde{q} &=& \sqrt{\alpha}~ \overline{\eta}+\overline{\theta} - \alpha \chiR \overline{\sigma}
  \label{eq:asymptotic_eqn}
\end{eqnarray}
Here  $\overline{\sigma}=
\lim_{\tau\to\infty}\tau^{-1}\sum_{t\leq\tau} \int\!Dz~ \sigma[\tilde{q}
t,z]$ and $Dz=(2\pi)^{-1/2}\rme^{-z^2/2}\rmd z$.  Given the properties of $\sigma[q]$, in cases where $\tilde{q}\neq 0$ we must
have $\overline{\sigma}=\sgn[\tilde{q}].\sigma[\infty]$;
for $\tilde{q}=0$ we know that $|\overline{\sigma}|\leq
\sigma[\infty]$. We inspect the possible solutions
`fickle' ($\tilde{q}=0$) versus
`frozen' ($\tilde{q}\neq 0$), noting
that in both cases we must have
$\sgn[\chiR\overline{\sigma}]=\sgn[\overline{\eta}\sqrt{\alpha}+\overline{\theta}]$. For the fickle solution the situations
is clear:
\begin{eqnarray}\hspace*{-5mm}
{\rm `fickle':}&~~~~~~
 \frac{\overline{\sigma}}{\sigma[\infty]} = \frac{\sqrt{\alpha}~ \overline{\eta}+\overline{\theta}}{\alpha \chiR\sigma[\infty]}
  ~~~~~~~~& {\rm
  exists~if~~~~}|\overline{\eta}\sqrt{\alpha}+\overline{\theta}|\leq
\sigma[\infty]|\chiR| \alpha
~~~\label{eq:condition_fickle}
\end{eqnarray}
For the frozen solution
we must distinguish between the cases $\chiR>0$ versus $\chiR<0$, representing negative versus positive feedback in the effective
 agent equation:
\begin{itemize}
\item $\chiR>0$:
\\[1mm]
Here, due to the absence of positive feedback in the system, we can write directly
\begin{eqnarray}
\hspace*{-15mm}
|\overline{\eta}\sqrt{\alpha}+\overline{\theta}|>
\sigma[\infty]\chiR \alpha:
&~~~& {\rm `frozen'~solution},~~~
\overline{\sigma}=\sigma[\infty]~\sgn[\overline{\eta}\sqrt{\alpha}+\overline{\theta}]
\label{eq:condition_frozen_chipos}
\\
\hspace*{-15mm}
|\overline{\eta}\sqrt{\alpha}+\overline{\theta}|\leq
\sigma[\infty]\chiR \alpha:
&~~~& {\rm `fickle'~solution},~~~~~
\overline{\sigma} = [\sqrt{\alpha}~ \overline{\eta}+\overline{\theta}]/\alpha \chiR
\label{eq:condition_fickle_chipos}
\end{eqnarray}
\item $\chiR<0$:
\\[1mm]
As was found earlier in other MG versions with positive feedback, the effective agent equation now allows for remanence effects,
leading to the potential for multiple solutions:
\begin{eqnarray}
\hspace*{-15mm}
|\overline{\eta}\sqrt{\alpha}+\overline{\theta}|>
\sigma[\infty]|\chiR| \alpha:
&~~~& {\rm `frozen'~solution},~~~
\overline{\sigma}=\sigma[\infty]~\sgn[\overline{\eta}\sqrt{\alpha}+\overline{\theta}]
\label{eq:condition_frozen_chineg}
\\
\hspace*{-15mm}
|\overline{\eta}\sqrt{\alpha}+\overline{\theta}|\leq
\sigma[\infty]|\chiR| \alpha:
&~~~& \left\{\begin{array}{ll}
{\rm `frozen' ~solutions},& ~~~\overline{\sigma}=\pm \sigma[\infty]
\\
{\rm `fickle'~solution},&~~~
\overline{\sigma} = [\sqrt{\alpha}~ \overline{\eta}+\overline{\theta}]/\alpha \chiR
\end{array}
\right.
\label{eq:condition_fickle_chineg}
\end{eqnarray}
For $|\overline{\eta}\sqrt{\alpha}+\overline{\theta}|\leq
\sigma[\infty]|\chiR| \alpha$ there are two stable solutions $\overline{\sigma}=\pm \sigma[\infty]$
separated by the unstable `fickle' one. The two branches $\overline{\sigma}=-\sigma[\infty]~\sgn[\overline{\eta}\sqrt{\alpha}
+\overline{\theta}]$ in this region must be remanent ones, and the  Maxwell construction tells us to choose in the stationary
state
\begin{eqnarray}
\chiR<0:
&~~~& {\rm `frozen'~solution},~~~
\overline{\sigma}=\sigma[\infty]~\sgn[\overline{\eta}\sqrt{\alpha}+\overline{\theta}]
\label{eq:chineg}
\end{eqnarray}
\end{itemize}
\vsp

\noindent
The variance of $\overline{\eta}$ is simply equal to $S^2_0$, since
\begin{eqnarray}
\bra \overline{\eta}^2 \ket
&=&\lim_{\tau\to\infty}\frac{1}{\tau^2}\sum_{tt^\prime=0}^{\tau}\Sigma_{tt^\prime}
= \lim_{t\to\infty}\Sigma(t)=S^2_0
\label{eq:persistent_eta}
\end{eqnarray}
We can now calculate equations for the persistent order parameters, sending $\overline{\theta}\to 0$
 as soon as possible.
As before we have to distinguish between $\chiR>0$ and $\chiR<0$:
\begin{itemize}
\item $\chiR>0$:
\\[1mm]
The persistent correlations follow from
$c=\bra\overline{\sigma}^2 \ket_\star$:
\begin{eqnarray}
\hspace*{-5mm}
c&=& \int\!\rmd\overline{\eta} ~P(\overline{\eta})\left\{
\theta\Big[\chiR\sigma[\infty]\sqrt{\alpha}\!-\!
|\overline{\eta}|\Big]
\frac{\overline{\eta}^2}{\alpha \chiR^2} +\theta\Big[
|\overline{\eta}|\!-\!
\chiR\sigma[\infty]\sqrt{\alpha}\Big] \sigma^2[\infty]
\right\} \nonumber
\\
\hspace*{-5mm}
&=&
\frac{2 S^2_0}{\alpha \chiR^2} \Big\{\frac{1}{2}{\rm Erf}[\frac{\chiR\sigma[\infty]\sqrt{\alpha}}{\sqrt{2S^2_0}}]-
\frac{\chiR\sigma[\infty]\sqrt{\alpha}}{\sqrt{2\pi S^2_0}}\rme^{-\frac{\alpha \chiR^2\sigma^2[\infty]}{2S^2_0}}
\Big\}
\nonumber
\\
\hspace*{-5mm}
&&
\hspace*{20mm}
+2\sigma^2[\infty]\Big\{\frac{1}{2}-\frac{1}{2}{\rm Erf}[\frac{\chiR\sigma[\infty]\sqrt{\alpha}}{\sqrt{2S^2_0}}]\Big\}
\label{eq:batch_c}
\end{eqnarray}
The frozen fraction $\phi$ and the susceptibility $\chi$ are calculated similarly:
\begin{eqnarray}
\phi&=& \int\!\rmd\overline{\eta} ~P(\overline{\eta})
\theta\Big[
|\overline{\eta}|\!-\!
\chiR\sigma[\infty]\sqrt{\alpha}\Big]  =1-{\rm Erf}[\frac{\chiR\sigma[\infty]\sqrt{\alpha}}{S_0\sqrt{2}}]
\\
\chi&=& \int\!\rmd\overline{\eta} ~P(\overline{\eta})\frac{\partial}{\partial\overline{\theta}} \overline{\sigma}
 =\frac{1}{\alpha\chiR}
{\rm Erf}[\frac{\chiR\sigma[\infty]\sqrt{\alpha}}{S_0\sqrt{2}}]
\end{eqnarray}
In terms of the usual short-hand $v=\chiR\sigma[\infty]\sqrt{\alpha}/S_0\sqrt{2}$,
we then  arrive at
\begin{eqnarray}
c&=&\sigma^2[\infty]\Big\{
 1+\frac{1-2v^2}{2v^2} {\rm Erf}[v]
-\frac{1}{v\sqrt{\pi}} \rme^{-v^2}\Big\}
\label{eq:c}
\\
\phi&=& 1-{\rm Erf}[v]
\label{eq:phi}
\\
\chi&=&
{\rm Erf}[v]/\alpha\chiR
\label{eq:chi}
\end{eqnarray}
\item $\chiR<0$:
\\[1mm]
This situation is simpler:
one has $c=\bra
\overline{\sigma}^2 \ket_\star=\sigma^2(\infty)$, and $\phi=1$.
The susceptibility $\chi$ for $\overline{\theta}=0$ becomes
\begin{eqnarray}
\chi&=& \lim_{\overline{\theta}\to 0}\int\!\rmd\overline{\eta} ~P(\overline{\eta})\frac{\partial}{\partial\overline{\theta}}
\overline{\sigma}
~=~ \frac{2\sigma[\infty]}{\sqrt{2\pi\alpha S_0^2}}
\label{eq:chiabnormal}
\end{eqnarray}
\end{itemize}
\vsp

\noindent
If $\chiR<0$ the system is fully frozen, and nothing further happens.
To close our persistent order parameter equations for $c$ and $\phi$ if $\chiR>0$, we need the
 ratio $\chiR/S_0$; to get also $\chi$ we need $\chiR$ and $S_0$.  We now take $A_e(t)=A_e$, and
define the asymptotic time averages $\overline{A}=\lim_{\tau\to\infty}\tau^{-1}\sum_{t\leq \tau}A(t)$ and $\overline{F}=
\lim_{\tau\to\infty}\tau^{-1}\sum_{t\leq \tau}F[A(t)]$. This allows us to write
\begin{eqnarray}
&&
\chiR=\partial\bra \overline{F}\ket/\partial A_e,~~~~~~~~
S_0=2\bra
\overline{F}^2\ket
\end{eqnarray}

\subsection{Analysis of the effective overall bid process}

Closing our stationary state equations requires extracting the values of $\chiR$ and $S_0$ from the effective process
(\ref{eq:bidproces1},\ref{eq:bidproces2}) for the overall bids.
We separate in the bid noise $\xi(t)$ and the bids $A(t)$ the persistent from the non-persistent terms:
\begin{eqnarray}
&& A(t)=\overline{A}+\tilde{A}(t),~~~~~~\xi(t)=z\sqrt{\frac{1}{2}(1+c)}+\tilde{\xi}(t)
\end{eqnarray}
Here $z$ is a zero-average unit-variance frozen Gaussian variable, and $\tilde{\xi}(t)$ is also a zero-average Gaussian variable, uncorrelated with $z$
and with covariances $\bra\tilde{\xi}(t)\tilde{\xi}(t^\prime)\ket=\frac{1}{2}\tilde{C}(t-t^\prime)$. Here $\tilde{C}(t)=C(t)-c$.
Our bid process (\ref{eq:bidproces1}) now becomes
\begin{eqnarray}
&&
\overline{A}= A_e-\chi \overline{F}+z\sqrt{\frac{1}{2}(1+c)}
\label{eq:bid_eqn_1}
\\
&&
\tilde{A}(t)= \tilde{\xi}(t) -\sum_{s>0}G(s)\Big\{F[\overline{A}+\tilde{A}(t-s)]-\overline{F}\Big\}
\label{eq:bid_eqn_2}
\end{eqnarray}
The two quantities $\overline{A}$ and $\overline{F}$ are both
 parametrized
by $z$, so we write $\overline{A}(z)$ and $\overline{F}(z)$. Also the non-persistent bid parts $\tilde{A}(t)$
depend on $z$ since $\overline{A}$ occurs in (\ref{eq:bid_eqn_2}), so we write $\tilde{A}(t,z)$.
Our static objects $S_0$ and $\chiR$ are known once we have $\overline{F}(z)$. In (\ref{eq:bid_eqn_1}) we have already one relation for the
two objects, so we need one more equation connecting $\overline{A}(z)$ to $\overline{F}(z)$ to obtain closed formulae.
To get this second relation we work out $\overline{F}(z)$:
\begin{eqnarray}
\hspace*{-10mm}
\overline{F}(z)&=& \int\!\rmd\tilde{A}~W(\tilde{A}|z)F[\overline{A}(z)\!+\!\tilde{A}],~~~~~~
W(a|z)=\lim_{\tau\to\infty}\frac{1}{\tau}\sum_{t\leq\tau}\delta[a\!-\!\tilde{A}(t,z)]
\end{eqnarray}
So far our analysis is direct and fully exact.
 What is left is to find the statistics $W(a|z)$ of the non-frozen bid contributions, which requires ans\"{a}tze.

To calculate $W(a|z)$ we assume the response function $G$ to decay much more slowly than the times over which the
$\tilde{A}(t,z)$ are correlated, so that to the time summation in (\ref{eq:bid_eqn_2}) we can apply the central limit theorem.
This tells us that also the $\tilde{A}(t,z)$ must be zero-average Gaussian.
Let us define the covariance matrix $\Xi_{tt^\prime}(z)=\bra \tilde{A}(t,z)\tilde{A}(t^\prime,z)\ket$, it must be time-translation
invariant, so we write $\Xi(t,z)=\Xi_{s+t,s}(z)$.  In \ref{app:messypart} we show that
\begin{eqnarray}
\Xi(t,z)&=&\frac{1}{2}\tilde{C}(t)+\order(\tau_C/\tau_G)
\end{eqnarray}
 If, as in earlier MG analyses \cite{Book2}, we can rely on $\lim_{N\to\infty}\tau_C/\tau_G=0$ in the ergodic regime (this will be our present ansatz)
then we have simply $\Xi(t,z)=
\frac{1}{2}\tilde{C}(t)$, and in particular $\Xi(0,z)=\frac{1}{2}(1-c)$, which closes our equations:
\begin{eqnarray}
\hspace*{-10mm}
\overline{A}(z)= A_e\!-\!\chi \overline{F}(z)\!+\!z\sqrt{\frac{1}{2}(1\!+\!c)},~~~~~~
\overline{F}(z)= \int\!Dx~F[\overline{A}(z)\!+\!x\sqrt{\frac{1}{2}(1\!-\!c)}]
\label{eq:Abar}
\end{eqnarray}
with the shorthand $D x =(2\pi)^{-1/2}\rme^{-x^2/2}\rmd x$. We eliminate $\overline{A}(z)$ and get $\overline{F}(z)=f(z,A_e)$, where $f(z,A_e)$ is the solution of the fixed-point equation
\begin{eqnarray}
f= \int\!Dx~F\Big[A_e-\chi f+z\sqrt{\frac{1}{2}(1+c)}+x\sqrt{\frac{1}{2}(1-c)}\Big]
\end{eqnarray}
The shows that $\overline{F}(z)=\Phi(A_e+z\sqrt{\frac{1}{2}(1+c)})$, where $\Phi(u)$ is to be solved from
\begin{eqnarray}
\Phi(u)= \int\!Dx~F\Big[u-\chi\Phi(u)+x\sqrt{\frac{1}{2}(1-c)}\Big]
\label{eq:PhiMap}
\end{eqnarray}
Clearly $\Phi(u)$ is anti-symmetric, since $F[A]$ is anti-symmetric.
Finally, we will solve the nonlinear functional equation (\ref{eq:PhiMap}). To do this we define a new function $\Delta(z)$ via the identity
$\Phi(u)= [u-\Delta^{-1}(u)]/\chi$,
insertion into (\ref{eq:PhiMap}) of which gives after some rewriting
\begin{eqnarray}
\Delta(z)&=& z+\chi\int\!Dx~F\Big[z+x\sqrt{\frac{1}{2}(1-c)}\Big]
\label{eq:Delta2}
\end{eqnarray}
Equation (\ref{eq:Delta2}) is explicit, but shows  that in the case of
positive feedback (as for the majority game, corresponding to $F[A]=-A$) there is again the possibility of multiple solutions,
which would here take the form of non-invertibility of the function $\Delta(z)$. Non-invertibility is signaled by finding $\Delta^\prime(z)=0$
for finite $z$, i.e. by
\begin{eqnarray}
 1+\frac{\chi\sqrt{2}}{\sqrt{1-c}}\int\!Dx~xF\Big[z+x\sqrt{\frac{1}{2}(1-c)}\Big]= 0
 \label{eq:invertibility}
\end{eqnarray}

\subsection{Equations for $\chiR$ and $S_0$ - closure of the stationary state theory}

Given the solution of (\ref{eq:PhiMap}),  we obtain
closure of our stationary state equations:
\begin{eqnarray}
\chiR&=&\frac{\sqrt{2}}{\sqrt{1+c}}\int\!Dz~ z\Phi\Big(A_e+z\sqrt{\frac{1}{2}(1+c)}\Big)
\label{eq:ChiR1}
\\
S_0^2&=&2\int\!Dz~\Phi^2\Big(A_e+z\sqrt{\frac{1}{2}(1+c)}\Big)
\label{eq:S01}
\end{eqnarray}
When expressing these equations in terms of the function $\Delta^{-1}(z)$ (which is parametrized by $c$ and $\chi$) we find that both are
expressed in terms of the following Gaussian integrals:
\begin{eqnarray}
I_0(c,\chi)&=&\int\!Dz~\Delta^{-1}\Big(A_e+z\sqrt{\frac{1}{2}(1+c)}\Big)
\label{eq:I0}
\\
I_1(c,\chi)&=&\int\!Dz~z\Delta^{-1}\Big(A_e+z\sqrt{\frac{1}{2}(1+c)}\Big)
\label{eq:I1}
\\
I_2(c,\chi)&=&\int\!Dz~\Big[\Delta^{-1}\Big(A_e+z\sqrt{\frac{1}{2}(1+c)}\Big)\Big]^2
\label{eq:I2}
\end{eqnarray}
We note that the validity of the last step depends crucially on
non-invertibility issues being absent or resolved. Substitution of (\ref{eq:Delta2}) into (\ref{eq:ChiR1},\ref{eq:S01}), followed by
re-arrangements, gives
\begin{eqnarray}
\chiR&=&\frac{1}{\chi}\Big[1-\frac{\sqrt{2}}{\sqrt{1\!+\!c}}I_1(c,\chi)\Big]
\label{eq:ChiR2}
\\
S_0^2
&=& \frac{2}{\chi^2}\Big[A^2_e-2A_e I_0(c,\chi)+\frac{1}{2}(1\!+\!c)+I_2(c,\chi)
 -\sqrt{2(1\!+\!c)}I_1(c,\chi) \Big]
 \label{eq:S02}
\end{eqnarray}
For $A_e=0$ one has $I_0(c,\chi)=0$ and the above integrals simplify.
The equations (\ref{eq:ChiR2},\ref{eq:S02}) are to be solved in combination with
(\ref{eq:c},\ref{eq:chi}) for $c$ and $\chi$. If this results in $\chiR>0$, the problem is solved and the observable $\phi$ follows from
(\ref{eq:phi}). As soon as $\chiR\leq 0$, we enter the fully frozen state $c=\phi=1$ induced by positive feedback in the
valuation dynamics.
From (\ref{eq:Abar}) we can also extract an expression for a static overall bid
susceptibility $\chi_{A}(z)=\partial \overline{A}(z)/\partial A_e$.

\subsection{Simple model examples}
\label{sec:simple_examples}

At this stage it is appropriate to inspect specific choices for $F[A]$, to serve as tests.
 We set $A_e=0$ for simplicity; the external bids were needed to calculate the overall bid susceptibility
 $\chiR$, but are no longer essential.
\begin{itemize}
\item $F[A]=A$, the standard Minority Game:
\\[1mm]
Here $\Delta(z)=z(1\!+\!\chi)$, so $\Delta^{-1}(z)=z/(1\!+\!\chi)$
and the Gaussian integrals $I_\ell(c,\chi)$ become
\begin{eqnarray}
&&
I_0(c,\chi)=0,~~~~~~
I_1(c,\chi)=\frac{\sqrt{1\!+\!c}}{\sqrt{2}(1\!+\!\chi)},~~~~~~
I_2(c,\chi)=\frac{1\!+\!c}{2(1\!+\!\chi)^2}
\end{eqnarray}
This then reproduces the correct relations
\begin{eqnarray}
&&
\chiR=1/(1\!+\!\chi),~~~~~~~~
S_0
= \sqrt{1\!+\!c}/(1\!+\!\chi)
\end{eqnarray}
The onset of non-invertibility is according to (\ref{eq:invertibility}) marked
by $1+\chi=0$ (which never happens in the ergodic phase of the standard MG, where $\chi\geq 0$).
Since always $\chiR>0$ we never enter the fully frozen (remanent) state obtained via the Maxwell construction.

\item $F[A]=-A$, the standard Majority Game:
\\[1mm]
Here $\Delta(z)= z(1\!-\!\chi)$, so $\Delta^{-1}(z)=z/(1\!-\!\chi)$
and the Gaussian integrals $I_\ell(c,\chi)$ become
\begin{eqnarray}
&&
I_0(c,\chi)=0,~~~~~~
I_1(c,\chi)=\frac{\sqrt{1\!+\!c}}{\sqrt{2}(1\!-\!\chi)},~~~~~~
I_2(c,\chi)=\frac{1\!+\!c}{2(1\!-\!\chi)^2}
\end{eqnarray}
The result for $\chiR$ and $S_0$ is
\begin{eqnarray}
&&
\chiR=1/(\chi\!-\!1),~~~~~~~~
S_0
= \sqrt{1\!+\!c}/|\chi\!-\!1|
\label{eq:maj}
\end{eqnarray}
If $\chiR>0$ (no remanence) one extracts from (\ref{eq:chi}) and (\ref{eq:maj}) that $\chi={\rm Erf}(v)/[{\rm Erf}(v)\!-\!\alpha]$,
so for $\alpha<1$ we can be sure that $\chi<0$ and run into the contradiction $\chiR<0$. Apparently, the general scenario is that where $\chiR<0$ and
 the system is in the fully frozen remanent state. Hence $\chi<1$, and its value is given by formula (\ref{eq:chiabnormal}), which reduces to
\begin{eqnarray}
\chi&=&\Big[ 1+\frac{\sqrt{2\pi\alpha(1\!+\!c)}}{2\sigma[\infty]}\Big]^{-1}\in(0,1)
\end{eqnarray}
The condition for leaving the frozen remanent state, viz. $\chi=1$, is seen to coincide with the condition for having non-invertibility for
 the overall bid process, but this condition will clearly never be met; the system is {\em always} in the fully frozen remanent state.
\end{itemize}

\section{Applications - greedy versus cautious contrarians}

We now apply our theory to specific choices for the function $F[A]$,
all corresponding to models that so far could only be studied via numerical simulations. For simplicity we choose $A_e=0$.

\subsection{Preparation}

Due to $A_e=0$, the equations (\ref{eq:ChiR2},\ref{eq:S02}) that close our equations for persistent order
 parameters simplify to
\begin{eqnarray}
\chi\chiR&=& 1-\sqrt{2}~I_1(c,\chi)/\sqrt{1\!+\!c}
\label{eq:ChiR3}
\\
|\chi| S_0
&=&\sqrt{1+c+2I_2(c,\chi)
 -2\sqrt{2(1\!+\!c)}I_1(c,\chi)}
 \label{eq:S03}
\end{eqnarray}
with
\begin{eqnarray}
&&\hspace*{-15mm}
I_1(c,\chi)=\int\!Dz~z\Delta^{-1}(z\sqrt{\frac{1}{2}(1\!+\!c)}),
~~~~~~
I_2(c,\chi)=\int\!Dz~\Big[\Delta^{-1}(z\sqrt{\frac{1}{2}(1\!+\!c)})\Big]^2
\label{eq:I1I2new}
\end{eqnarray}
and with $\Delta(z)$ as given in (\ref{eq:Delta2}). It will prove efficient to define the function
\begin{eqnarray}
c(v)&=&\sigma^2[\infty]\Big\{
 1+\frac{1-2v^2}{2v^2} {\rm Erf}(v)
-\frac{1}{v\sqrt{\pi}} \rme^{-v^2}\Big\}
\label{eq:cv}
\end{eqnarray}
Combining our relations so far then allows us to conclude that
ergodic solutions with $\chiR>0$ follow from solving the following two coupled equations for the basic unknowns $(v,\chi)$, after which the
order parameters $\phi$ and $c$ follow via $\phi=1-{\rm Erf}(v)$ and $c=c(v)$ (so always $v\geq 0$):
\begin{eqnarray}
&&\hspace*{-15mm}
\frac{\sigma[\infty]\sqrt{{\rm Erf}(v)}}{v\sqrt{2[1+c(v)]}}
=\sgn(\chi)\Big\{\frac{1+c(v)+2I_2(c(v),\chi)
 -2\sqrt{2(1\!+\!c(v))}I_1(c(v),\chi)}{1+c(v)-\sqrt{2[1\!+\!c(v)]}I_1(c(v),\chi)}\Big\}^{\frac{1}{2}}
\\
&&\hspace*{-15mm}
\alpha= \frac{{\rm Erf}(v)}
{1-\sqrt{2}I_1(c(v),\chi)/\sqrt{1\!+\!c(v)}}
\end{eqnarray}
Only $\chi> 0$ is possible, so we are allowed to introduce a further function $U(v)\geq 0$,
\begin{eqnarray}
U(v)&=&\frac{\sigma^2[\infty]{\rm Erf}(v)}{2[1\!+\!c(v)]v^2}
\label{eq:Uv}
\end{eqnarray}
and compactify our equations for $(v,\chi)$, describing solutions with $\chi>0$ and $\chiR>0$, to
\begin{eqnarray}
U(v)
&=& \frac{1+\frac{I_2(c(v),\chi)}{[1+c(v)]/2}
 -2\frac{I_1(c(v),\chi)}{\sqrt{[1+c(v)]/2}}}{1-\frac{I_1(c(v),\chi)}{\sqrt{[1+c(v)]/2}}}
~~~~~~~~
\alpha= \frac{{\rm Erf}(v)}
{1-\frac{I_1(c(v),\chi)}{\sqrt{[1+c(v)]/2}}}
\label{eq:compact_with_U}
\end{eqnarray}
One proves from the definitions of $I_{1,2}(c,\chi)$ that always $I_2(c,\chi)
\geq  2|I_1(c,\chi)|\sqrt{[1+c(v)]/2}$, so the numerator of the first of our compact equations is always nonnegative.
To have a solution with $\chiR>0$ we must demand that the denominator is also positive.

In contrast, in states with $\chiR<0$ one has  $c=\sigma^2(\infty)$, $\phi=1$ i.e. the system is always fully frozen. Here the only order
parameter left to be calculated is $\chi$, which follows from
\begin{eqnarray}
\hspace*{-10mm}
\frac{2\sigma[\infty]}{\sqrt{2\pi\alpha[1+\sigma^2(\infty)]}}&=& \sgn(\chi)\sqrt{1+\frac{I_2(\sigma^2(\infty),\chi)}{[1\!+\!\sigma^2(\infty)]/2}
-2\frac{I_1(\sigma^2(\infty),\chi)}{\sqrt{[1\!+\!\sigma^2(\infty)]/2}}}
\end{eqnarray}
Again only $\chi>0$ is possible (which we may rely upon in the rest of this paper), hence
 \begin{eqnarray}
 \frac{2\sigma^2[\infty]}{\pi\alpha[1+\sigma^2(\infty)]}&=& 1+\frac{I_2(\sigma^2(\infty),\chi)}{[1\!+\!\sigma^2(\infty)]/2}
 -2\frac{I_1(\sigma^2(\infty),\chi)}{\sqrt{[1\!+\!\sigma^2(\infty)]/2}}
 \end{eqnarray}
  To confirm that indeed $\chiR<0$ we must verify the outcome of
\begin{eqnarray}
\chiR&=& \chi^{-1}\Big[1-\frac{\sqrt{2}}{\sqrt{1\!+\!\sigma^2(\infty)}}I_1(\sigma^2(\infty),\chi)\Big]
\end{eqnarray}

\subsection{Invertible overall bid impact functions}

\begin{figure}[h]
\vspace*{0mm} \hspace*{-30mm} \setlength{\unitlength}{0.52mm}
\begin{picture}(250,100)

  \put(110,0){\includegraphics[height=100\unitlength,width=140\unitlength]{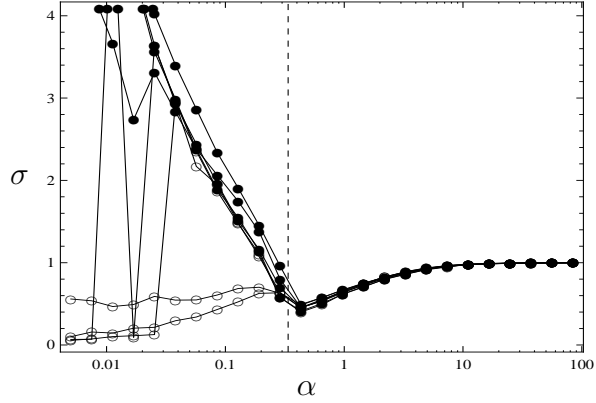}}
  \put(103,50){\large\here{$\sigma$}}  \put(174,-6){\large$\alpha$}

\end{picture}
 \vspace*{5mm}
\caption{Simulation results for the volatility $\sigma$ of MG-type
models with $F[A]=\sgn[A]|A|^\gamma$, for $\gamma = 0,1,2,3$. Vertical dashed line:
predicted location of the  $\chi\to\infty$ transition.
 Empty/full markers correspond to biased/unbiased initial conditions. The location of the phase transition seems indeed independent of $\gamma$.
 Even for $\alpha>\alpha_c$ the volatility appears
 only weakly dependent upon the greed exponent $\gamma$. However, for values of $\gamma\geq 4$ (excessive agent greed)
 the  ergodic region is destroyed and the efficient phase of the market vanishes
 (see fig \ref{fig:gammalarge}).
}\label{fig:invertible1}
\end{figure}
\begin{figure}[h]
\vspace*{2mm} \hspace*{-9mm} \setlength{\unitlength}{0.52mm}
\begin{picture}(300,100)

  \put(10,0){\includegraphics[height=100\unitlength,width=120\unitlength]{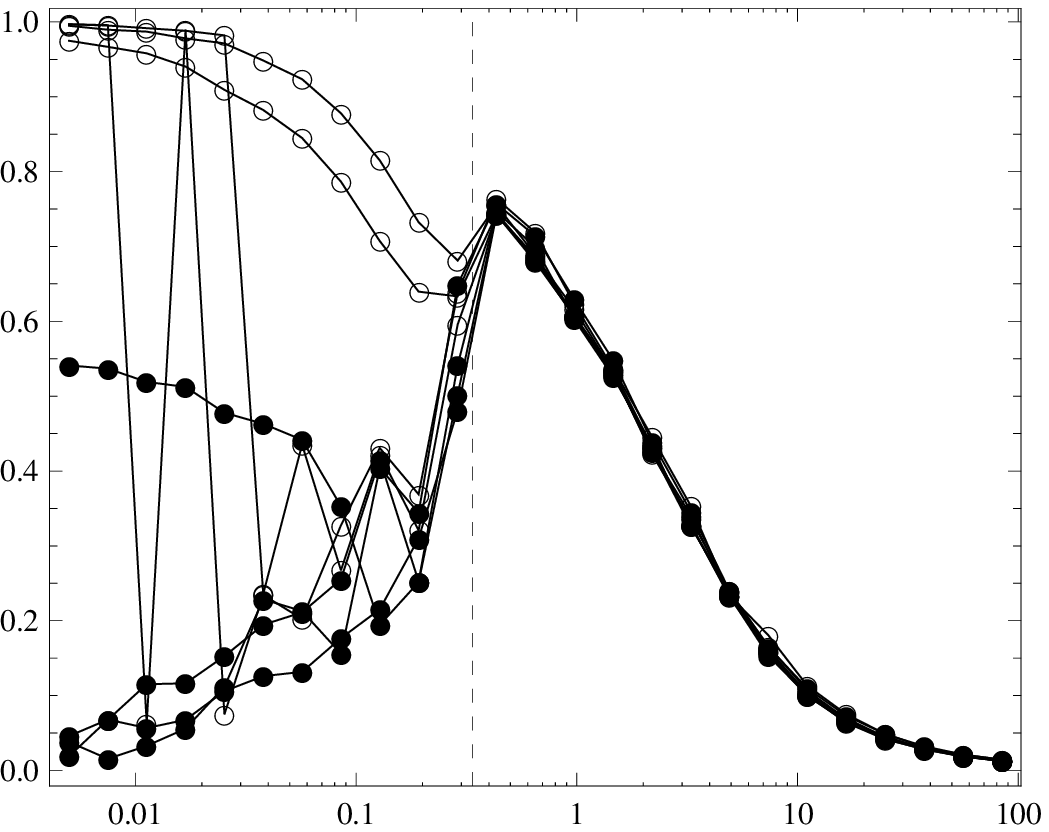}}
  \put(3,55){\large\here{$c$}}  \put(64,-6){\large$\alpha$}

  \put(150,0){\includegraphics[height=100\unitlength,width=120\unitlength]{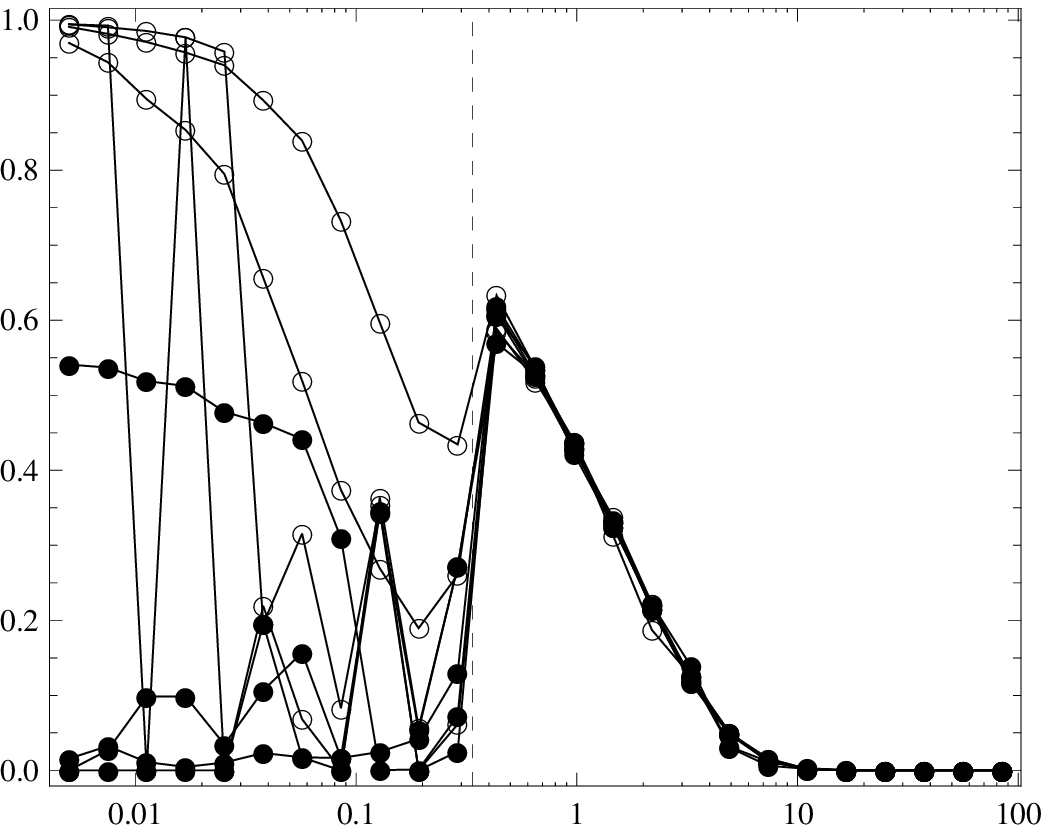}}
  \put(143,55){\large\here{$\phi$}}  \put(204,-6){\large$\alpha$}

\end{picture}
 \vspace*{1mm}
\caption{Simulation results for the fraction of frozen agents $\phi$ and correlations $c$ for MG-type
models with $F[A]=\sgn[A]|A|^\gamma$, for $\gamma = 0,1,2,3$. Vertical dashed line:
predicted location of the $\chi\to\infty$ transition.
 Empty/full markers correspond to biased/unbiased initial conditions. As observed for the volatility, also $\phi$ and $c$  appear
to be only weakly dependent on $\gamma$ in the ergodic region $\alpha<\alpha_c$.
}
\label{fig:invertible2}
\end{figure}
\begin{figure}[h]
\vspace*{0mm} \hspace*{-10mm} \setlength{\unitlength}{0.52mm}
\begin{picture}(300,100)

  \put(10,0){\includegraphics[height=100\unitlength,width=120\unitlength]{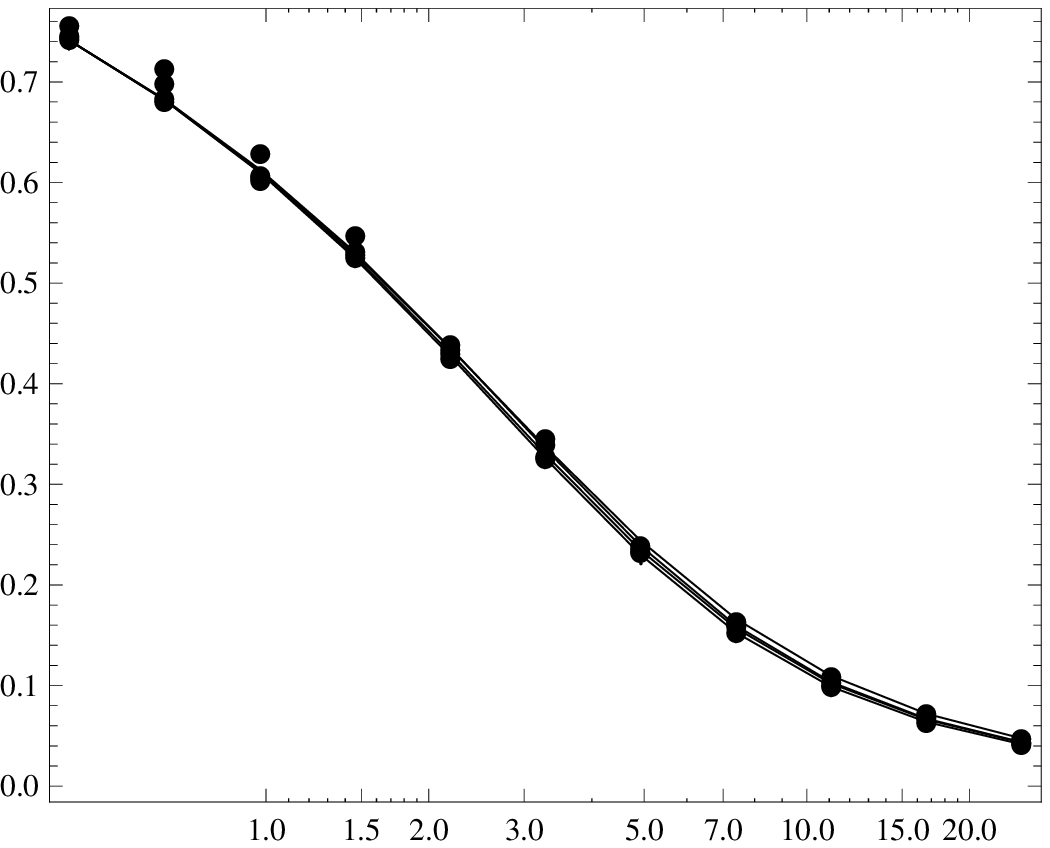}}
  \put(4,55){\large\here{$c$}}  \put(64,-11){\large$\alpha$}

  \put(150,0){\includegraphics[height=100\unitlength,width=120\unitlength]{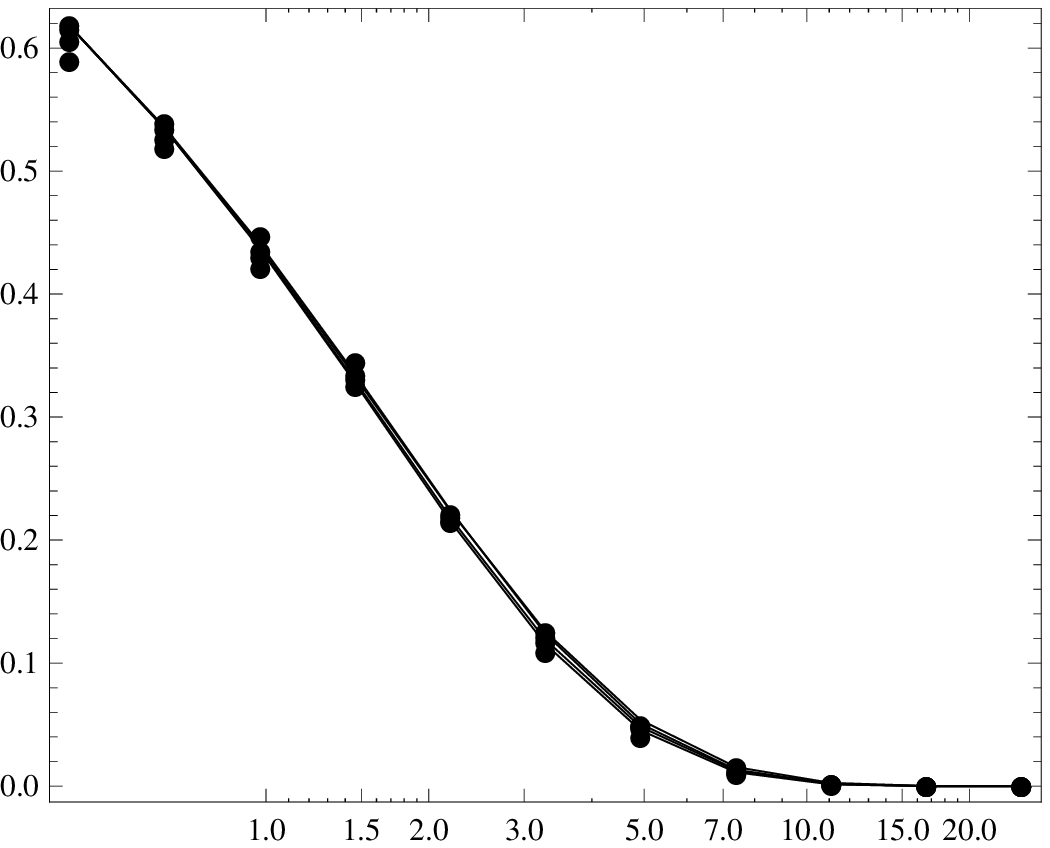}}
  \put(145,55){\large\here{$\phi$}}  \put(204,-11){\large$\alpha$}

\end{picture}
 \vspace*{3mm}
\caption{Theory versus simulation results for MG-type models with $F[A]=\sgn[A]|A|^\gamma$, for $\gamma = 0,1,2,3$.
Observed versus predicted values of the persistent correlations $c$ and the fraction of frozen agents $\phi$, for the ergodic regime.
Lines: theoretical
prediction; markers: simulation results. There is clearly excellent agreement between the theoretical predictions and the computer simulations.
}
\label{fig:invertible3}
\end{figure}
\begin{figure}[h]
\vspace*{4mm} \hspace*{-1mm} \setlength{\unitlength}{0.52mm}
\begin{picture}(300,190)

  \put(12,-1){\includegraphics[height=90\unitlength,width=110\unitlength]{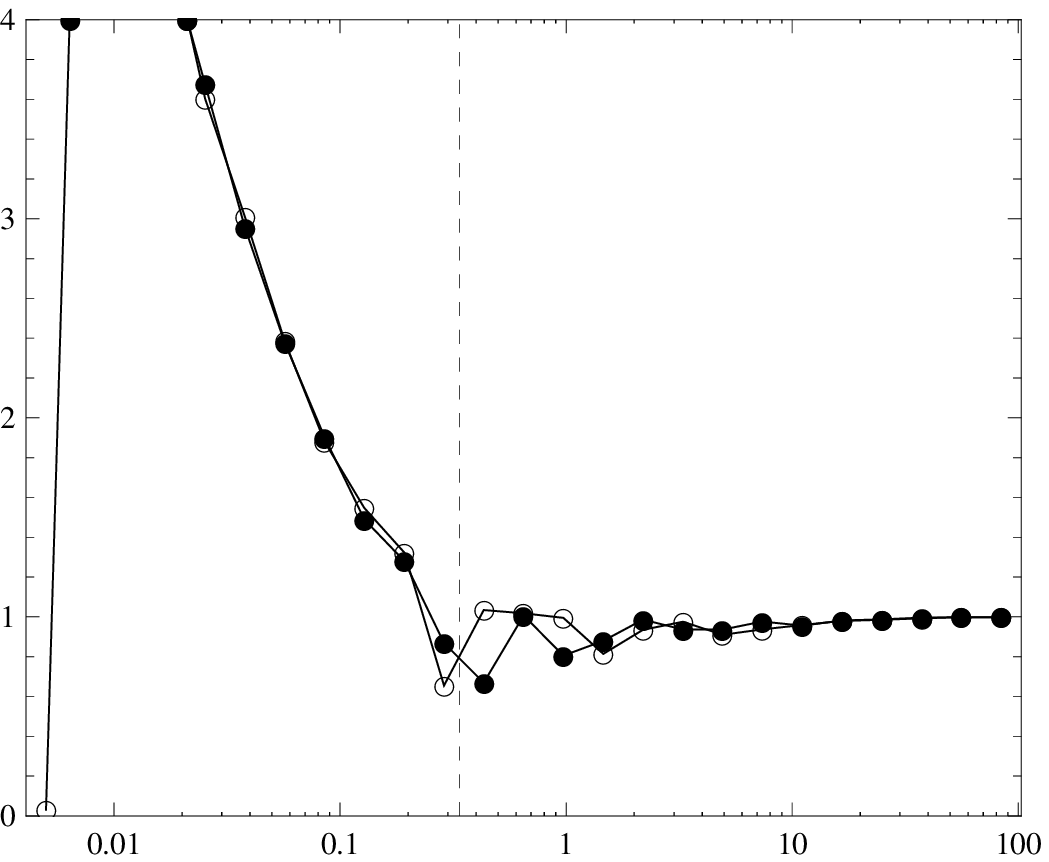}}
  \put(5,44){\large\here{$\sigma$}}  \put(64,-8){\large$\alpha$}

  \put(142,-1){\includegraphics[height=90\unitlength,width=110\unitlength]{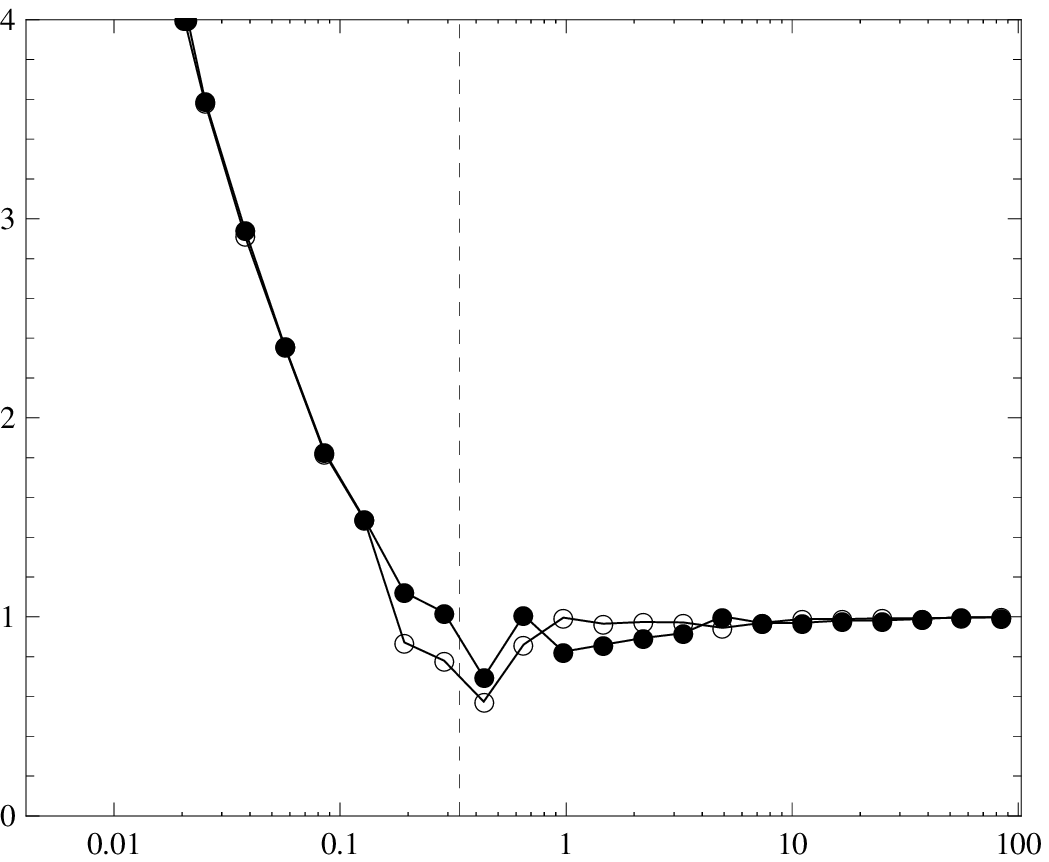}}
  \put(135,44){\large\here{$\sigma$}}  \put(194,-8){\large$\alpha$}

  \put(10,100){\includegraphics[height=90\unitlength,width=110\unitlength]{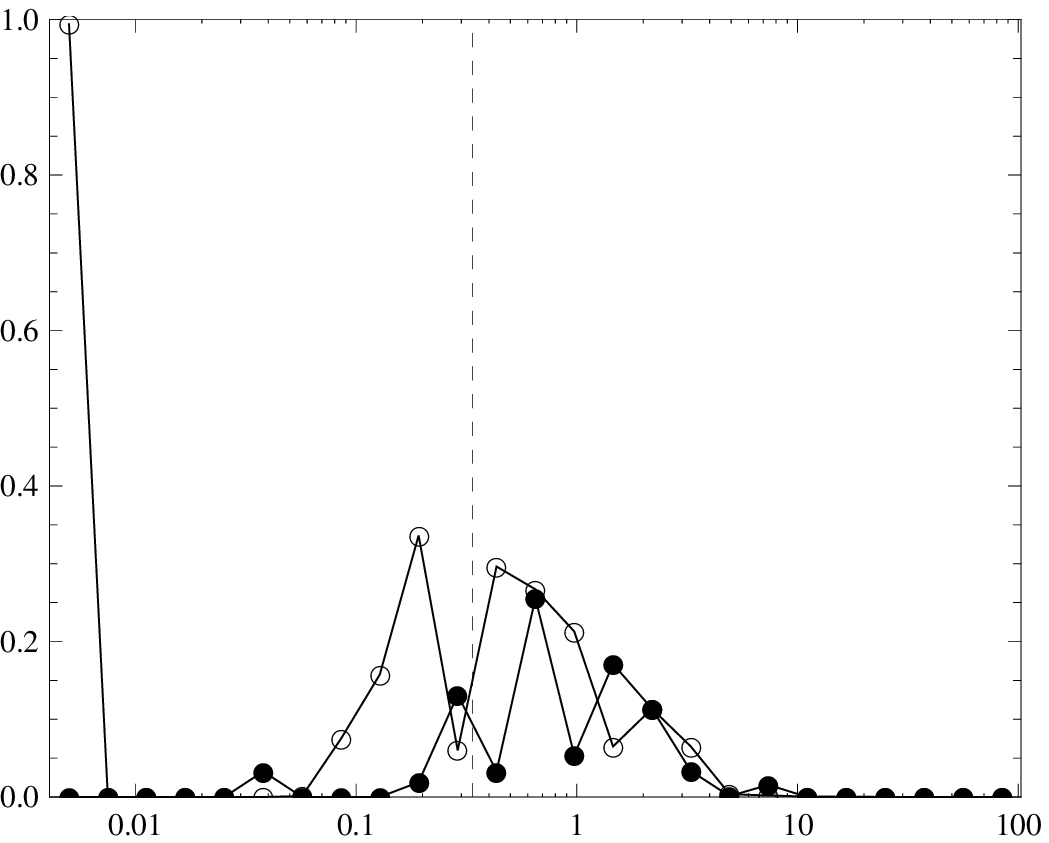}}
  \put(5,145){\large\here{$\phi$}}  \put(64,93){\large$\alpha$}

  \put(140,100){\includegraphics[height=90\unitlength,width=110\unitlength]{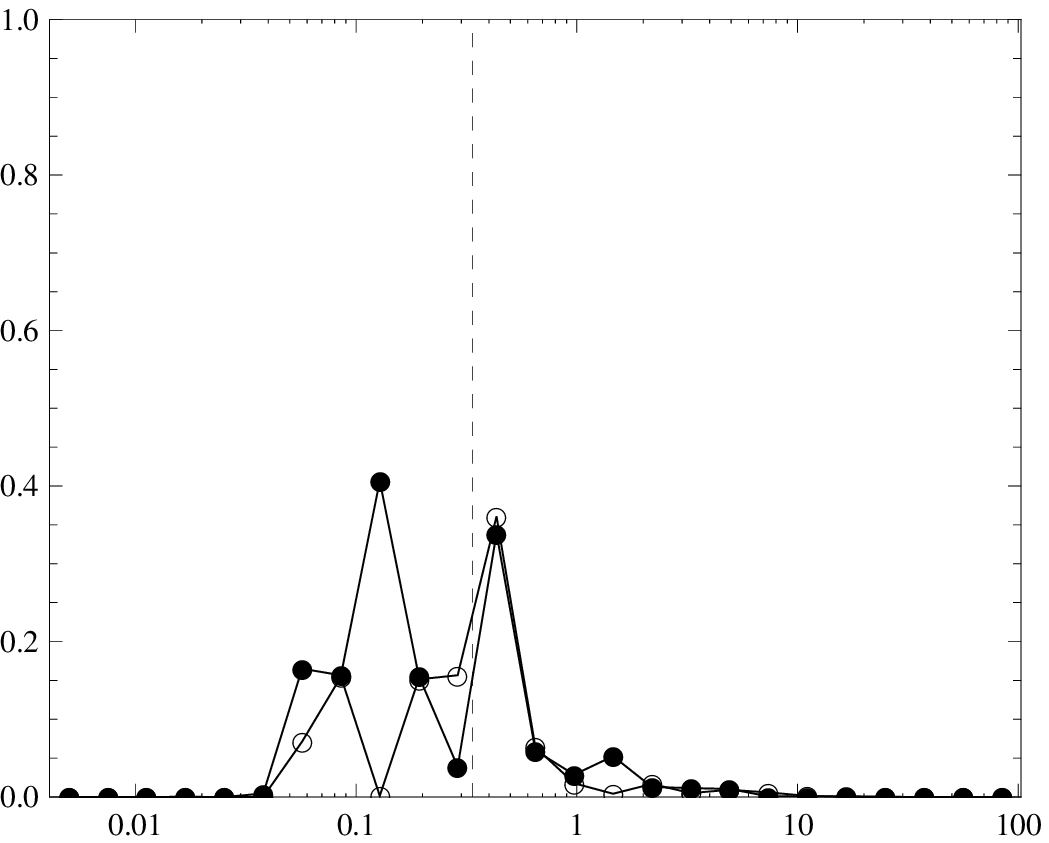}}
  \put(135,145){\large\here{$\phi$}}  \put(194,93){\large$\alpha$}

  \put(64,191){\large\here{$\gamma=4$}}  \put(185,191){\large$\gamma=5$}

\end{picture}
 \vspace*{1mm}
\caption{Fraction of frozen agents $\phi$ and volatility $\sigma$, measured in simulations of MG-type
systems with $F[A]=\sgn[A]|A|^\gamma$, for $\gamma = 4,5$. Dashed line:
predicted location of the $\chi\to\infty$ transition.
Empty/full markers correspond to biased/unbiased initial conditions. For $\gamma\geq 4$ we no longer appear to have the usual ergodic phase, and more extensive simulations with different durations and system sizes are required to
determine the nature of the macroscopic state(s).
}\label{fig:gammalarge}
\end{figure}

Here we focus on those models, which includes the original MG, in which $F[A]$ is monotonically increasing.
 The ansatz $\chi>0$ now guarantees that $\Delta(z)$ is invertible. These models have a remarkable universality property: all exhibit an ergodicity-breaking transition, marked by $\chi\to\infty$, exactly at
the value $\alpha_c$ of the standard minority game, irrespective of the precise form of the function $F[A]$.
To demonstrate this we first rewrite (\ref{eq:Delta2}):
\begin{eqnarray}
\frac{u-\Delta^{-1}(u)}{\chi}&=&\int\!Dx~F\big[\Delta^{-1}(u)+x\sqrt{\frac{1}{2}(1-c)}\big]
\end{eqnarray}
From this we conclude that, for all $z\in\R$,
\begin{eqnarray}
\lim_{\chi\to\infty}\int\!Dx~F\big[\Delta^{-1}(z)+x\sqrt{\frac{1}{2}(1-c)}\big]=0
\end{eqnarray}
Since $F[A]$ is monotonic, the obvious solution $\Delta^{-1}(z)=0$ (guaranteed by the anti-symmetry of $F[A]$) must be unique, and hence we
know that generally $\lim_{\chi\to\infty}\Delta^{-1}(z)=0$.
It now follows that $\lim_{\chi\to\infty}I_1(c,\chi)=\lim_{\chi\to\infty}I_2(c,\chi)=0$, and therefore via (\ref{eq:ChiR3},\ref{eq:S03}) we find
\begin{eqnarray}
\hspace*{-15mm}
\lim_{\chi\to\infty}\chi\chiR= 1,~~~~~~~~
\lim_{\chi\to\infty}|\chi| S_0=\sqrt{1\!+\!c},~~~~~~~~\lim_{\chi\to\infty}\frac{S_0}{\chiR}=\sgn(\chiR)\sqrt{1\!+\!c}
\end{eqnarray}
The ansatz $\chi>0$ guarantees $\sgn(\chiR)=1$ (with $\chiR\downarrow 0$ as $\chi\to\infty$), and our persistent order parameter equations
(\ref{eq:c},\ref{eq:phi},\ref{eq:chi}) close {\em at} the transition $\chi\to\infty$ exactly in the same way as they would for the conventional
 MG, i.e.
\begin{eqnarray}
c&=&\sigma^2[\infty]\Big\{
 1+\frac{1-2v^2}{2v^2} {\rm Erf}[v]
-\frac{1}{v\sqrt{\pi}} \rme^{-v^2}\Big\}
\\
\alpha &=&{\rm Erf}[v],~~~~~~~~v=\sigma[\infty]\sqrt{\alpha}/\sqrt{2(1\!+\!c)}
\end{eqnarray}
So all systems with monotonically increasing anti-symmetric $F[A]$ will, provided they have an ergodic regime, {\em always} exhibit a $\chi\!\to\!\infty$ transition
at the conventional MG value $\alpha_c\!\approx\! 0.3374$, irrespective of the actual form of $F[A]$.
\vsp

Away from the transition one should expect a dependence of the values of the persistent order parameters on the choice
made for $F[A]$. However, for `sensible' choices of $F[A]$ one generally finds this dependence to be weak. To illustrate
this we now focus on a particular class of models, with monotonic valuation update functions  of the form
\begin{eqnarray}
F[A]=\sgn[A]|A|^\gamma, ~~~~~~~~\gamma\geq 0
\label{eq:invertible_models}
\end{eqnarray}
These models are relatively straightforward generalizations of the original MG (which corresponds to $\gamma=1$),
nevertheless for $\gamma\neq 1$ the standard solution route (i.e. generating functional analysis of the strategy selection process, without including
the overall bid dynamics explicitly in the formalism), would not have worked.
For $\gamma>1$ the agents with (\ref{eq:invertible_models}) place more importance on making money by exploiting large fluctuations ($|A|>1$), so can be described as greedy high-risk contrarians, whereas for $\gamma<1$ they prefer to exploit small market fluctuations, and operate as cautious low-risk contrarians.

We know from the above argument that their
$\chi\to\infty$ ergodicity breaking transition point $\alpha_c$ will be identical to that of the standard MG. Let us now calculate the
persistent order parameters in the ergodic region $\alpha>\alpha_c$.
On physical grounds one does not expect a negative susceptibility and $F[A]$ increases monotonically, hence $\Delta(z)$ will be invertible,
and we can transform the integrals (\ref{eq:I1I2new}) via the substitution
$z=\Delta(x)\sqrt{2}/\sqrt{1+c}$, and find
\begin{eqnarray}
I_1(c,\chi)&=&\int\!\frac{\rmd x}{\sqrt{2\pi}}~\rme^{-\Delta^2(x)/(1+c)}
\label{eq:I1invertible}
\\
I_2(c,\chi)&=&\frac{\sqrt{2}}{\sqrt{1\!+\!c}}\int\!\frac{\rmd x}{\sqrt{2\pi}}~\rme^{-\Delta^2(x)/(1+c)} x^2 \Delta^\prime(x)
\label{eq:I2new}
\end{eqnarray}
This removes the need for inversion of $\Delta(z)$.
 For the models (\ref{eq:invertible_models})
 one then finds
\begin{eqnarray}
\Delta(z)&=& z+\chi\int\!Dx~\sgn\Big[z+x\sqrt{\frac{1}{2}(1\!-\!c)}\Big]\Big|z+x\sqrt{\frac{1}{2}(1\!-\!c)}\Big|^\gamma
\end{eqnarray}
For non-integer $\gamma$ the integral has to be calculated numerically. For integer $\gamma$ one finds
\begin{eqnarray}
\hspace*{-10mm}
\gamma=0:&~~~~&\Delta(z)= z+\chi~ {\rm Erf}\Big[\frac{z}{\sqrt{1\!-\!c}}\Big]
\\
\hspace*{-10mm}
\gamma=1:&~~~&\Delta(z)= z+\chi z
\\
\hspace*{-10mm}
\gamma=2:&~~~&\Delta(z)= z+\big[z^2\!+\!\frac{1}{2}(1\!-\!c)\big]\chi ~ {\rm Erf}\Big[\frac{z}{\sqrt{1\!-\!c}}\Big]
+\frac{\chi z\sqrt{1\!-\!c}}{\sqrt{\pi}}\rme^{-z^2/(1-c)}
\\[-1mm]
\hspace*{-10mm}
\gamma=3:&~~~&\Delta(z)= z+\chi[z^3+\frac{3}{2}(1\!-\!c)z]
\end{eqnarray}
Since on physical grounds one does not expect a negative susceptibility, invertibility is expected to hold and the present family of models
should behave qualitatively as the ordinary MG. One also expects that as the susceptibility goes to zero for large alpha the behavior of the
system should be almost independent of the value of $\gamma$.
For $\gamma\leq 3$ these predictions are borne out by numerical simulations (all carried out with $N=4097$, and measured during 1000 steps after an 2000 steps equilibration period), as shown in figures \ref{fig:invertible1}-\ref{fig:gammalarge}. For large $\gamma$ (excessive agent greed), however, in spite of the agents still operating as contrarians, the ergodic phase appears to be destroyed by their increasing focus on big-gain big-risk decisions, and there is no longer an efficient market regime with low volatility.

\section{Applications - dynamic switching between contrarian trading and trend following }

 Next we inspect a class of models in which agents switch from trend-following to contrarian behaviour, dependent on the absolute value $|A|$ of
 the overall market bid.  The rationale is to create more realistic agent behaviour, based on interpreting $A$ as a measure of the price returns in the  market. One example was proposed in \cite{demartino2}, corresponding to $F[A]=\epsilon A^3-A$, with $\epsilon>0$.
 This model, in which agents are trend-follows for $|A|<1/\sqrt{\epsilon}$ but contrarians for $|A|>1/\sqrt{\epsilon}$, can not be solved
 analytically using the standard generating functional analysis route.
 Here we generalize their model slightly, allowing also for the reverse tendency, where agents become trend-followers for {\em large} instead of
  small $|A|$, and analyze the case
\begin{eqnarray}
F[A]=\tau A(1-A^2/A_0^2),~~~~~~{\rm with}~~\tau=\pm 1,~A_0>0
\label{eq:giardina_type}
\end{eqnarray}
The model of \cite{demartino2} corresponds to $\tau=-1$ and $A_0=1/\sqrt{\epsilon}$.
For $\tau=1$, in contrast, agents behave as contrarians for modest deviations of the returns from their average value,
 but switch to herding when they perceive the market to be anomalous, i.e. for $|A|>A_0$.
For $A_0\to \infty$, the model (\ref{eq:giardina_type}) reduces either to the standard MG (for $\tau=1$) or to a majority type
game (for $\tau=-1$); for $A_0\to 0$ one anticipates the opposite.

\subsection{Non-invertible overall bid impact functions}

\begin{figure}[t]
\vspace*{0mm} \hspace*{25mm} \setlength{\unitlength}{0.5mm}
\begin{picture}(200,90)

  \put(0,0){\includegraphics[height=90\unitlength]{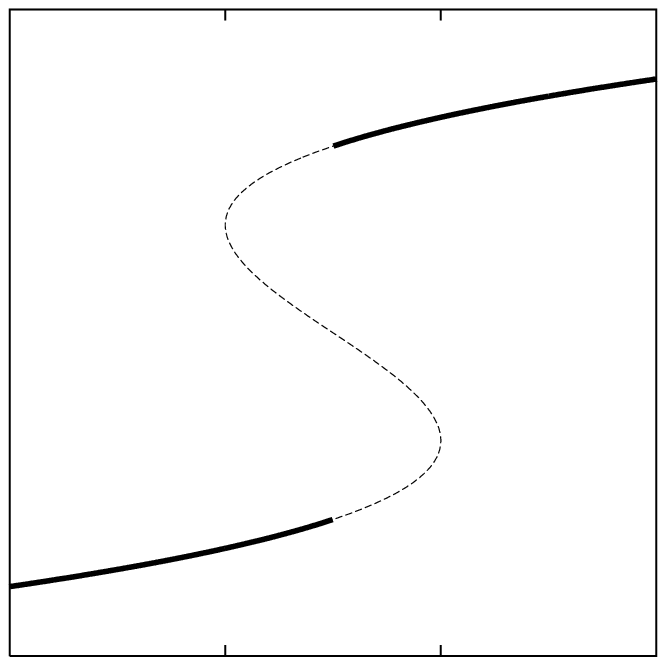}}
 \put(-11,60){$z(r)$}
 \put(20,-5){\small $-|q|^{3/2}$}  \put(56,-5){\small $|q|^{3/2}$}
 \put(8,79){\small\sl Maxwell option}

  \put(115,0){\includegraphics[height=90\unitlength]{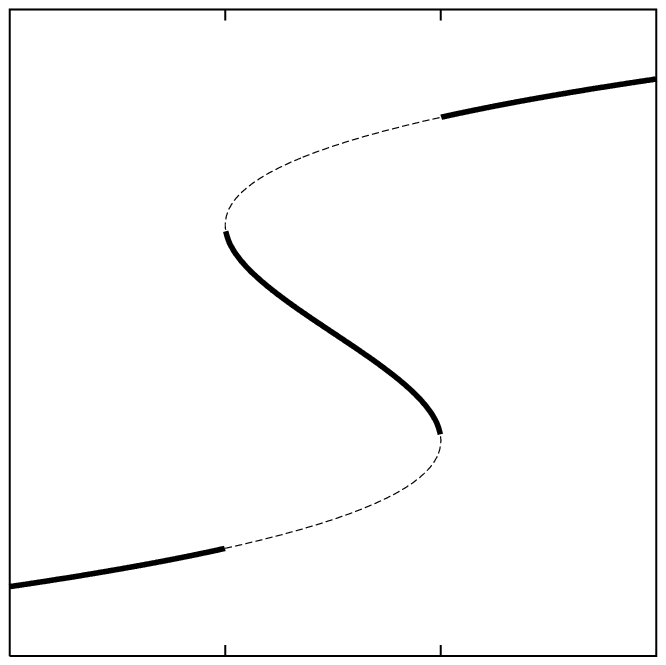}}
 \put(104,60){$z(r)$}
 \put(135,-5){\small $-|q|^{3/2}$}  \put(171,-5){\small $|q|^{3/2}$}
 \put(123,79){\small\sl alternative option}
\end{picture}
\vspace*{4mm}
\caption{
The options for constructing an anti-symmetric solution $z(r)$ of the cubic equation $z^3+3qz-2r=0$ for $q<0$,
shown as a function of $r$.
 Dashed: all solutions.
 Left: the Maxwell construction, which assumes the middle solution for $|r|<|q|^{3/2}$ to be unstable, and the others to reflect
 remanence. Right: the consequence of assuming the
middle solution to be stable. Only the one on the right turns out to have the correct  $A_0\to\infty$ limit
(see main text). }
\label{fig:solutions}
\end{figure}

For (\ref{eq:giardina_type}) one finds
\begin{eqnarray}
\Delta(z)&=& z\big[1+\tilde{\chi} -\frac{3}{2}(1-c)\tilde{\chi}/A_0^2\big]-\tilde{\chi} z^3/A_0^2
\end{eqnarray}
where $\tilde{\chi}=\tau\chi$. Whether $\Delta(z)$ is invertible will depend on $A_0$.
If $A_0^2(1+\tilde{\chi}^{-1})>\frac{3}{2}(1-c)$, there will always be a region of non-invertibility, with three solutions $z$
of the equation $\Delta(z)=\Delta$. The latter are the roots of a cubic equation and can therefore be calculated analytically,
\begin{eqnarray}
z^3- z\big[A_0^2(1+\tilde{\chi})/\tilde{\chi}-\frac{3}{2}(1-c)\big]+ \frac{A_0^2 \Delta}{\tilde{\chi}}=0
\end{eqnarray}
Following \cite{AbraSteg} we write the cubic equation in the form $z^3+3qz-2r=0$ by defining
\begin{eqnarray}
&& q=\frac{1}{2}(1-c)-A_0^2(1+\tilde{\chi})/3\tilde{\chi},~~~~~~r=-A_0^2 \Delta/2\tilde{\chi}
\label{eq:ASnotation}
\end{eqnarray}
We can then classify the solution(s) of the equation $\Delta(z)=\Delta$ as follows:
\begin{eqnarray}
\hspace*{-20mm}
q^3+r^2>0:&~{\rm one~soln,}~    & z=[r+\sqrt{q^3+r^2}]^{1/3}+[r-\sqrt{q^3+r^2}]^{1/3}
\label{eq:z_unique}
\\
\hspace*{-20mm}
q^3+r^2<0:&~{\rm three~solns,~} & z_1=2|q|^{\frac{1}{2}}\cos(\frac{1}{3}{\rm arccos}(\frac{r}{\sqrt{|q|^3}}))
\\
\hspace*{-20mm}
&&\hspace*{-15mm} z_{2}=-|q|^{\frac{1}{2}}\cos(\frac{1}{3}{\rm arccos}(\frac{r}{\sqrt{|q|^3}}))
+\sqrt{3}|q|^{\frac{1}{2}}\sin(\frac{1}{3}{\rm arccos}(\frac{r}{\sqrt{|q|^3}}))
\\
&&\hspace*{-15mm} z_{3}=-|q|^{\frac{1}{2}}\cos(\frac{1}{3}{\rm arccos}(\frac{r}{\sqrt{|q|^3}}))
- \sqrt{3}|q|^{\frac{1}{2}}\sin(\frac{1}{3}{\rm arccos}(\frac{r}{\sqrt{|q|^3}}))
\end{eqnarray}
 Our solution must respect the symmetry $\Delta^{-1}(-z)=-\Delta^{-1}(z)$, which translates into searching for a root with $z(-r)=-z(r)$. Inspection reveals that for $q^3+r^2>0$ the solution (\ref{eq:z_unique}) has the desired symmetry.
For $q^3+r^2<0$ we find that $z_2(-r)=-z_2(r)$, and that $z_3(-r)=-z_1(r)$. Hence $z_{1,3}(r)$ represent two `extremal' solution branches
(related to each other by inversion symmetry), and $z_2(r)$ represents a middle branch. The region of multiple solutions is defined
by $|r|<|q|^{3/2}$, where we have infinitely many options for assigning a value to $\Delta^{-1}(z)$. The conventional one is the
 Maxwell construction, based on assuming the multiplicity of solutions to be caused by remanence. Here the middle branch $z_2(r)$
is taken to be dynamically unstable, and
one takes for $0<r<|q|^{3/2}$ the continuation $z_1(r)$ of the $r>|q|^{3/2}$ solution, and   for $-|q|^{3/2}<r<0$ the continuation $z_3(r)$
of the  $r<-|q|^{3/2}$ solution, with a discontinuity at $r=0$. Our second option implies assuming the middle branch to be stable, i.e.
choosing $z_2(r)$ for $|r|<|q|^{3/2}$.
The two options are illustrated in figure \ref{fig:solutions}.
Assessing which of these choices (if any) is correct would in principle require a stability analysis of the asymptotic solution of the
overall bid dynamics.

Upon translating the above arguments into the language of $\Delta^{-1}(z)$, the resulting picture is the following, where still
$q=\frac{1}{2}(1\!-\!c)-\frac{1}{3}A_0^2(1\!+\!\tilde{\chi})/\tilde{\chi}$ but now with
$r(z)=-\frac{1}{2}A_0^2 z/\tilde{\chi}$:
\begin{eqnarray}
\hspace*{-20mm}
q\!>\!0~~{\rm or~~} q\!<\!0,~|r(z)|\!>\!|q|^{3/2}:&~~~~&
\Delta^{-1}(z)=[r(z)\!+\!\sqrt{q^3\!+\!r^2(z)}]^{\frac{1}{3}}+[r(z)\!-\!\sqrt{q^3\!+\!r^2(z)}]^{\frac{1}{3}}
\nonumber\\
\hspace*{-20mm}
\label{eq:correct_sol0}
\end{eqnarray}
Here $\Delta(z)$ is fully invertible. In the alternative scenario we have the Maxwell option:
\begin{eqnarray}
\hspace*{-10mm}
q<0,~|r(z)|<|q|^{3/2}:&~~~& \Delta^{-1}(z)=2~\sgn[r(z)]|q|^{\frac{1}{2}}\cos(\frac{1}{3}{\rm arccos}(\frac{|r(z)|}{\sqrt{|q|^3}}))
\label{eq:Maxwell}
\end{eqnarray}
But we also have the non-Maxwell solution, which can be rewritten as:
\begin{eqnarray}
\hspace*{-10mm}
q<0,~|r(z)|\!<\!|q|^{3/2}:&~~~& \Delta^{-1}(z)=-2|q|^{\frac{1}{2}}
\sin(\frac{1}{3}{\rm arcsin}(\frac{r(z)}{\sqrt{|q|^3}}))
\label{eq:non-Maxwell}
\end{eqnarray}
Only for $A_0\to\infty$ can we decide between our candidate solutions for $q<0$ and $|r(z)|<|q|^{3/2}$ without analyzing dynamic
stability in the underlying bid process. There
(\ref{eq:giardina_type}) reduces to the standard MG for $\tau=1$ and to the standard majority game for $\tau=-1$, both of which we analyzed in
section \ref{sec:simple_examples}. The correct solution must reproduce $\lim_{A_0\to \infty}\Delta^{-1}(z)=z/(1+\tilde{\chi})$.
For $A_0\to\infty$, $\tilde{\chi}\notin(-1,0)$,  and finite $z$ one has $q<0$ and $|r(z)|<|q|^{3/2}$
(in fact $\lim_{A_0\to\infty}r(z)/|q|^{3/2}=0$), so we do indeed probe the region of ambiguity. Our test reveals, using
$\sqrt{3}\sin(\frac{1}{3}{\rm arccos}(x))-\cos(\frac{1}{3}{\rm arccos}(x))=-\frac{2}{3}x+\order(x^3)$, that for $A_0\to\infty$:
\begin{eqnarray}
\hspace*{-10mm}
{\rm Maxwell~soln:}&~~~ \Delta^{-1}(z)=A_0|1+\tilde{\chi}^{-1}|^{1/2}\sgn(z)+\order(A_0^0),~~~& {\rm incorrect}
\\
\hspace*{-10mm}
{\rm alternative~soln:}&~~~\Delta^{-1}(z)=z/(1+\tilde{\chi})+\order(A_0^{-1}),~~~& {\rm correct}
\end{eqnarray}
For $\tilde{\chi}\in(-1,0)$ we have $q>0$, so there is no ambiguity: $\Delta^{-1}(z)$ is given
 by (\ref{eq:correct_sol0}), which for $A_0\to\infty$ reproduces correctly
 $\Delta^{-1}(z)=z/(1+\tilde{\chi})+\order(A_0^{-1})$.
 Hence, for $A_0\to\infty$ the only acceptable solution is
(\ref{eq:correct_sol0},\ref{eq:non-Maxwell})
\footnote{It is surprising that in this problem the Maxwell construction is not always the correct way to handle the multiplicity of
solutions, given its track record in physical many-particle systems. However, minority games do not obey
 detailed balance, so intuition developed on the basis of bifurcation analyses obtained via free energy minimization in equilibrium
 systems may be misleading.}. We resolve the remaining ambiguity as  follows: since the pure majority and minority games
 always exhibit $\lim_{z\to 0}\frac{\rmd}{\rmd z}\Delta^{-1}(z)>0$, we take this property to hold generally. This means that for weak
 random bids the effect of agent interaction is to push these bids back to zero (contrarian action) or to amplify them (trend-following),
 but there is no change of sign. For $\tau=-1$ we now have either
$\frac{3}{2}(1-c)/A_0^2>1-\chi^{-1}$ and the unambiguous solution (\ref{eq:correct_sol0}), or we have $\frac{3}{2}(1-c)/A_0^2<1-\chi^{-1}$
and the Maxwell option (\ref{eq:Maxwell}).
For $\tau=1$
we must always demand $\frac{3}{2}(1-c)/A_0^2<1+\chi^{-1}$ and the non-Maxwell option (\ref{eq:non-Maxwell}).

\subsection{Conventional $\chi\to\infty$ transitions}

 We can now calculate transition lines. In the present model we can have a traditional $\chi\to\infty$
 transition, a transition where $\chiR$ changes sign (switching from a minority to a majority-type game), and a transition
 marking the emergence of jumps in $\Delta^{-1}(z)$. We start with the calculation of the $\chi\to\infty$ transitions.

For $\tau=1$, $\chi\to\infty$ and finite $z$ one has $r(z)=\order(\chi^{-1})$ and
$q=\frac{1}{2}(1\!-\!c)-\frac{1}{3}A_0^2+\order(\chi^{-1})$, from which it follows that
if $\lim_{\chi\to\infty}\frac{3}{2}(1\!-\!c)\neq A_0^2$ then
\begin{eqnarray}
\Delta^{-1}(z)&=&-\frac{z}{\chi}\Big[\frac{3}{2}(1\!-\!c)/A_0^2-1\Big]^{-1}+\order(\chi^{-2})
\end{eqnarray}
From this one extracts $\lim_{\chi\to\infty}I_1(c,\chi)=\lim_{\chi\to\infty}I_2(c,\chi)=0$, and via the same arguments
that applied to models with invertible bid impact functions one is led to
\begin{eqnarray}
\hspace*{-15mm}
\lim_{\chi\to\infty}\chi\chiR= 1,~~~~~~
\lim_{\chi\to\infty}\chi S_0=\sqrt{1\!+\!c},~~~~~~\lim_{\chi\to\infty}\frac{S_0}{\chiR}=\sgn(\chiR)\sqrt{1\!+\!c}
\end{eqnarray}
If  $\chiR>0$, our equations (\ref{eq:c},\ref{eq:phi},\ref{eq:chi}) close for $\chi\to\infty$ exactly
 as they would for the conventional MG, with the standard $\chi\to\infty$ transition at the value $\alpha_c\approx 0.3374$.
 If $\chiR<0$, on the other hand, no $\chi\to\infty$ transition is possible. Taking into account the requirement
 $\lim_{z\to 0}\frac{\rmd}{\rmd z}\Delta^{-1}(z)>0$ implies that for $\tau=1$ the $\chi\to\infty$ transition exists only if
$A_0^2>\frac{3}{2}(1-c)$. Here $q<0$ and $r(z)=\order(\chi^{-1})$, so we are in the remanent region.

For $\tau=-1$, $\chi\to\infty$ and finite $z$ one still has $r(z)=\order(\chi^{-1})$ and
$q=\frac{1}{2}(1\!-\!c)-\frac{1}{3}A_0^2+\order(\chi^{-1})$. So if $\lim_{\chi\to\infty}\frac{3}{2}(1\!-\!c)\neq A_0^2$ then for $q<0$
we will have $|r(z)|\ll |q|^{3/2}$. Hence
\begin{eqnarray}
A_0^2<\frac{3}{2}(1\!-\!c):&~~~&\Delta^{-1}(z)=\frac{z}{\chi}\Big[\frac{3}{2}(1\!-\!c)/A_0^2-1\Big]^{-1}\!+\order(\chi^{-2})
\\
A_0^2>\frac{3}{2}(1\!-\!c):&~~~& \Delta^{-1}(z)=\sgn(z)\sqrt{A_0^2-\frac{3}{2}(1\!-\!c)}
+\order(\chi^{-1})
\end{eqnarray}
If $A_0^2<\frac{3}{2}(1\!-\!c)$ we return as expected to the conventional MG transition line at $\alpha_c\approx 0.3374$; however, in
contrast to $\tau=1$ this line is now in the non-remanent region.
If $A_0^2>\frac{3}{2}(1\!-\!c)$, on the other hand, the $\chi\to\infty$ line will be in the remanent region. With some
foresight we now write $A_0^2=\frac{3}{2}(1\!-\!c)+\frac{1}{2}(1\!+\!c)\Xi^2$ with $\Xi\geq 0$, and obtain
\begin{eqnarray}
I_1(c,\chi)&=& \Xi\sqrt{\frac{1\!+\!c}{\pi}}
+\order(\chi^{-1}),~~~~~~
I_2(c,\chi)=\frac{1}{2}(1\!+\!c)\Xi^2
+\order(\chi^{-1})
\end{eqnarray}
which leads to
\begin{eqnarray}
U(v)-1
&=&
\frac{\Xi^2
 -\Xi\sqrt{2/\pi}}{1-\Xi\sqrt{2/\pi}}
,~~~~~~
\alpha
={\rm Erf}(v)/[1-\Xi\sqrt{2/\pi}]
\end{eqnarray}
Upon solving the first equation for $\Xi$,
 \begin{eqnarray}
 \Xi_\pm(v)&=& \frac{1}{\sqrt{2\pi}}\Big\{
 2\!-\!U(v)\pm \sqrt{[2\!-\!U(v)]^2+2\pi[U(v)\!-\!1]}\Big\}
 \end{eqnarray}
 we then arrive at a convenient parametrization of the transition line in the $(\alpha,A_0)$ plane, with $v\geq 0$ as a parameter.
 The line turns out to have two branches (indicated by $\pm$):
\begin{eqnarray}
A_0^\pm(v)&=&\sqrt{\frac{3}{2}[1-c(v)]+\frac{1}{2}[1+c(v)]\Xi^2_\pm(v)}
\\
\alpha_\pm(v)&=& {\rm Erf}(v)/[1-\Xi_\pm(v)\sqrt{2/\pi}]
\end{eqnarray}
We must demand $0\leq\Xi_\pm(v)\leq \sqrt{\pi/2}$, to guarantee $\chiR>0$, and $[2\!-\!U(v)]^2+2\pi[U(v)\!-\!1]\geq 0$,
to ensure $\Xi_\pm(v)\in\R$.
We have now found the $\chi\to\infty$ transition line for both
$A_0^2<\frac{3}{2}(1-c)$ (in the non-remanent region, where it is independent of $A_0$), and for
$\frac{3}{2}(1-c)<A_0^2<\frac{3}{2}(1\!-\!c)+\frac{1}{4}\pi(1\!+\!c)$ (in the remanent region, where it depends on $A_0$).
At the moment where $A_0^2=\frac{3}{2}(1\!-\!c)+\frac{1}{4}\pi(1\!+\!c)$, the bid susceptibility $\chiR$ goes through zero, marking a switch
to majority game behavior; we find below that this occurs for the $+$ branch at $v=0$.

To aid and test numerical evaluation it will be helpful to assess the limits $v\to 0$ and $v\to \infty$ of the above parametrized branches.
For small $v$ one finds, using ${\rm Erf}(v)=(2v/\sqrt{\pi})[1-\frac{1}{3}v^2+\order(v^4)]$ and the fact that $\Xi(v)$ cannot be negative,
\begin{eqnarray}
\hspace*{-15mm}
c(v)&=&\sigma^2[\infty][ 1\!-\!\frac{4v}{3\sqrt{\pi}}]+\order(v^3),~~~~~~~~
U(v)=\frac{\sigma^2[\infty]}{v[1\!+\! \sigma^2[\infty]]\sqrt{\pi}}+\order(1)
\label{eq:Usmallv}
\\
\hspace*{-15mm}
\Xi_+(v)&=& \sqrt{\frac{\pi}{2}}-
\frac{v[1\!+\! \sigma^2[\infty]]\pi(\pi\!-\!2)}{2\sigma^2[\infty]\sqrt{2}}
+\order(v^2)
\end{eqnarray}
Hence
\begin{eqnarray}
\lim_{v\to 0} A_0^+(v)&=&\sqrt{\frac{3}{2}[1-\sigma^2[\infty]]+\frac{\pi}{4}[1+\sigma^2[\infty]]}
\\
\lim_{v\to 0} \alpha_{+}(v)&=&
\frac{4\sigma^2[\infty]}{[1\!+\! \sigma^2[\infty]]\pi(\pi\!-\!2)}
\end{eqnarray}
For $\sigma[\infty]=1$ (no decision noise) this gives
$\lim_{v\to 0}\alpha_+(v)=2/\pi(\pi-2)\approx 0.5577$ and $\lim_{v\to 0}A_0^+(v)=\sqrt{\pi/2}\approx 1.2533$.
For $v\to\infty$, in contrast, we observe due to $\lim_{v\to\infty}U(v)=0$ that
the $\Xi_{\pm}(v)$ are no longer real-valued; both branches terminate and meet at the value $v_{\rm max}$ such that
$[U(v_{\rm max})-2]^2=2\pi[1-U(v_{\rm max})]$, i.e. where
\begin{eqnarray}
U(v_{\rm max})&=& \sqrt{\pi-2}[ \sqrt{\pi}-\sqrt{\pi-2}]
\end{eqnarray}

\subsection{Onset of remanence-induced discontinuities in the $\chiR>0$ region}

It is clear that discontinuities emerge in our equations as soon as $q\leq 0$. Even if it is not yet clear which precise observables
 will be affected by these discontinuities, it implies that there exists an alternative transition marked by the condition $q=0$, i.e. by
 \begin{eqnarray}
 c=1-\frac{2}{3}A_0^2(1+\tilde{\chi}^{-1})
 \label{eq:newtransition}
 \end{eqnarray}
When (\ref{eq:newtransition}) holds, we find upon expanding for small $q$ that
$\lim_{q\to 0}\Delta^{-1}(z)= -(A_0^2 z/\tilde{\chi})^{1/3}$, and so along the line (\ref{eq:newtransition})
the integrals (\ref{eq:I1I2new}) reduce to gamma functions:
\begin{eqnarray}
\hspace*{-15mm}
I_1(c,\chi)&= - \Big(\frac{A_0^2 \sqrt{2(1\!+\!c)}}{2\tilde{\chi}}\Big)^{1/3} \int\!Dz~|z|^{4/3}
&=- \Big(\frac{A_0^2 \sqrt{1\!+\!c}}{2\tilde{\chi}\sqrt{2}}\Big)^{1/3} \frac{1}{3\sqrt{\pi}}\Gamma(\frac{1}{6})
\\
\hspace*{-15mm}
I_2(c,\chi)&= \Big(\frac{A_0^2 \sqrt{2(1\!+\!c)}}{2\tilde{\chi}}\Big)^{2/3}\int\!Dz~|z|^{2/3}
&=\Big(\frac{A_0^2 \sqrt{1\!+\!c}}{\tilde{\chi}}\Big)^{2/3} \frac{1}{\sqrt{\pi}}\Gamma(\frac{5}{6})
\end{eqnarray}
This
leads to
\begin{eqnarray}
U(v)-1
&=& \frac{6\Gamma(\frac{5}{6})\Xi^2
 +\Gamma(\frac{1}{6})\Xi}
 {3\sqrt{\pi}+\Gamma(\frac{1}{6})\Xi}
~~~~~~~~
\alpha= {\rm Erf}(v)/[1+\Gamma(\frac{1}{6})\Xi/3\sqrt{\pi}]
\end{eqnarray}
in which now $\Xi=[A_0^2/\tilde{\chi}[1+c(v)]]^{1/3}$.
Solving the first equation for $\Xi$ gives
\begin{eqnarray}
\hspace*{-10mm}
\Xi_\pm(v)&=& \frac{\Gamma(\frac{1}{6})}{12\Gamma(\frac{5}{6})}\left\{ U(v)\!-\!2
\pm \sqrt{[2\!-\!U(v)]^2-72[1\!-\!U(v)]\sqrt{\pi}\Gamma(\frac{5}{6})/\Gamma^2(\frac{1}{6})}\right\}
\end{eqnarray}
Again, after combination with the $q=0$ condition,  we thereby arrive at a parametrization of the transition line in the $(\alpha,A_0)$ plane,
with $v\geq 0$ as a parameter, with two $\pm$ branches:
\begin{eqnarray}
A_0^\pm(v)&=&\sqrt{\frac{3}{2}[1-c(v)]-[1+c(v)]\Xi^3_\pm(v)}
\\
\alpha_\pm(v)&=&{\rm Erf}(v)/\Big[1+\frac{\Gamma(\frac{1}{6})}{3\sqrt{\pi}}\Xi_\pm(v)\Big]
\end{eqnarray}
The corresponding value of $\chi$ then follows from
\begin{eqnarray}
\chi&=& \tau\Big[\frac{3[1-c(v)]}{2[1+c(v)]\Xi^3_\pm(v)}-1\Big]
\end{eqnarray}
To have $\Xi_\pm(v)\in\R$ we must demand
$[U(v)+ 2(\xi-1)]^2\geq 4\xi(\xi-1)$, where $\xi=18\sqrt{\pi}\Gamma(\frac{5}{6})/\Gamma^2(\frac{1}{6})\approx 1.1623$.
Since $U(v)\geq 0$ this implies $U(v)\geq 2\sqrt{\xi\!-\!1}(\sqrt{\xi}\!-\!\sqrt{\xi\!-\!1})\approx 0.5441$.
Secondly, to have $A_0\in\R$ and $\alpha>0$ we must demand, respectively,
 $\Xi_\pm(v)\leq [\frac{3}{2}[1-c(v)]/[1+c(v)]]^{1/3}$ and $\Xi_\pm(v)\geq -3\sqrt{\pi}/\Gamma(\frac{1}{6})$. The third
 condition is $\chi\geq 0$, which translates into: $\Xi_\pm(v)\geq 0$ for $\tau=1$, and $\Xi_\pm(v)\leq 0$ for $\tau=-1$.
The second and third conditions can be combined into
\begin{eqnarray}
\tau=1:&~~~& 0\leq \Xi_\pm(v)\leq [\frac{3}{2}[1-c(v)]/[1+c(v)]]^{1/3}
\\
\tau=-1:&~~~& -3\sqrt{\pi}/\Gamma(\frac{1}{6})\leq \Xi_\pm(v)\leq 0
\end{eqnarray}
For small $v$ we may use expansion (\ref{eq:Usmallv}), which tells us that $U(v)$ diverges at $v=0$, to find
\begin{eqnarray}
\Xi_+(v)&=& \frac{U(v)\Gamma(\frac{1}{6})}{6\Gamma(\frac{5}{6})}+\order(1)=
 \frac{\sigma^2[\infty]\Gamma(\frac{1}{6})}{6v[1\!+\! \sigma^2[\infty]]\sqrt{\pi}\Gamma(\frac{5}{6})} +\order(1)
\\
\Xi_-(v)&=&
 - \frac{3\sqrt{\pi}}{\Gamma(\frac{1}{6})}
+\frac{3\pi v[1\!+\! \sigma^2[\infty]]}{\sigma^2[\infty]\Gamma(\frac{1}{6})}
\Big[
\frac{18 \sqrt{\pi}\Gamma(\frac{5}{6})}{\Gamma^2(\frac{1}{6})}
- 1
\Big]
+\order(v^2)
\end{eqnarray}
Hence
\begin{eqnarray}
\lim_{v\to 0}\alpha_-(v)&=& \frac{2\sigma^2[\infty]\Gamma^2(\frac{1}{6})}{\pi[1\!+\! \sigma^2[\infty]]
\big(
18 \sqrt{\pi}\Gamma(\frac{5}{6})
- \Gamma^2(\frac{1}{6})
\big)}
\\
\lim_{v\to 0}A_0^{-}(v)&=&\sqrt{\frac{3}{2}[1-\sigma^2[\infty]]+27[1+\sigma^2[\infty]]\pi^{3/2}/\Gamma^{3}(\frac{1}{6})}
\end{eqnarray}
For $\sigma[\infty]=0$ (i.e. no decision noise) this  result gives
$\lim_{v\to 0}\alpha_-(v)= \Gamma^2(\frac{1}{6})/\pi[
18 \sqrt{\pi}\Gamma(\frac{5}{6})
- \Gamma^2(\frac{1}{6})]\approx 1.9611$ and
$\lim_{v\to 0}A_0^{-}(v)=3\sqrt{6}\pi^{3/4}/\Gamma^{3/2}(\frac{1}{6})\approx 1.3204$.
For the $`+'$ branch the limit $v\to 0$ does not exist; instead when $v$ is decreased the line terminates at
the value $v_{\rm min}$ such that $A_0^+(v_{\rm min})=0$. Finally, the two branches terminate and meet at the value $v_{\rm max}$ such that
$[U(v_{\rm max})-2]^2=2\lambda[1-U(v_{\rm max})]$, in which $\lambda=36\sqrt{\pi}\Gamma(\frac{5}{6})/\Gamma^2(\frac{1}{6})$,
i.e. where
\begin{eqnarray}
U(v_{\rm max})&=& \sqrt{\lambda-2}[ \sqrt{\lambda}-\sqrt{\lambda-2}]
\end{eqnarray}
(note: via the relation $\Gamma(z)\Gamma(1-z)=\pi/\sin(\pi z)$ one could simplify our expressions further).

\subsection{The $\chiR=0$ transition lines}

We saw earlier that $\chiR$ can change sign at the $\chi=\infty$ line. Here we inspect the possibility of having a $\chiR=0$ transition
for finite $\chi$ and finite $\alpha$. The effective agent equation (\ref{eq:asymptotic_eqn}) would now give
$\tilde{q}=\overline{\eta}\sqrt{\alpha}+\theta$, leading to $c=\sigma^2[\infty]$ and $\phi=1$. The susceptibility would become
\begin{eqnarray}
\chi&=&\frac{1}{\sqrt{\alpha}}\bra \frac{\partial}{\partial\overline{\eta}}\overline{\sigma} \ket_{\star}=~
\frac{2\sigma[\infty]}{\sqrt{2\pi\alpha S_0^2}}
 \end{eqnarray}
The $\chiR$ transition line is hence to
 be solved from the following coupled equations for $(\chi,\alpha,A_0)$:
\begin{eqnarray}
\hspace*{-5mm}
I_1(\sigma^2(\infty),\chi)&=& \sqrt{\frac{1}{2}[1\!+\!\sigma^2(\infty)]},~~~~~~~
\alpha= \frac{\sigma^2[\infty]/\pi}{I_2(\sigma^2(\infty),\chi)\!-\!\frac{1}{2}[1\!+\!\sigma^2(\infty)]}
 \end{eqnarray}
One can use $\chi>0$ as a parameter, solve the first equation for $A_0(\chi)$, and then calculate $\alpha(\chi)$ via the second.
For $\sigma(\infty)=1$ (no decision noise)
the two equations simplify further to
\begin{eqnarray}
&& 1=\int\!Dz~z\Delta^{-1}(z)
~~~~~~~~
1+\frac{1}{\alpha\pi}=
\int\!Dz~\Big[\Delta^{-1}(z)\Big]^2
\label{eq:chiR0_conditions}
\end{eqnarray}
We will explore this latter case, where $\Delta(z)=z(1+\tau\chi)-z^3/A_0^2$ and $r(z)=\frac{1}{2}z (A_0^2+3q)$,
in more detail.
This requires working out the function $\Delta^{-1}(z)$ for the two model instances.
\vsp

If $\tau=1$ we have $q=-\frac{1}{3}A_0^2(1\!+\!\chi^{-1})\leq -\frac{1}{3}A_0^2$; here we must in the remanent regime always select the
 non-Maxwell option (\ref{eq:non-Maxwell}). We replace $\chi$ as a parameter by $\lambda=2|q|^{3/2}/(3|q|-A_0^2)\geq 0$. This implies that
$r(z)/|q|^{3/2}=-z/\lambda$, and hence
\begin{eqnarray}
\hspace*{-15mm}
\Delta^{-1}(z)&=& \sgn(z)\sqrt{|q|}~\Omega(|z|/\lambda)
\\
\hspace*{-15mm}
\Omega(x)&=&
2\theta[1\!-\!x]\sin(\frac{1}{3}{\rm arcsin}(x))
-\theta[x\!-\!1]\Big\{
[x\!+\!\sqrt{x^2\!-\!1}]^{\frac{1}{3}}+[x\!-\!\sqrt{x^2\!-\!1}]^{\frac{1}{3}}
\Big\}
\end{eqnarray}
The first equation of
(\ref{eq:chiR0_conditions}) then gives $\sqrt{|q|}=[\int\!Dz~|z|\Omega(|z|/\lambda)]^{-1}$, and we obtain the following explicit
parametrization of the $\chiR=0$ line:
\begin{eqnarray}
A_0(\lambda)&=& \frac{1}{\int\!Dz~|z|\Omega(|z|/\lambda)}\sqrt{3-\frac{2}{\lambda\int\!Dz~|z|\Omega(|z|/\lambda)}}
\\
\alpha(\lambda)&=&\frac{1}{\pi}\Big[\frac{\int\!Dz~\Omega^2(|z|/\lambda)}{[\int\!Dz~|z|\Omega(|z|/\lambda)]^2}-1\Big]^{-1}
\end{eqnarray}
Let us determine the extremal points. For $\lambda\to\infty$ we  use
$\Omega(x)=\frac{2}{3}x+\frac{8}{81}x^3+\order(x^5)$ and get
\begin{eqnarray}
\lim_{\lambda\to\infty}A_0(\lambda)&=& \sqrt{3},~~~~~~~~
\lim_{\lambda\to\infty}\alpha^{-1}(\lambda)=0
\end{eqnarray}
For small $\lambda$ we note that the $\chiR=0$ line terminates at the value $\lambda_c$ where $A_0$ ceases to be real-valued.
 This value $\lambda_c$ is the solution of
\begin{eqnarray}
\frac{3}{2}\lambda\int\!Dz~|z|\Omega(|z|/\lambda)=1
\label{eq:lambdac}
\end{eqnarray}
and corresponds to the following point in the $(\alpha,A_0)$ plane:
\begin{eqnarray}
A_0=0,~~~~~~~~\alpha&=&\frac{1}{\pi}\Big[\frac{9}{4}\lambda_c^2\int\!Dz~\Omega^2(|z|/\lambda_c)-1\Big]^{-1}
\label{eq:lambdac_point}
\end{eqnarray}
\vsp

If $\tau=-1$
we have $q=-\frac{1}{3}A_0^2(1\!-\!\chi^{-1})$ and
$r(z)=\frac{1}{2}A_0^2 z/\chi$. Here $3q+A_0^2=A_0^2/\chi\geq 0$ so $q\geq -\frac{1}{3}A_0^2$, and in the remanent region we
select the Maxwell option (\ref{eq:Maxwell}). If now we define as our line parameter
$\lambda= 2|q|^{3/2}/(3q+A_0^2)\geq 0$ we will have
$r(z)/|q|^{3/2}= z/\lambda$, and hence
\begin{eqnarray}
\hspace*{-25mm}
&&
\Delta^{-1}(z)=~\sgn(z)\sqrt{|q|}~\Omega(|z|/\lambda)
\\[1mm]
\hspace*{-25mm}
q>0:&~~~& \Omega(x)=[x\!+\!\sqrt{1\!+\!x^2}]^{\frac{1}{3}}+[x\!-\!\sqrt{1\!+\!x^2}]^{\frac{1}{3}}
\\
\hspace*{-25mm}
q<0:&~~~& \Omega(x)= \theta[x\!-\!1]\Big\{[x\!+\!\sqrt{x^2\!-\!1}]^{\frac{1}{3}}+[x\!-\!\sqrt{x^2\!-\!1}]^{\frac{1}{3}}\Big\}
+2\theta[1\!-\!x]~\cos(\frac{1}{3}{\rm arccos}(x))
\end{eqnarray}
Again we find the first equation of
(\ref{eq:chiR0_conditions}) giving $\sqrt{|q|}=[\int\!Dz~|z|\Omega(|z|/\lambda)]^{-1}$, but now our parametrization becomes
\begin{eqnarray}
A_0(\lambda)&=& \frac{1}{\int\!Dz~|z|\Omega(|z|/\lambda)}\sqrt{\frac{2}{\lambda\int\!Dz~|z|\Omega(|z|/\lambda)}-3~\sgn(q)}
\\
\alpha(\lambda)&=&\frac{1}{\pi}\Big[\frac{\int\!Dz~\Omega^2(|z|/\lambda)}{[\int\!Dz~|z|\Omega(|z|/\lambda)]^2}-1\Big]^{-1}
\end{eqnarray}
 In the limit $\lambda\to\infty$ we use the expansions $\Omega(x)=\frac{2}{3}x-\frac{8}{81}x^3+\order(x^5)$ for $q>0$,
 and $\Omega(x)=\sqrt{3}+\frac{1}{3}x-\frac{\sqrt{3}}{18}x^2+\order(x^3)$ for $q<0$.
 We then find
\begin{eqnarray}
\hspace*{-10mm}
q>0:&~~~~ \lim_{\lambda\to 0}A_0(\lambda)= \sqrt{3}\approx 1.7321,~~~~&\lim_{\lambda\to\infty}\alpha^{-1}(\lambda)=0
\\
\hspace*{-10mm}
q<0:&~~~~ \lim_{\lambda\to\infty} A_0(\lambda)= \sqrt{\frac{\pi}{2}}\approx 1.2533,~~~~&
\lim_{\lambda\to\infty} \alpha(\lambda)=\frac{2}{\pi(\pi\!-\!2)}\approx 0.5576
\end{eqnarray}
The latter point coincides with the termination point of the $\chi\to \infty$ transition line for $\tau=-1$.
The $\chiR=0$ line for $\tau=-1$ can only have a finite $\lambda$ termination point with $A_0=0$ if $q>0$, i.e. in the remanent region.
In practice we find no such points for $\tau=-1$, irrespective of $q$.

\subsection{Phase diagrams}

\begin{figure}[t]
\vspace*{3mm} \hspace*{-8mm} \setlength{\unitlength}{0.52mm}
\begin{picture}(300,100)\label{fig:phasediagrams_nonergodic}

  \put(10,0){\includegraphics[height=100\unitlength,width=170\unitlength]{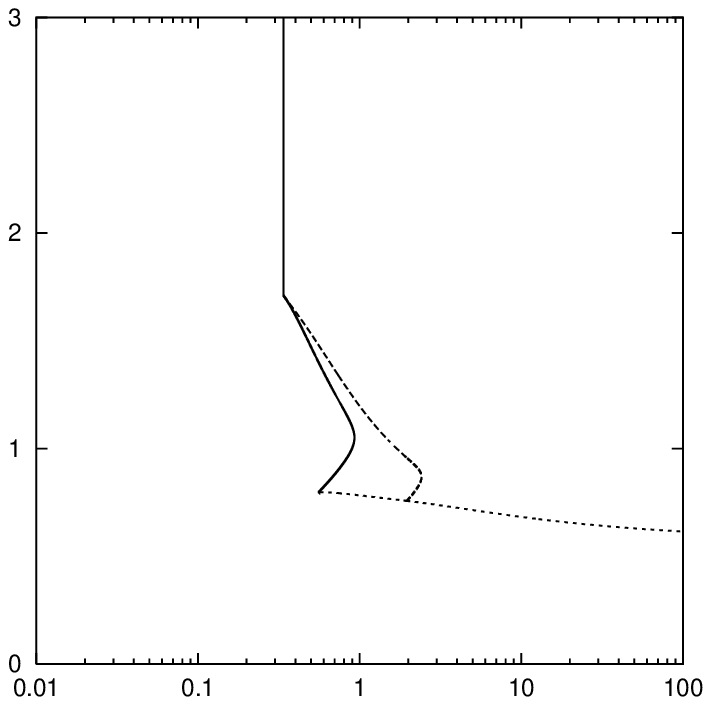}}
  \put(7,55){\large\here{$A_0^{-1}$}}  \put(65,-11){\large$\alpha$}
  \put(65,100){$\tau=-1$}
  \put(84,80){MinGame}  \put(87,72){$\chi<\infty$}
  \put(24,80){MinGame} \put(27,72){$\chi=\infty$}
  \put(54,10){MajGame}  \put(72,29){R}

  \put(145,0){\includegraphics[height=100\unitlength,width=170\unitlength]{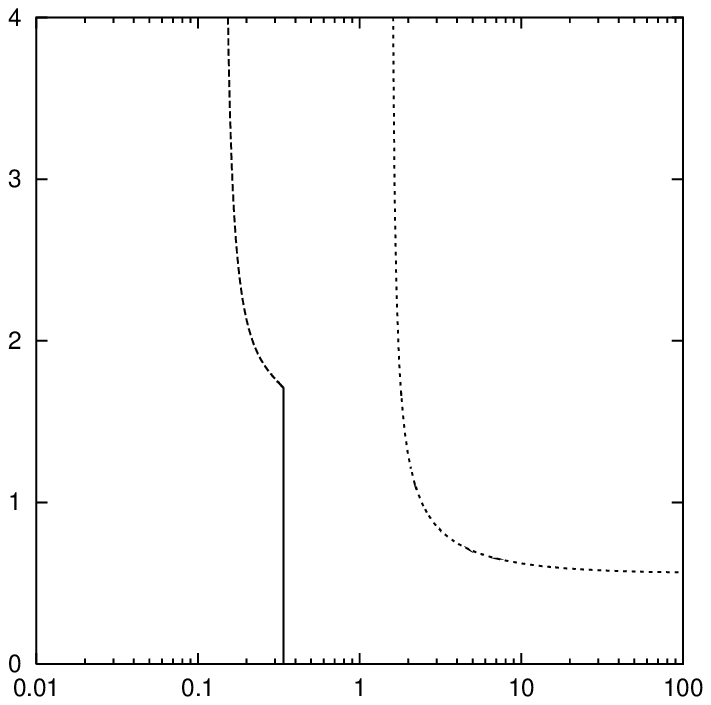}}
  \put(142,55){\large\here{$A_0^{-1}$}}  \put(202,-11){\large$\alpha$}
  \put(202,100){$\tau=1$}
  \put(160,20){MinGame} \put(163,12){$\chi=\infty$}
  \put(220,61){R-MajGame}
  \put(205,10){R-MinGame}
  \put(155,80){MajGame}

\end{picture}
 \vspace*{4mm}
\caption{Phase diagrams for the models with
$F[A]=\tau A[1-A^2/A_0^2]$, $\tau=\pm1$.
Solid lines: $\chi=\infty$ transition, dashed lines: remanence onset transition, dotted lines: $\chiR=0$ (minority-to-majority) transition.
The MinGame phase is characterized by $\chiR>0$ (contrarian trading), the MajGame phase by $\chiR<0$ (trend-following). R indicates remanence.
In the left picture ($\tau=-1$, giving trend-following for $|A|<A_0$ and contrarian trading for $|A|>A_0$), the phase diagram is reasonably complete and reliable. For small $A_0$ we have only the MinGame phase, with the standard $\chi\to\infty$ transition at $\alpha_c\!\approx\! 0.3374$. As $A_0$ is increased a small remanent region is formed, until we switch to a MajGame state for large $A_0$.
In the right picture ($\tau=1$, giving contrarian trading for $|A|<A_0$ and trend-following for $|A|>A_0$),
 we find the MinGame phase, with the standard $\chi\to\infty$ transition at $\alpha_c\!\approx\! 0.3374$, for large $A_0$ (as expected).
However, the overall picture is now more uncertain since most of the diagram is remanent. Since we can never be sure of picking the right solution in the remanent phase, the location of the $\chiR=0$ line cannot be taken as certain, and indeed intuition suggests that it should have connected to the point where the $\chi=\infty$ and the remanence onset lines meet. In both diagrams we  cannot draw the line separating the non-ergodic MinGame
phase from the MajGame phase for small $\alpha$; as this line involves non-ergodic phases, we have no access to it with a theory that assumes ergodicity.
}\label{fig:phasediagram}
\end{figure}
We can now summarize the picture obtained by analyzing time-translation invariant
states in terms of a phase diagram. The control parameters are $\alpha\geq 0$, $A_0\geq 0$ and $\tau=\pm 1$. We have so far identified
three transition types:  $\chi\to\infty$, $\chiR\to 0$ and $q\to 0$.
The result is shown in figure \ref{fig:phasediagram}, and discussed in detail in the corresponding figure caption.
Most of the technical subtleties in the present models are generated by remanence, so the phase diagram for $\tau=-1$ (giving the model of
\cite{demartino2}) is the most complete and reliable, since the remanent region is relatively contained.
The situation is different for $\tau=1$, where most of the phase diagram is remanent, and consequently we cannot be certain of the location of the $\chiR=0$ line. A proper resolution here would require going beyond the asymptotic limit of the effective agent and the effective overall bid processes, in order to determine dynamic stability and dependence on initial conditions of the values of the persistent order parameters.

\subsection{Simulations}

We have tested our predictions for the values of persistent order parameters and the locations of phase transition lines, for the models (\ref{eq:giardina_type}) and for both values of $\tau$.
All simulations were carried out without decision noise, for systems of size $N = 4097$ and for both unbiased ($q_i(0) =\pm 10^{-4}$)
and biased ($q_i(0) =\pm 1$) random initializations. Observables were measured during 1000 batch steps, following an equilibration stage of 2000
batch steps.

\begin{figure}[t]
\vspace*{-9mm} \hspace*{10mm} \setlength{\unitlength}{0.35mm}
\begin{picture}(350,330)

\put(0,000){\epsfxsize=110\unitlength\epsfbox{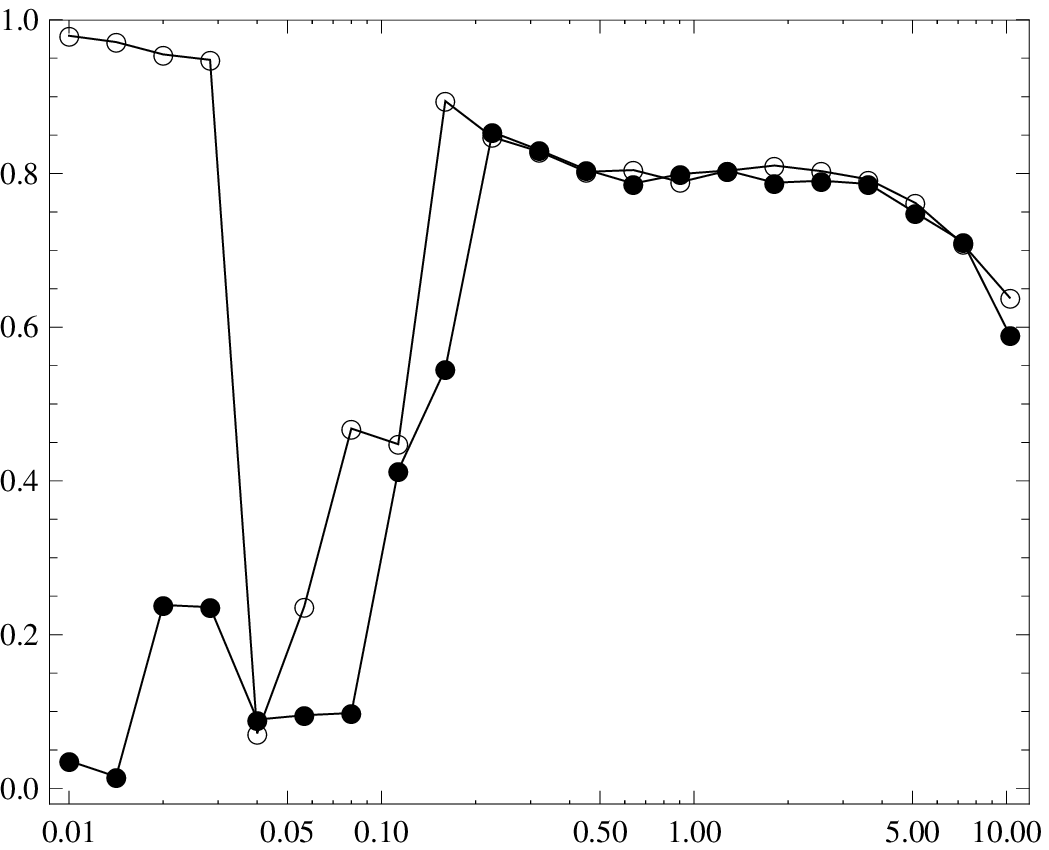}}
 \put(115,000){\epsfxsize=110\unitlength\epsfbox{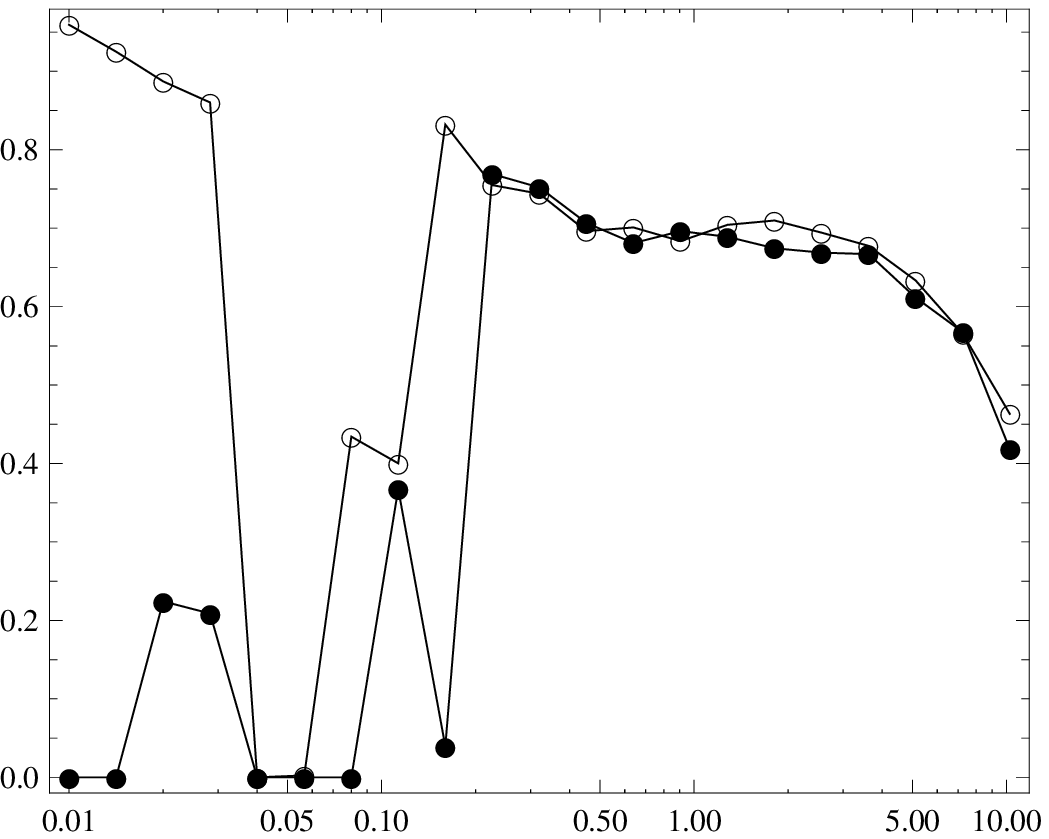}}
\put(230,000){\epsfxsize=110\unitlength\epsfbox{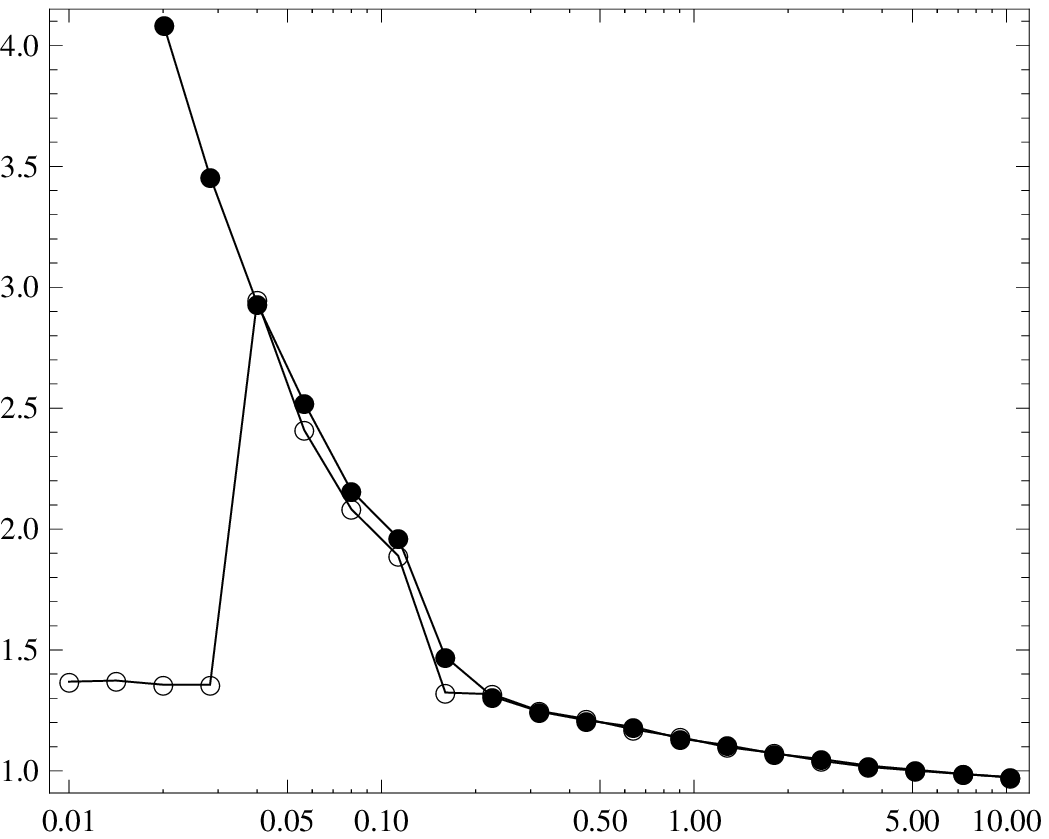}}

\put(0,100){\epsfxsize=110\unitlength\epsfbox{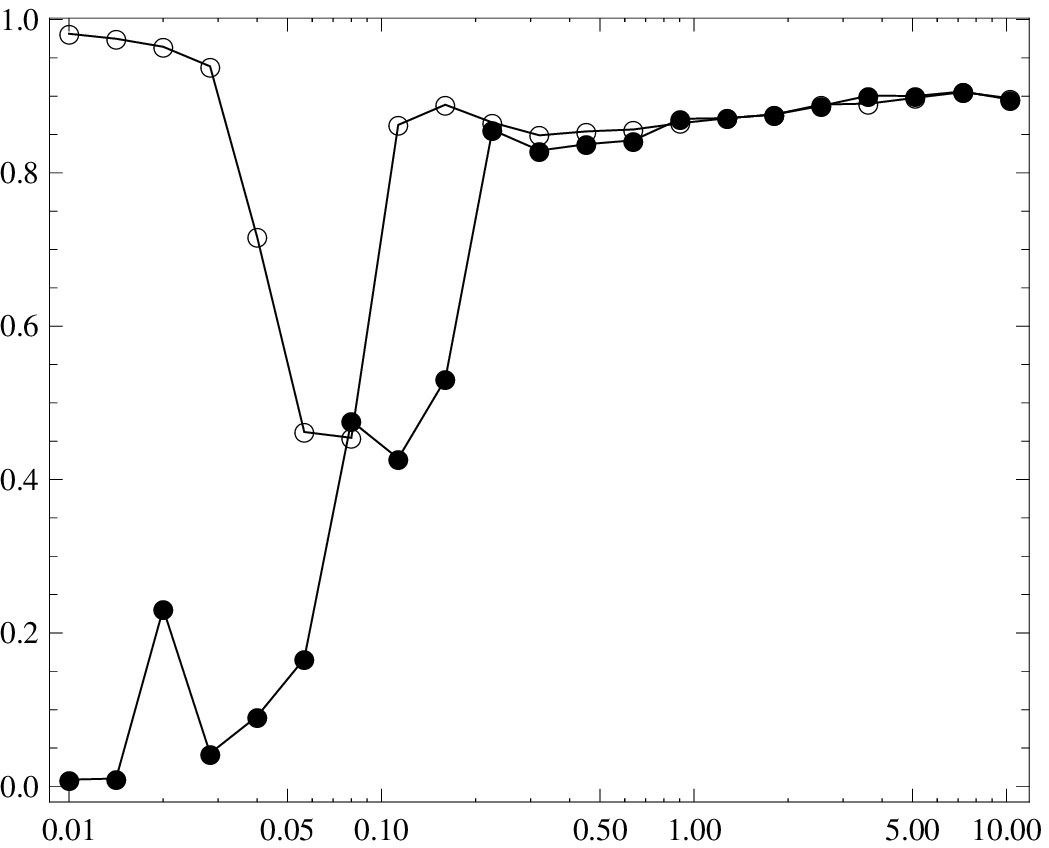}}
 \put(115,100){\epsfxsize=110\unitlength\epsfbox{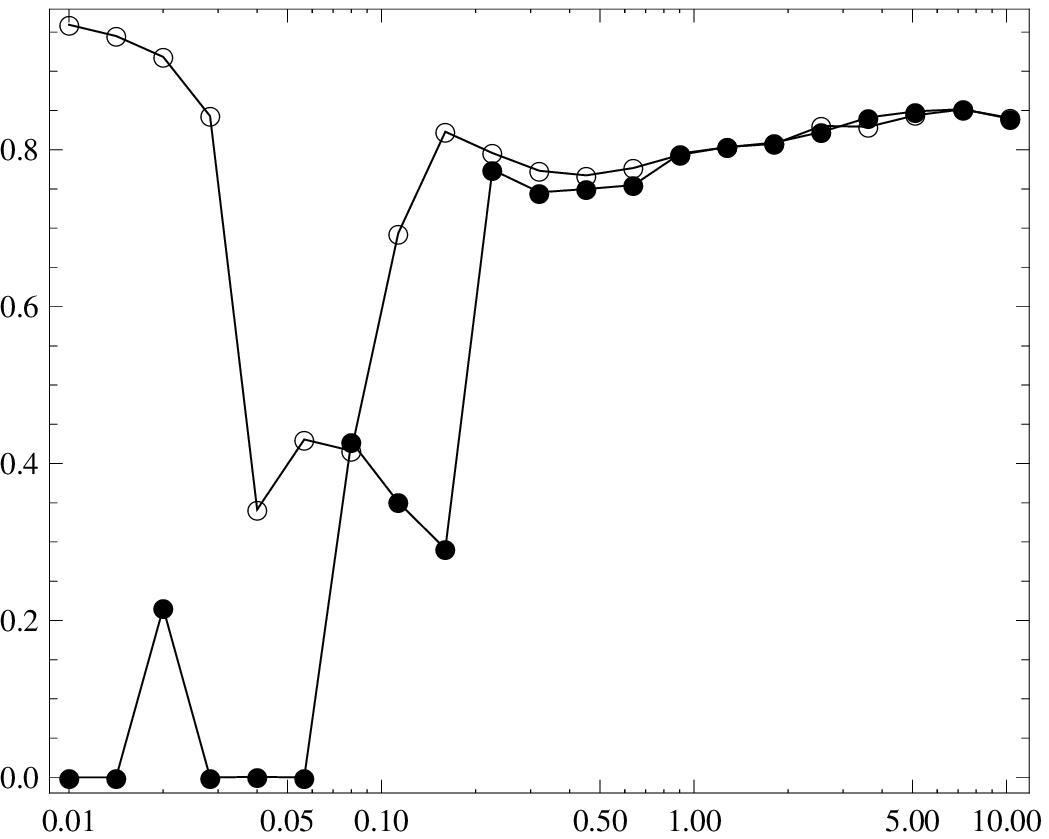}}
\put(230,100){\epsfxsize=110\unitlength\epsfbox{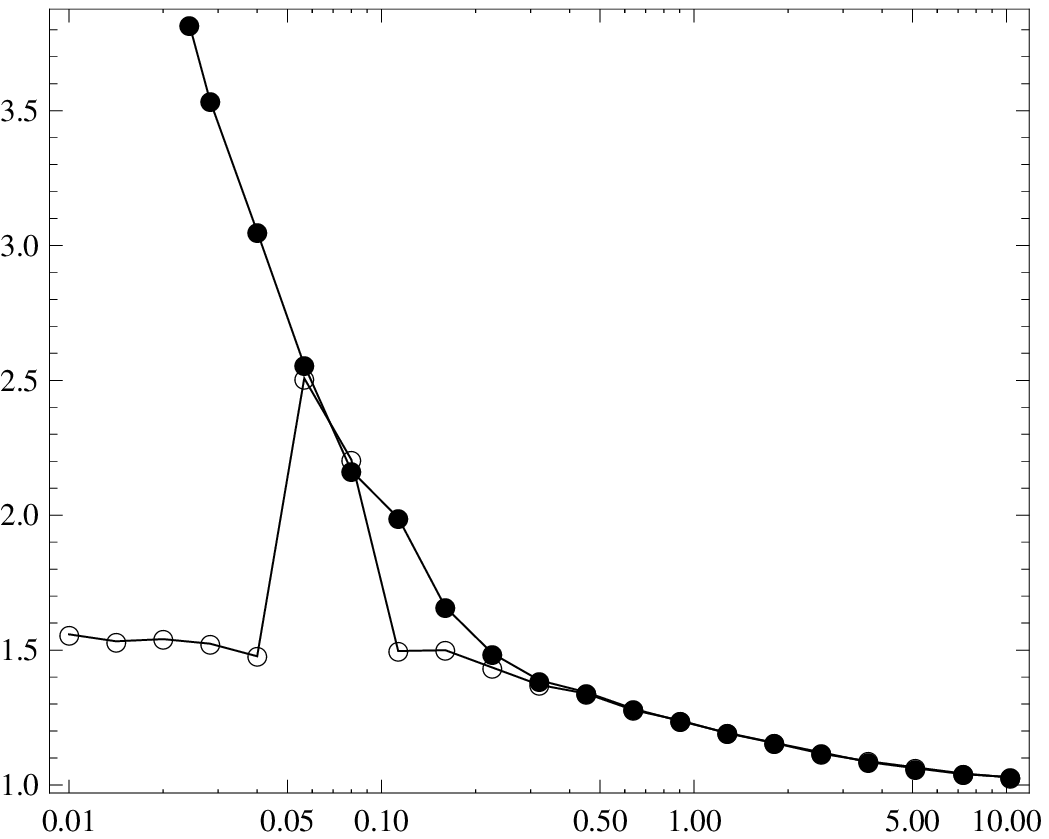}}

\put(0,200){\epsfxsize=110\unitlength\epsfbox{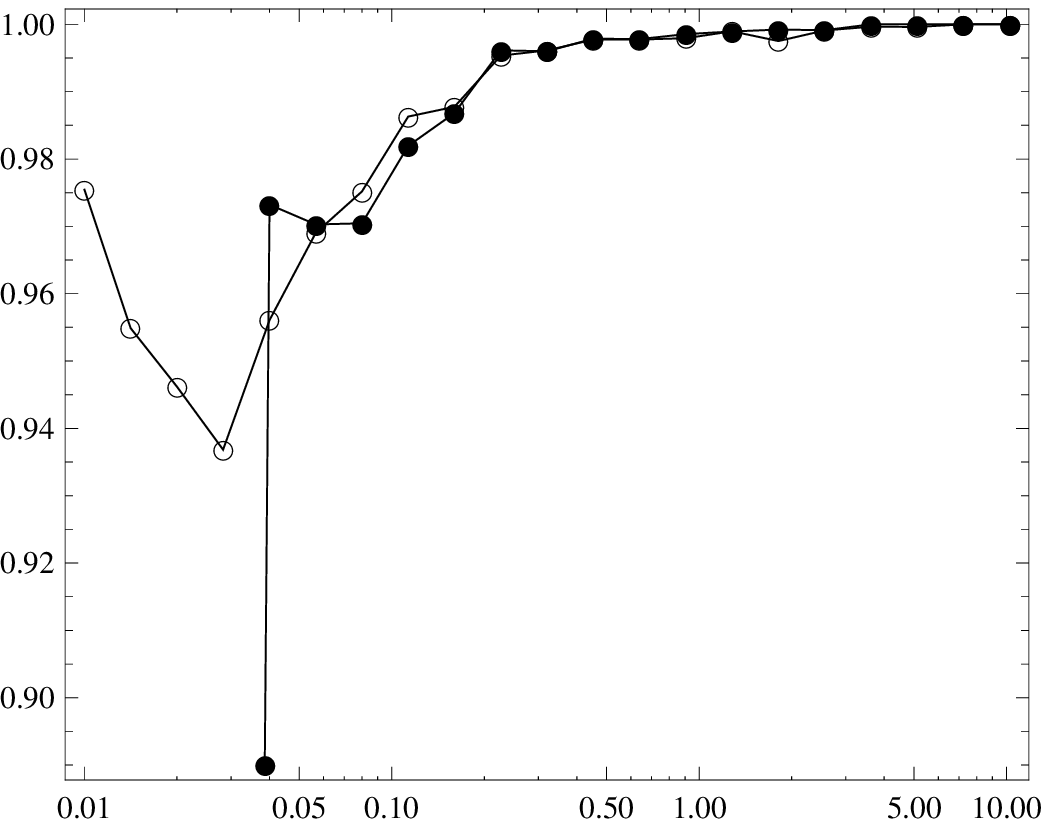}}
 \put(115,200){\epsfxsize=110\unitlength\epsfbox{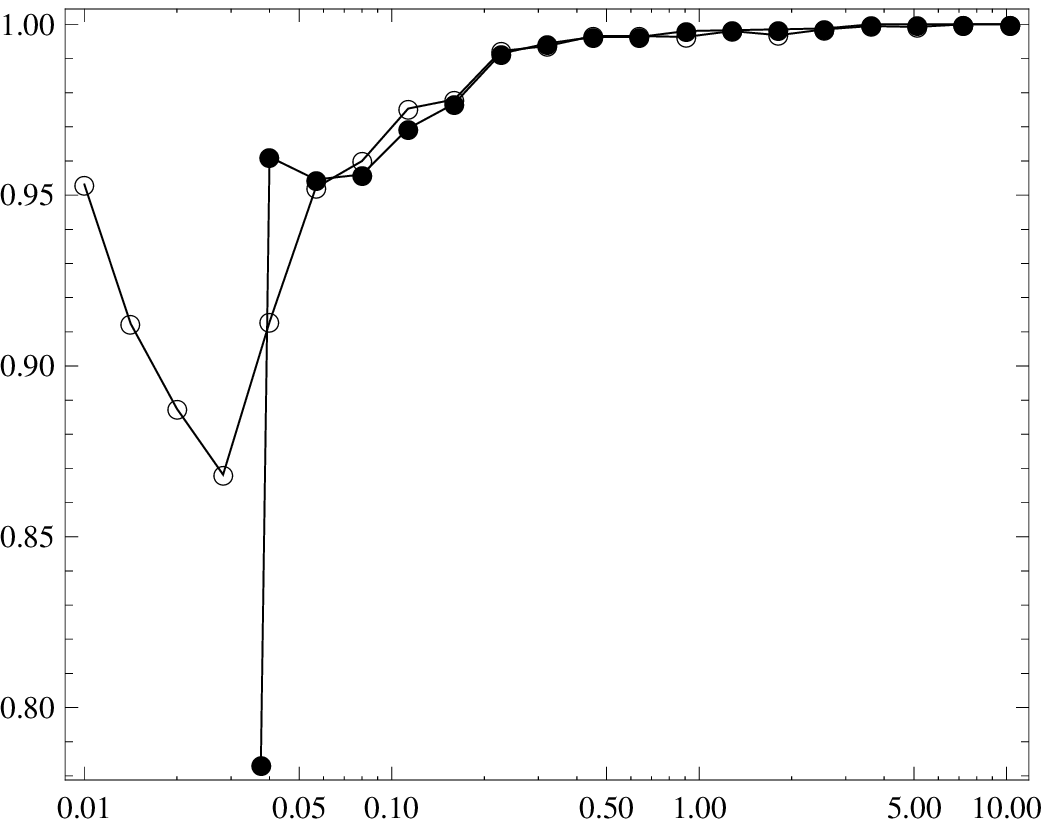}}
\put(233,200){\epsfxsize=106\unitlength\epsfbox{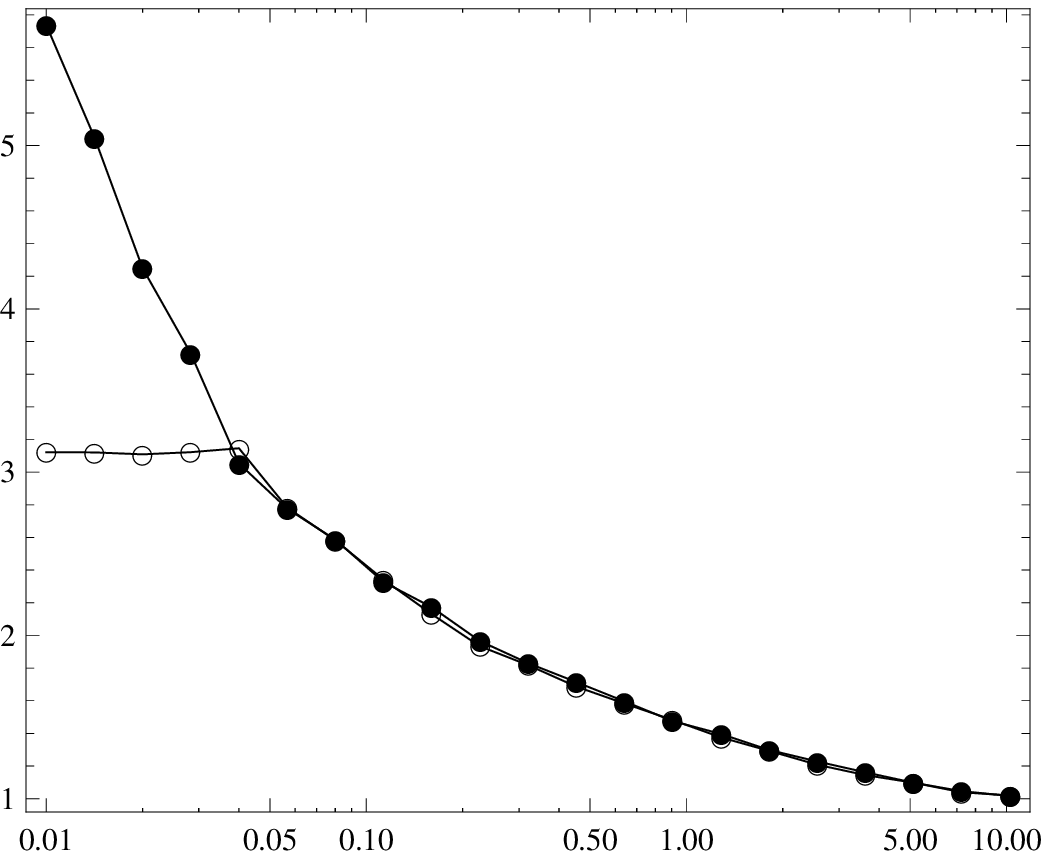}}

\put(55,-10){$\alpha$}\put(170,-10){$\alpha$}\put(288,-10){$\alpha$}

\put(50,295){$c$}
\put(170,295){$\phi$} \put(280,295){$\sigma$}
 \put(-55,50){$A_0^{-1}\!\approx 0.71$}
 \put(-55,150){$A_0^{-1}\!\approx 0.62$}
 \put(-55,250){$A_0^{-1}\!\approx 0.32$}
\end{picture}
\vspace*{1mm}
\caption{Simulation results for the persistent correlations $c$, the
fraction of frozen agents $\phi$, and the volatility $\sigma$, for $\tau=-1$ (i.e. $F[A]=-A+A^3/A_0^2$).
The chosen values of $A_0^{-1}$ should, according to the phase diagram, probe the MajGame phase of the model, without any $\chi=\infty$ transition.
Full markers: \emph{tabula rasa} initial conditions; empty markers: biased initial conditions.
}
\label{A0small}
\end{figure}

\begin{figure}[t]
\vspace*{-9mm} \hspace*{10mm} \setlength{\unitlength}{0.35mm}
\begin{picture}(350,330)

\put(0,000){\epsfxsize=110\unitlength\epsfbox{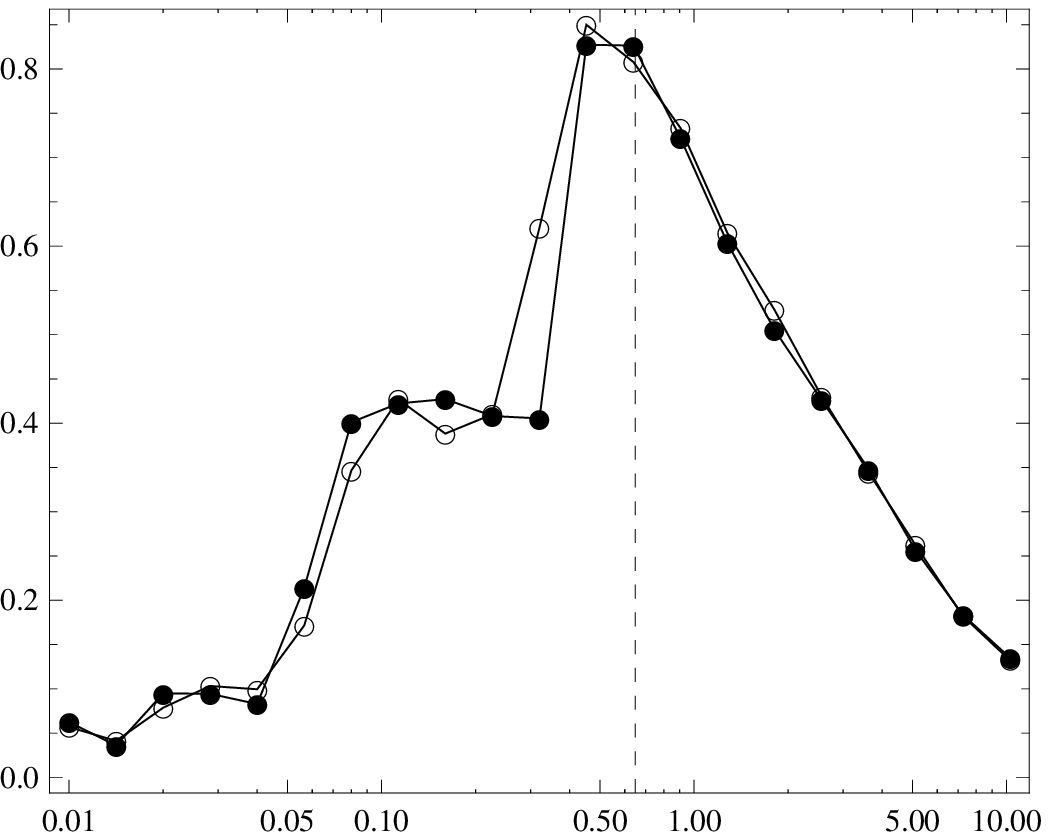}}
 \put(115,000){\epsfxsize=110\unitlength\epsfbox{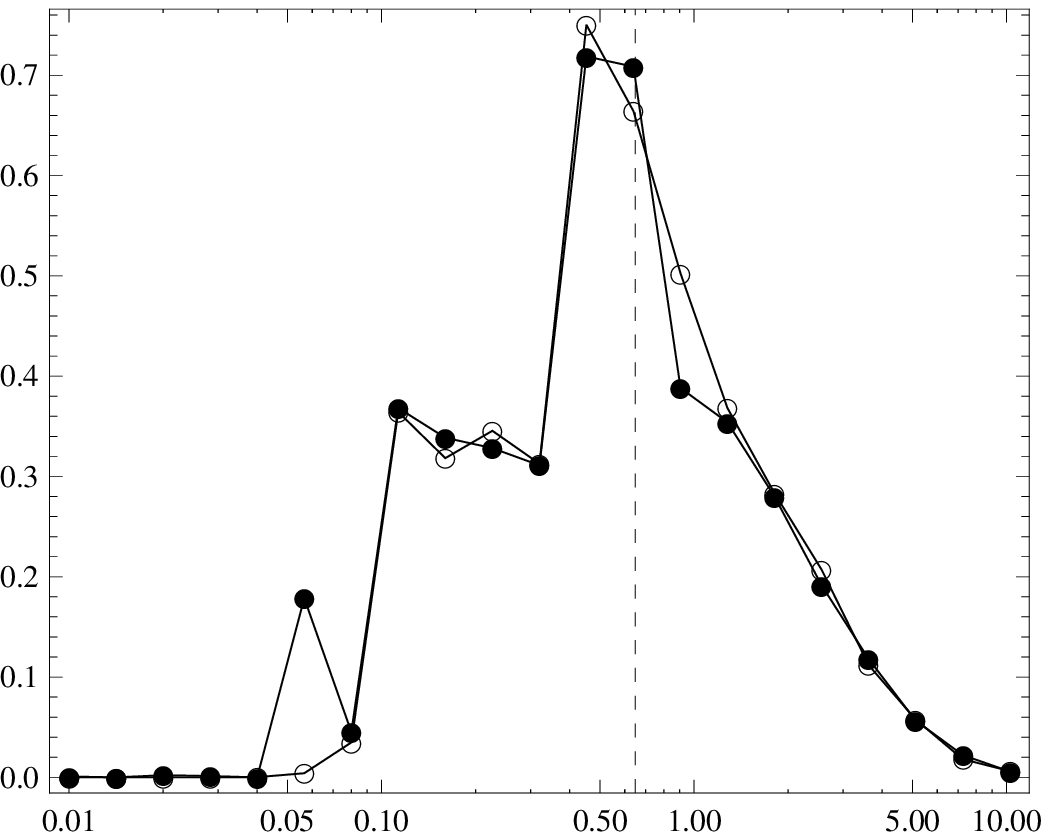}}
\put(230,000){\epsfxsize=108\unitlength\epsfbox{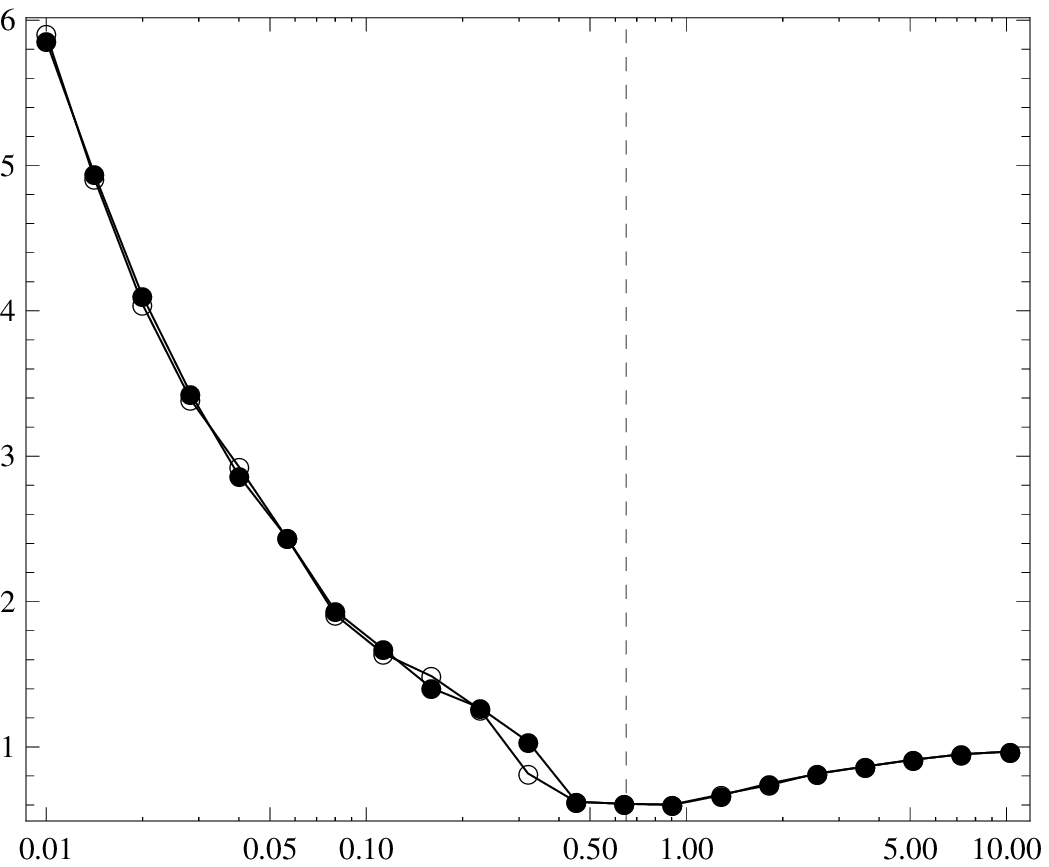}}

\put(0,100){\epsfxsize=110\unitlength\epsfbox{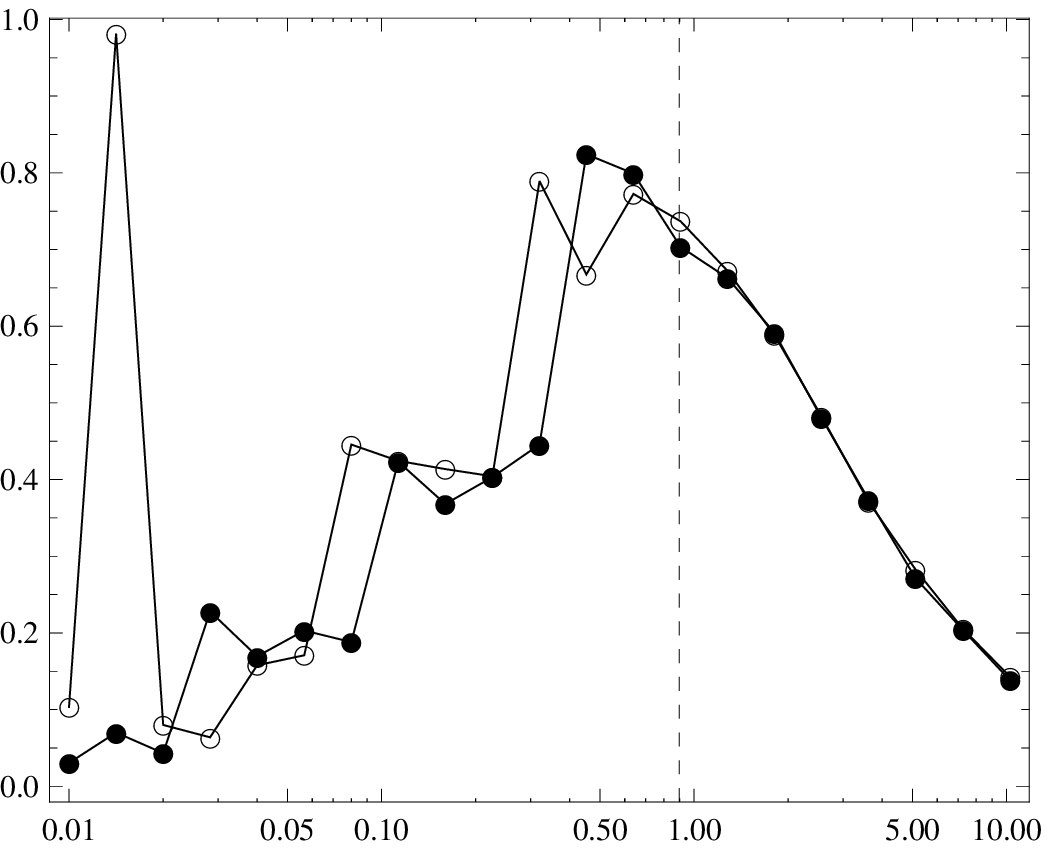}}
 \put(115,100){\epsfxsize=110\unitlength\epsfbox{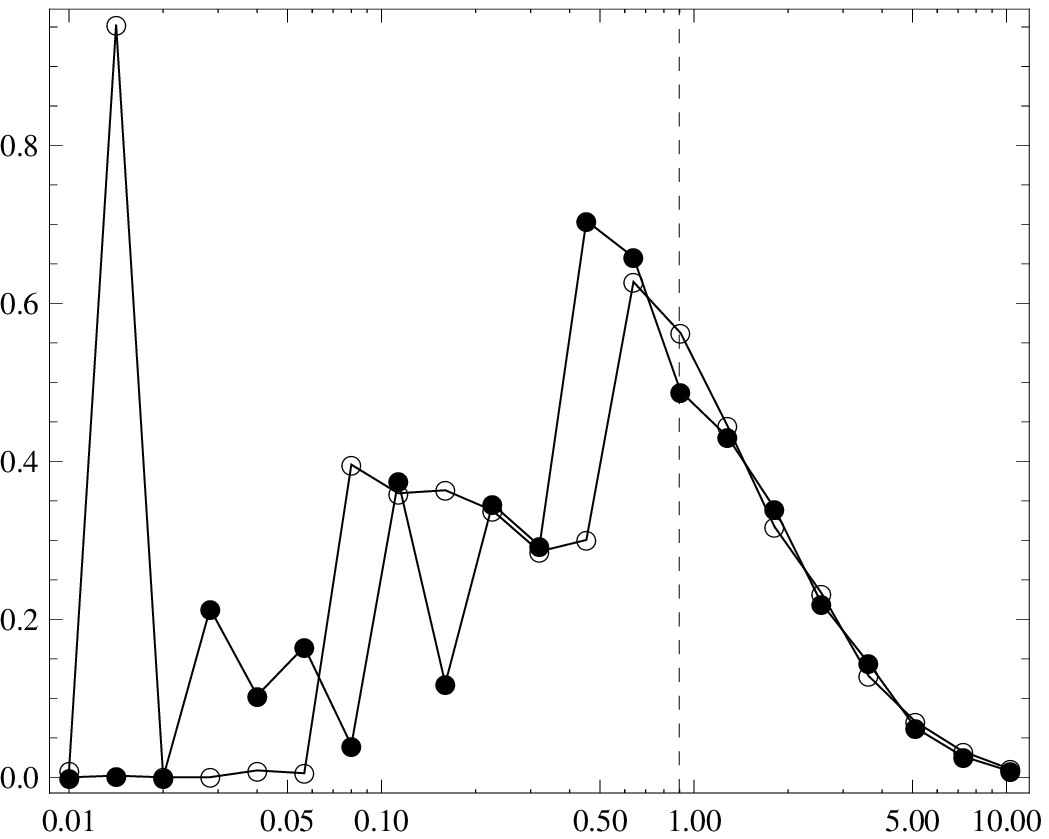}}
\put(230,100){\epsfxsize=108\unitlength\epsfbox{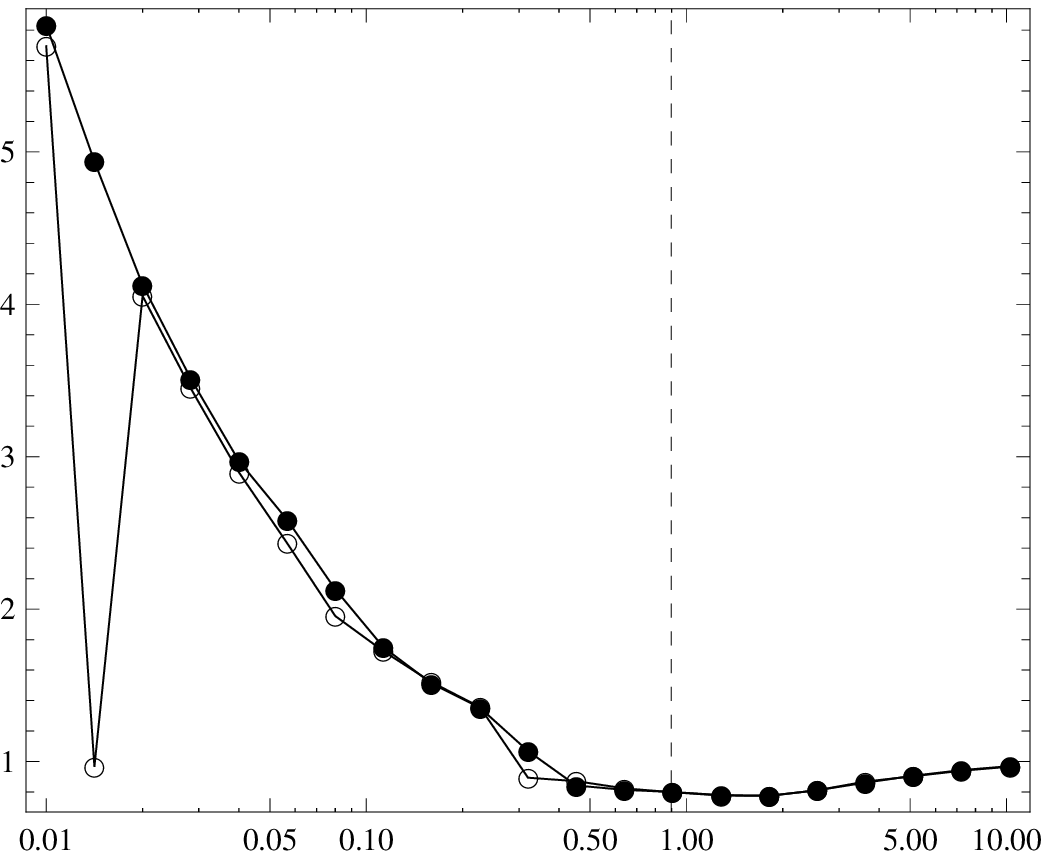}}

\put(0,200){\epsfxsize=110\unitlength\epsfbox{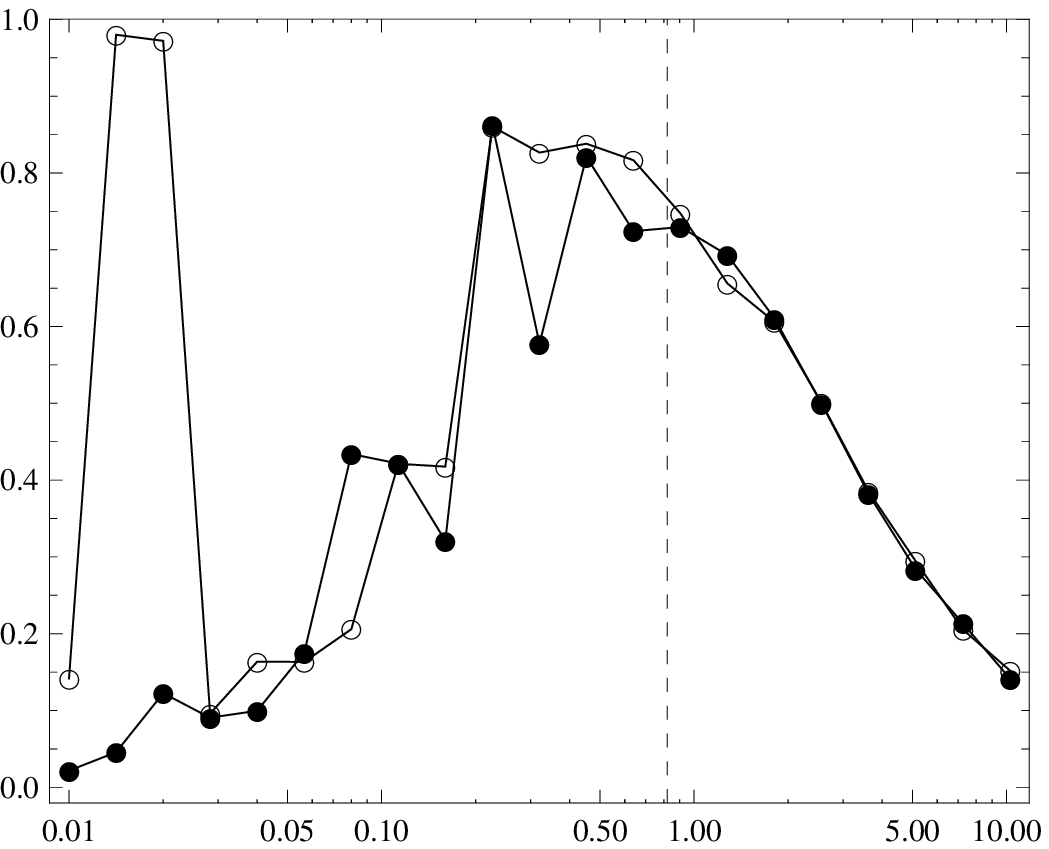}}
 \put(115,200){\epsfxsize=110\unitlength\epsfbox{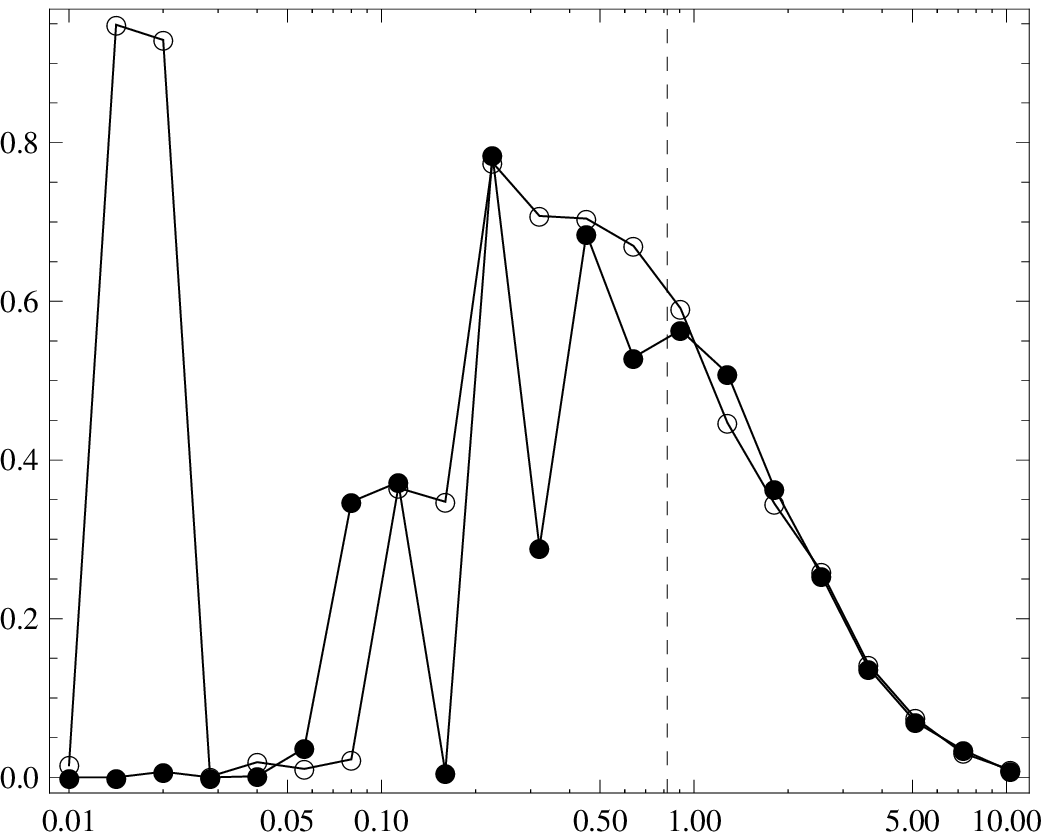}}
\put(230,200){\epsfxsize=108\unitlength\epsfbox{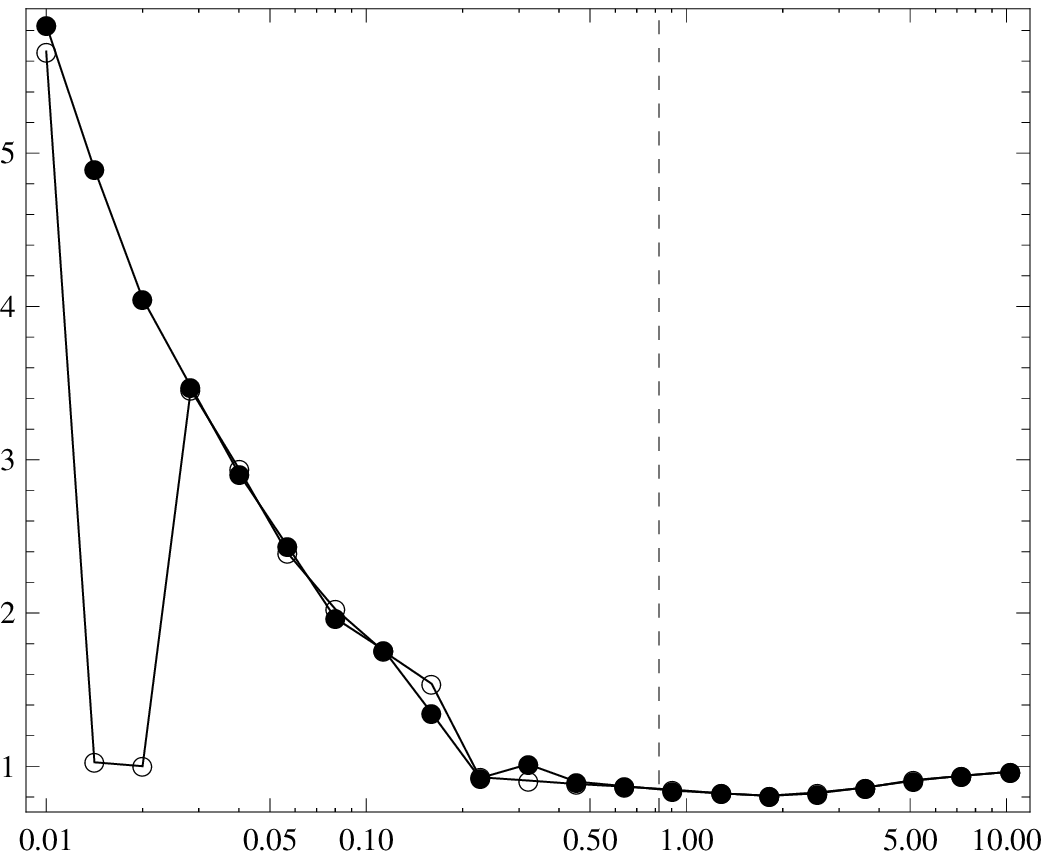}}

\put(55,-10){$\alpha$}\put(170,-10){$\alpha$}\put(288,-10){$\alpha$}

\put(50,295){$c$}
\put(170,295){$\phi$} \put(280,295){$\sigma$}
 \put(-55,50){$A_0^{-1}\!\approx 1.30$}
 \put(-55,150){$A_0^{-1}\!=1$}
 \put(-55,250){$A_0^{-1}\!\approx 0.95$}
\end{picture}
\vspace*{1mm}
\caption{
Simulation results for the persistent correlations $c$, the
fraction of frozen agents $\phi$, and the volatility $\sigma$, for $\tau=-1$ (i.e. $F[A]=-A+A^3/A_0^2)$.
The chosen values of $A_0^{-1}$ should, according to the phase diagram, probe the region close where a $A_0$-dependent $\chi=\infty$ transition
occurs (indicated with a vertical dashed line), with a MinGame phase for large $\alpha$.
Full markers: \emph{tabula rasa} initial conditions; empty markers: biased initial conditions.
}\label{A0medium}
\end{figure}

\begin{figure}[t]
\vspace*{-9mm} \hspace*{10mm} \setlength{\unitlength}{0.35mm}
\begin{picture}(350,330)

\put(0,000){\epsfxsize=110\unitlength\epsfbox{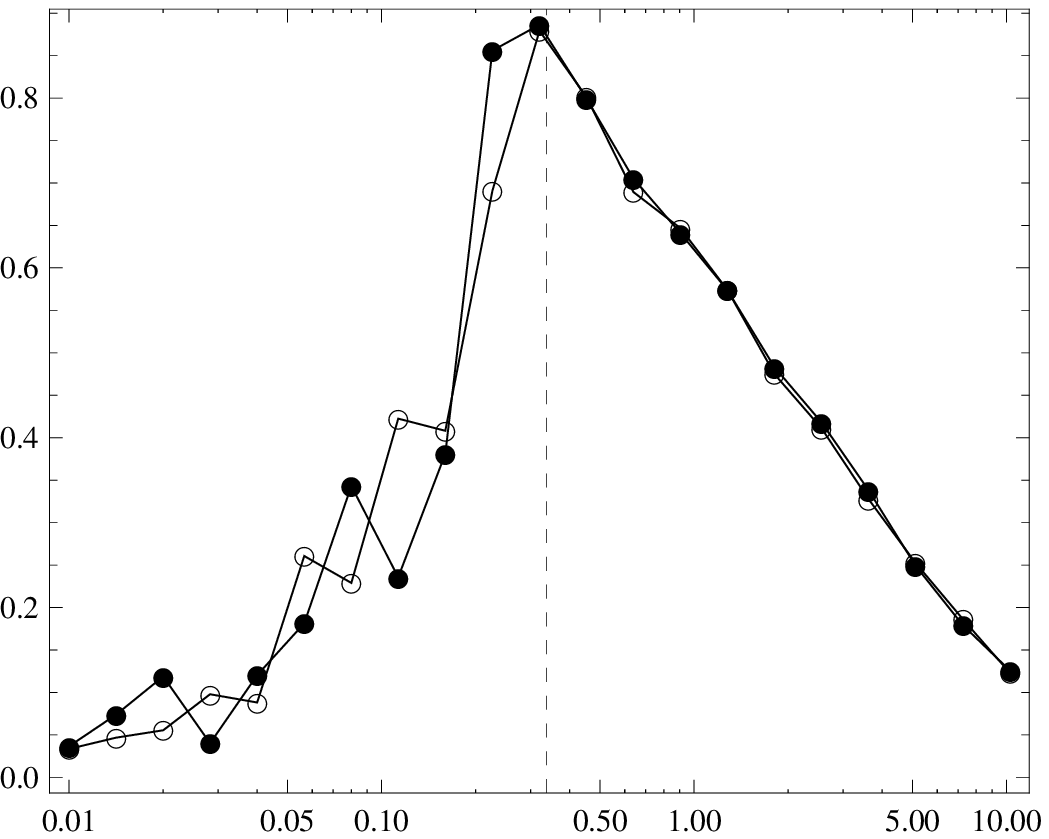}}
 \put(115,000){\epsfxsize=110\unitlength\epsfbox{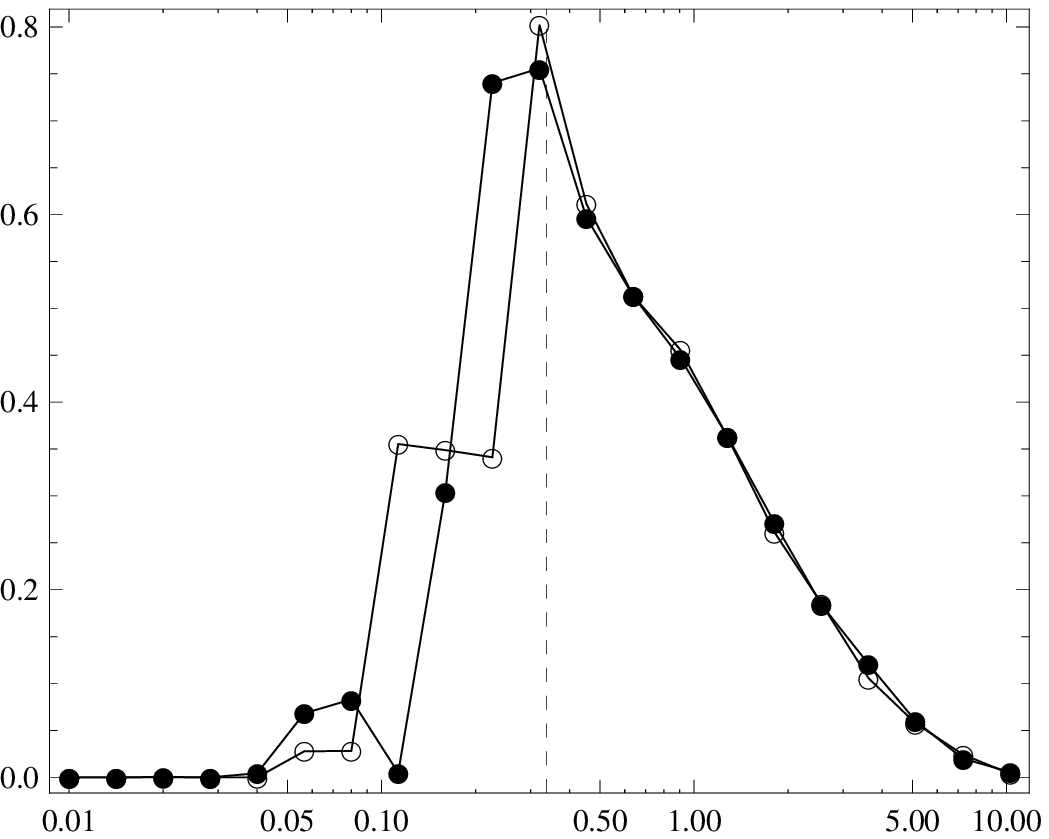}}
\put(230,000){\epsfxsize=108\unitlength\epsfbox{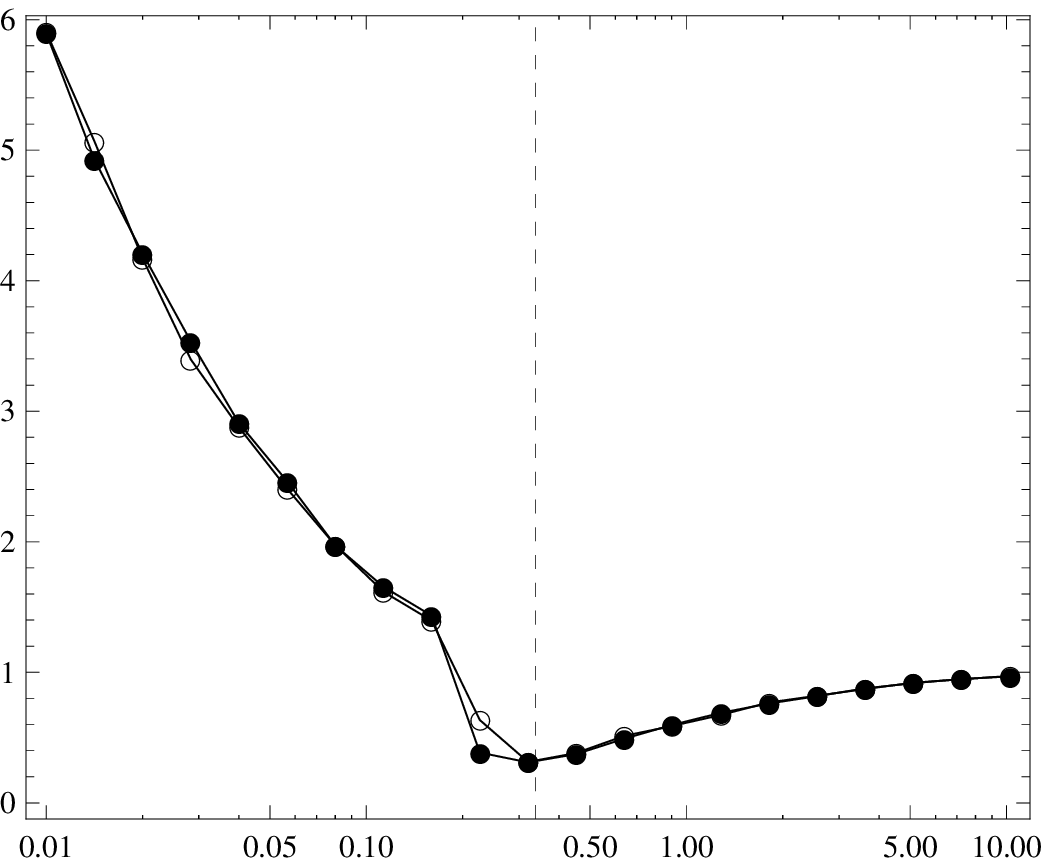}}

\put(0,100){\epsfxsize=110\unitlength\epsfbox{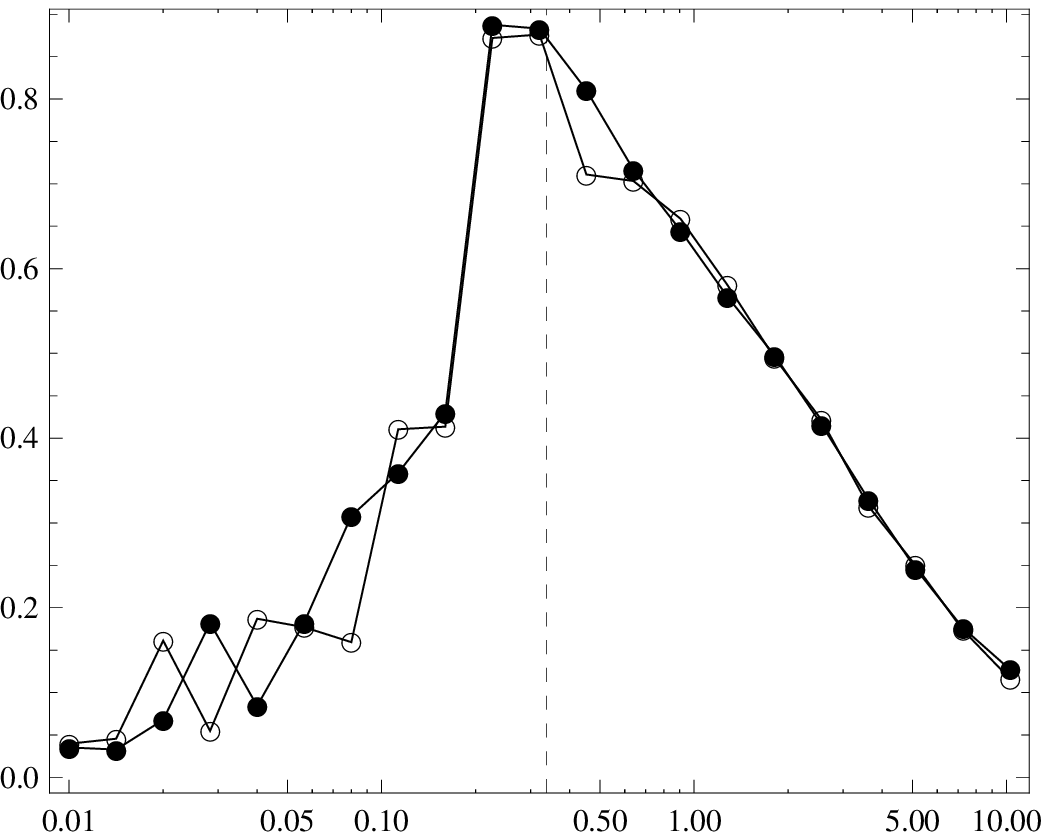}}
 \put(115,100){\epsfxsize=110\unitlength\epsfbox{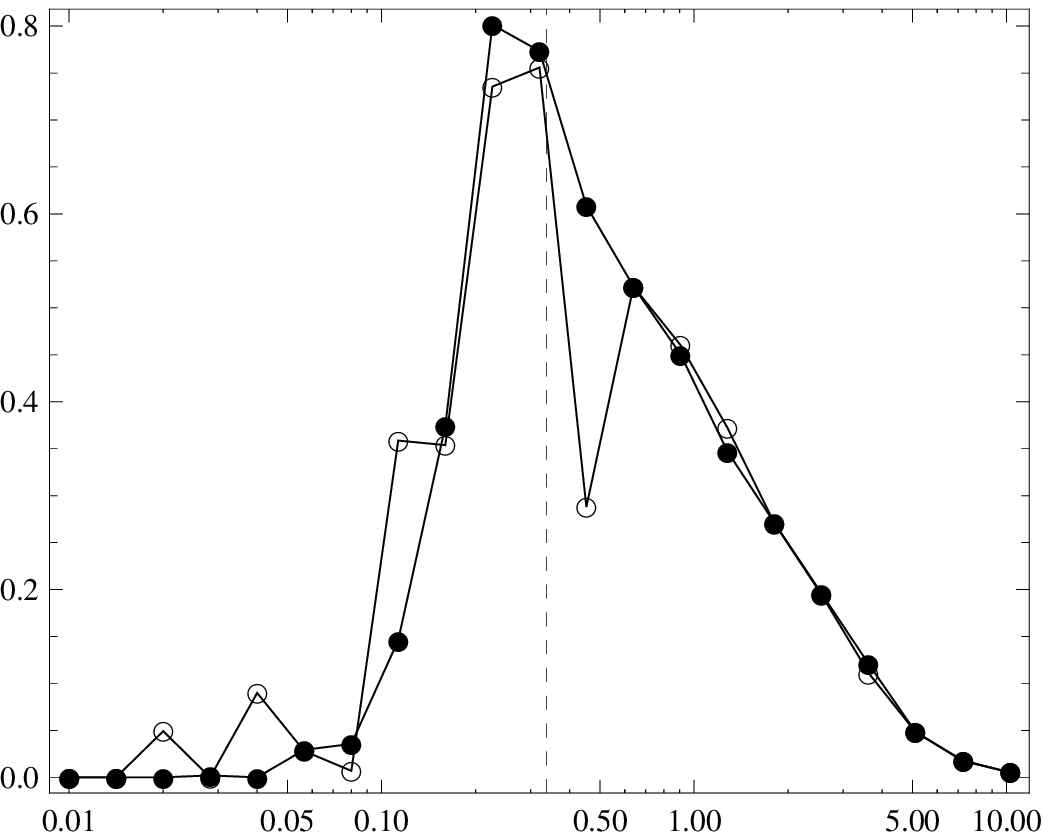}}
\put(230,100){\epsfxsize=108\unitlength\epsfbox{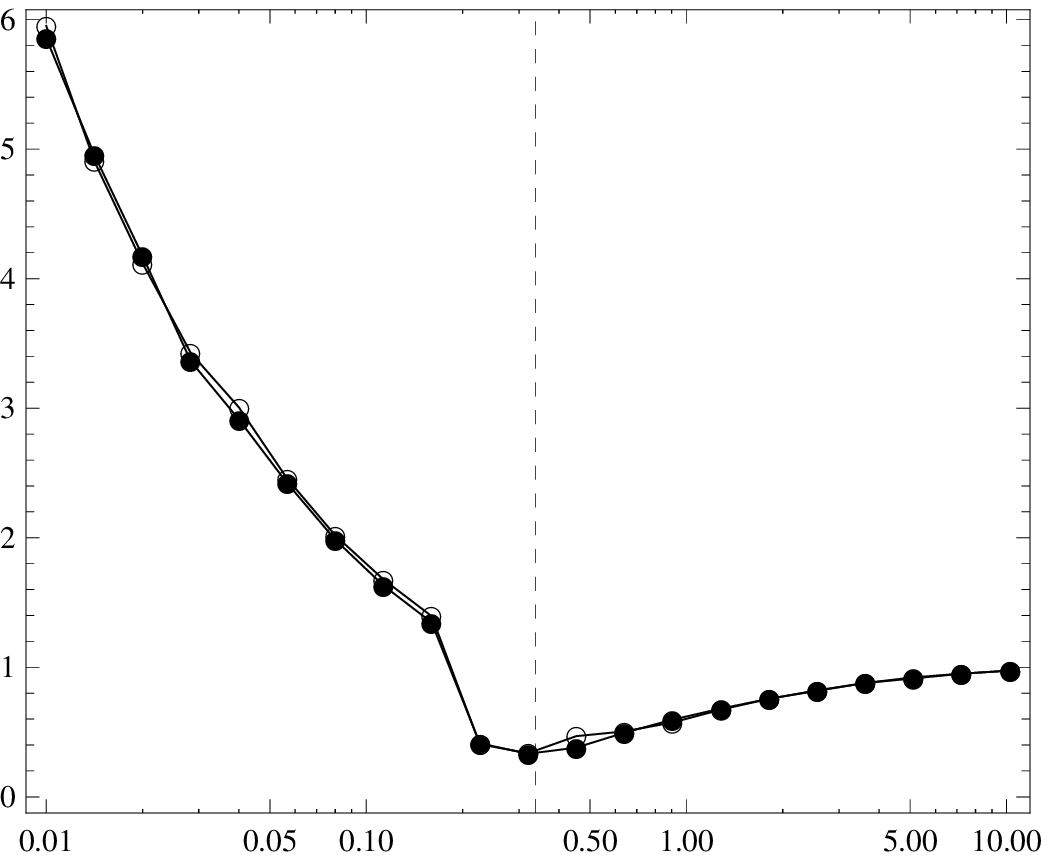}}

\put(0,200){\epsfxsize=110\unitlength\epsfbox{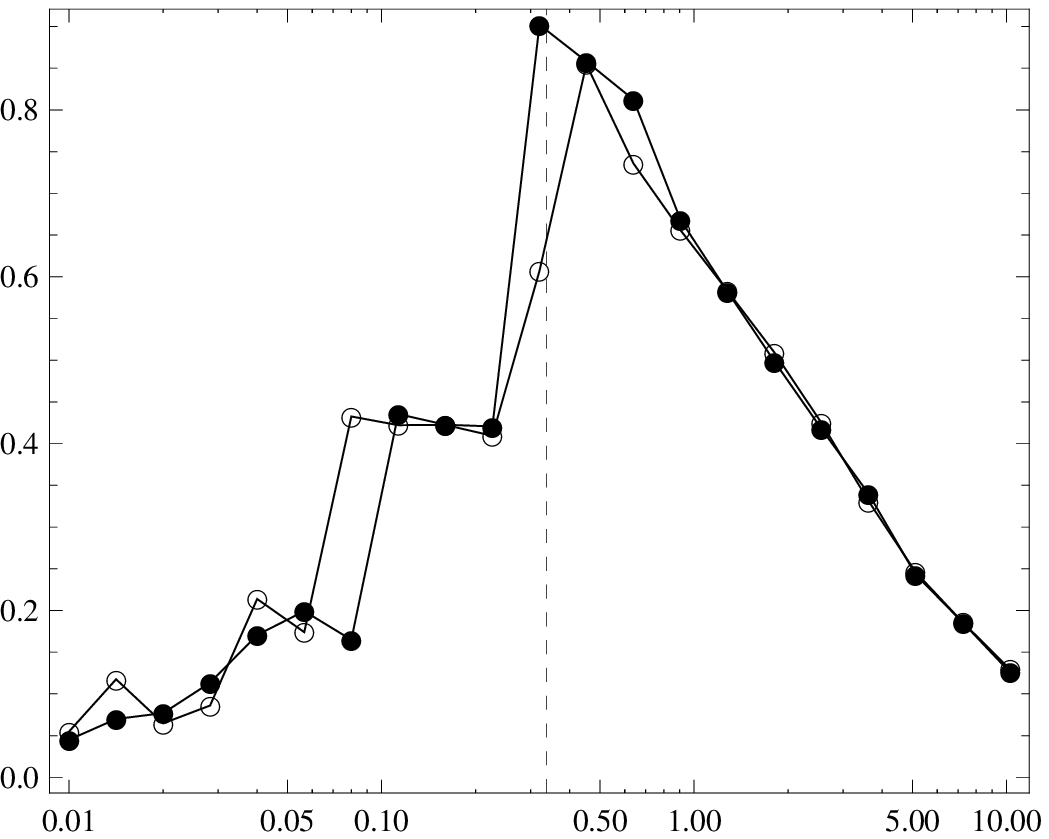}}
 \put(115,200){\epsfxsize=110\unitlength\epsfbox{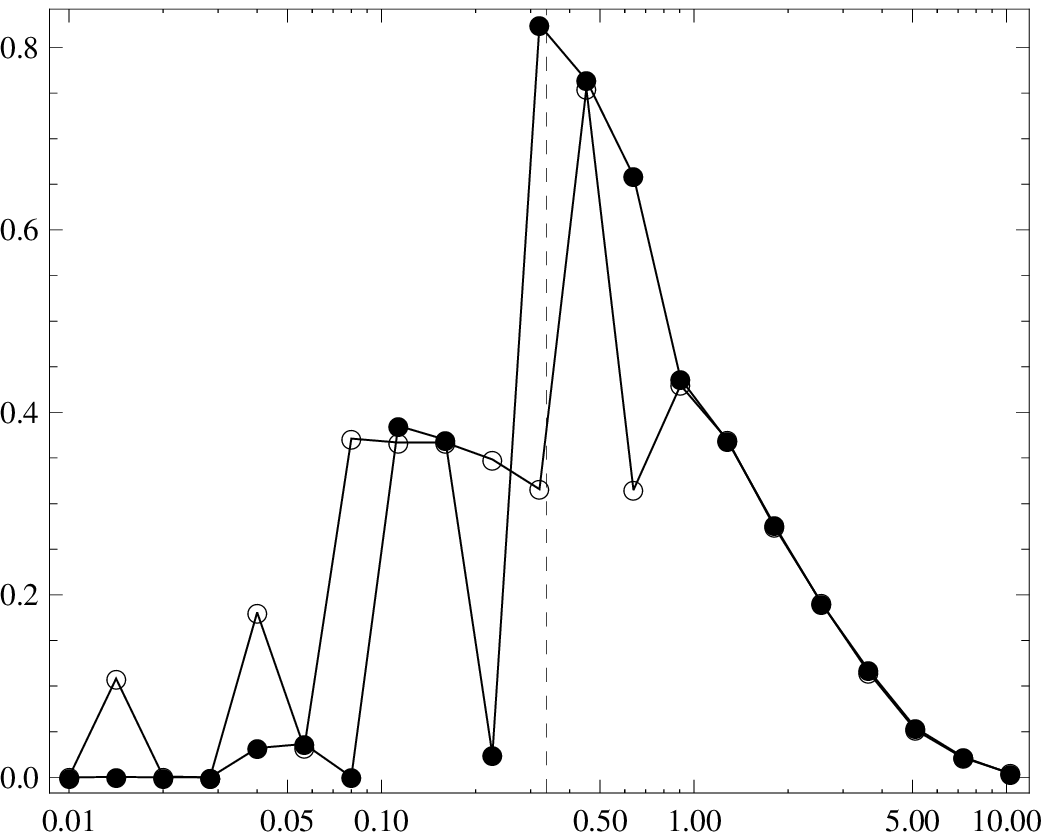}}
\put(230,200){\epsfxsize=108\unitlength\epsfbox{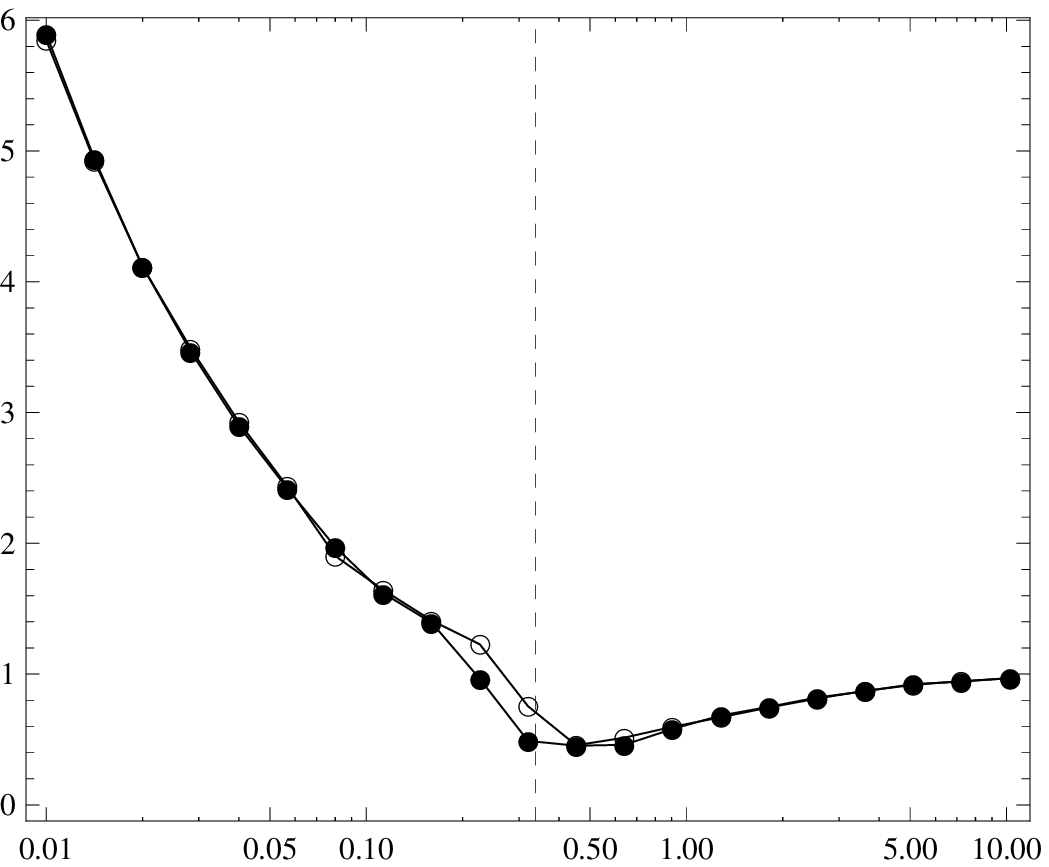}}

\put(55,-10){$\alpha$}\put(170,-10){$\alpha$}\put(288,-10){$\alpha$}

\put(50,295){$c$}
\put(170,295){$\phi$} \put(280,295){$\sigma$}
 \put(-55,50){$A_0^{-1}\!\approx 2.43$}
 \put(-55,150){$A_0^{-1}\!=2.25$}
 \put(-55,250){$A_0^{-1}\!\approx 1.70$}

\end{picture}
\vspace*{1mm}
\caption{
Simulation results for the persistent correlations $c$, the
fraction of frozen agents $\phi$, and the volatility $\sigma$, for $\tau=-1$ (i.e. $F[A]=-A+A^3/A_0^2)$.
The chosen values of $A_0^{-1}$ should, according to the phase diagram, probe the region close where a $\chi=\infty$ transition
should occur at the conventional MG value $\alpha_c\!\approx\! 0.3374$ (indicated with a vertical dashed line), with an ergodic MinGame phases for large $\alpha$ and a non-ergodic MinGame phase for small $\alpha$.
Full markers: \emph{tabula rasa} initial conditions; empty markers: biased initial conditions.
We note that the behavior in the non-ergodic regime is
not typical of the standard MG.}\label{A0large}
\end{figure}

We begin with $\tau=-1$ (the model of \cite{demartino2}). In figure \ref{A0small} we present results for the volatility $\sigma$, the persistent correlations $c$, and
the fraction of frozen agents $\phi$, for values of $A_0^{-1}$ corresponding to the MajGame regime, without remanence.
The simulations indeed reproduce a majority-game type state for large $A_0$ (top row), but as $A_0$ is reduced (middle and bottom rows)
the system exhibits behaviour that is less clear-cut than what is suggested by the phase diagram. We are still in the non-remanent region, so this cannot be explained by remanence effects.
Close to the remanent region R of the $\tau=-1$ phase diagram, the simulations are in rough agreement with the predictions, see figure \ref{A0medium}.
The vertical dashed line marks the $\chi=\infty$ transition, and seems to agree with simulations within the finite size limitations.
In figure \ref{A0large} we probe the region where we expect MG-type behaviour. This is borne out in the ergodic region, for $\alpha>\alpha_c$
(and the transition is found in the right place, in agreement with the phase diagram). However, for $\alpha<\alpha_c$ the behaviour is found to be far from typical; in contrast to the standard MG, the differences between the biased and unbiased initial conditions are small, and more likely to result of instabilities than long-term memory. In fact one notes similarities in behavior with figure \ref{fig:gammalarge}
for $F[A]=\sgn(A)|A|^{\gamma}$ with $\gamma=4,5$; this suggest as a  possible explanation that
if $A_0^{-1}$ gets larger, the first term of the present $F[A]=A^3/A_0^2\!-\!A$ plays a role similar to $\sgn(A)|A|^{\gamma}$ with $\gamma\geq 4$.

\begin{figure}[t]
\vspace*{-9mm} \hspace*{10mm} \setlength{\unitlength}{0.35mm}
\begin{picture}(350,430)

\put(0,000){\epsfxsize=110\unitlength\epsfbox{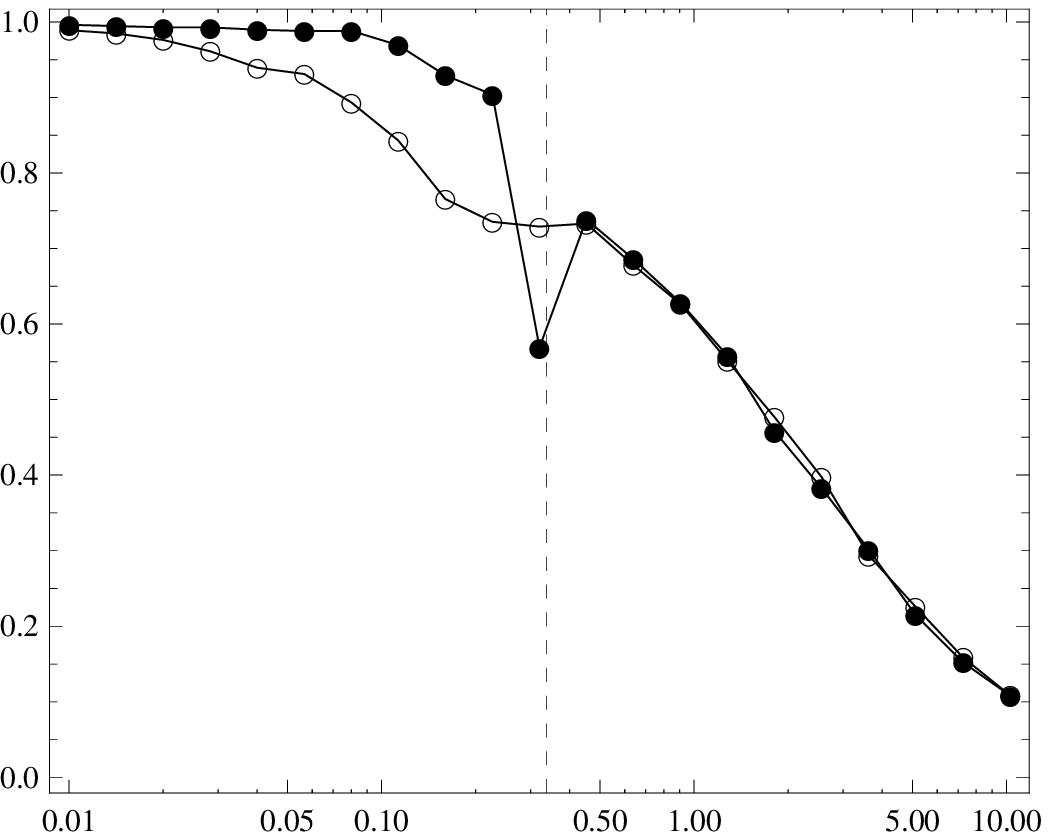}}
 \put(115,000){\epsfxsize=110\unitlength\epsfbox{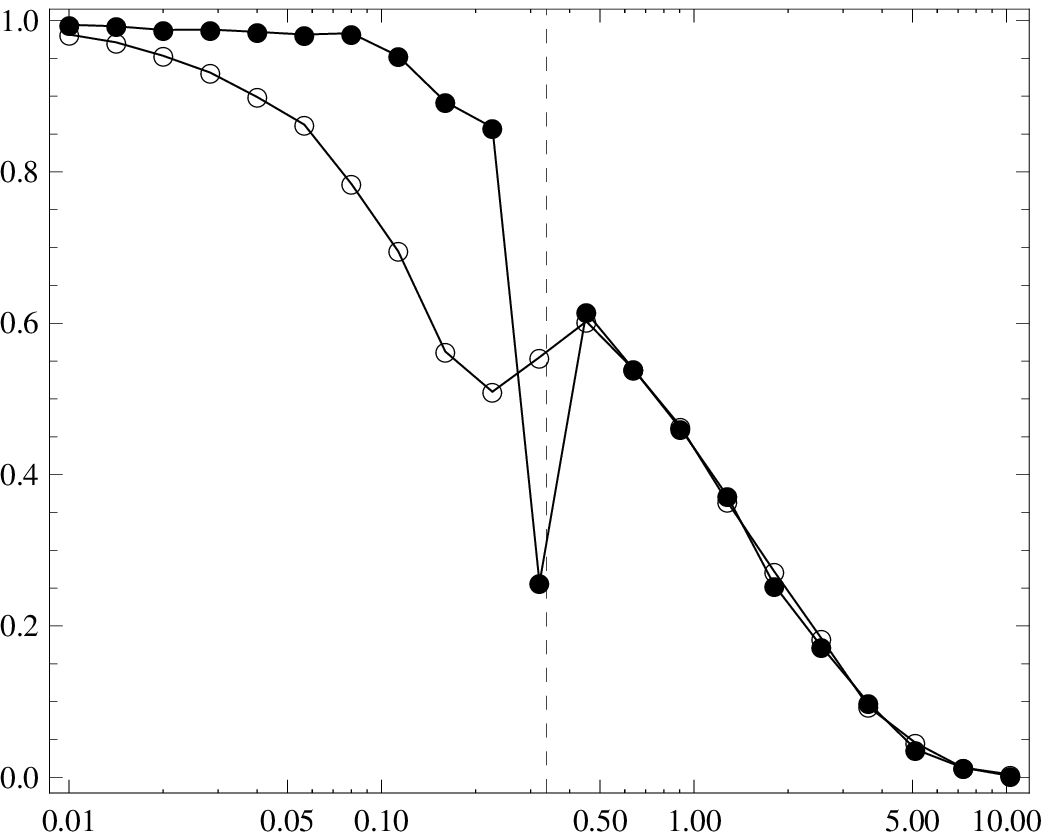}}
\put(230,000){\epsfxsize=108\unitlength\epsfbox{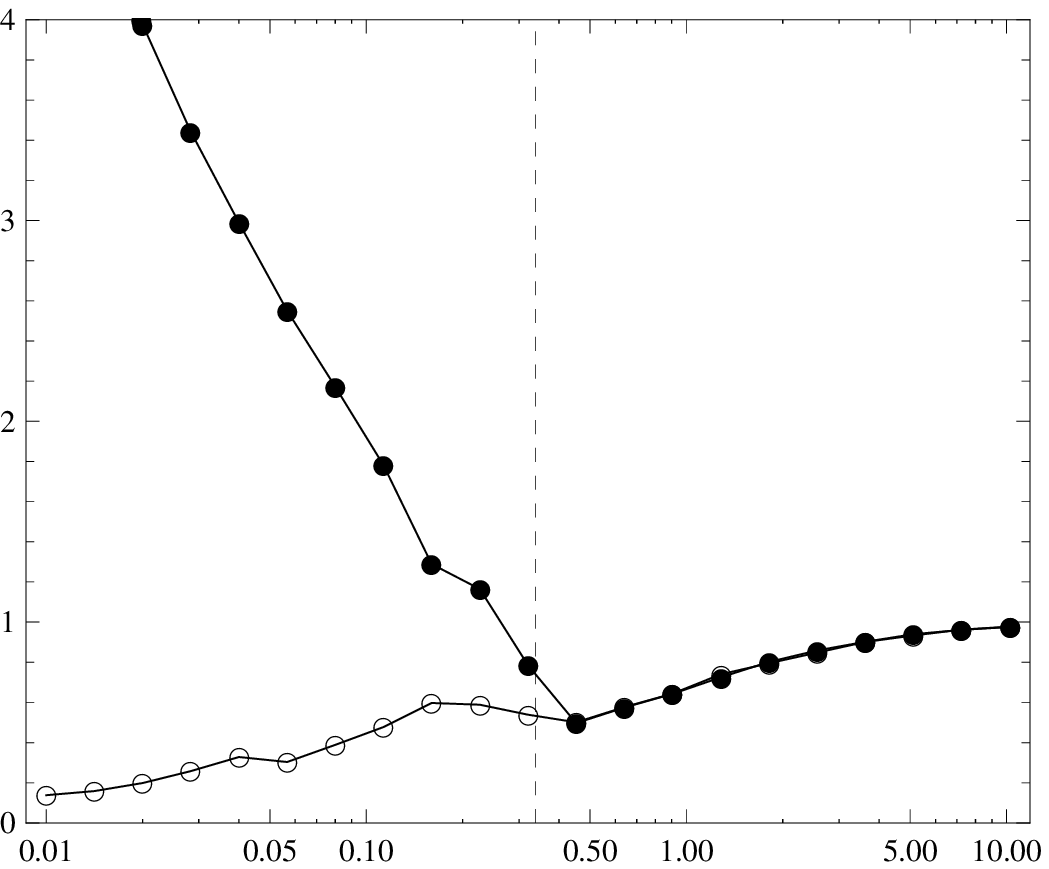}}

\put(0,100){\epsfxsize=110\unitlength\epsfbox{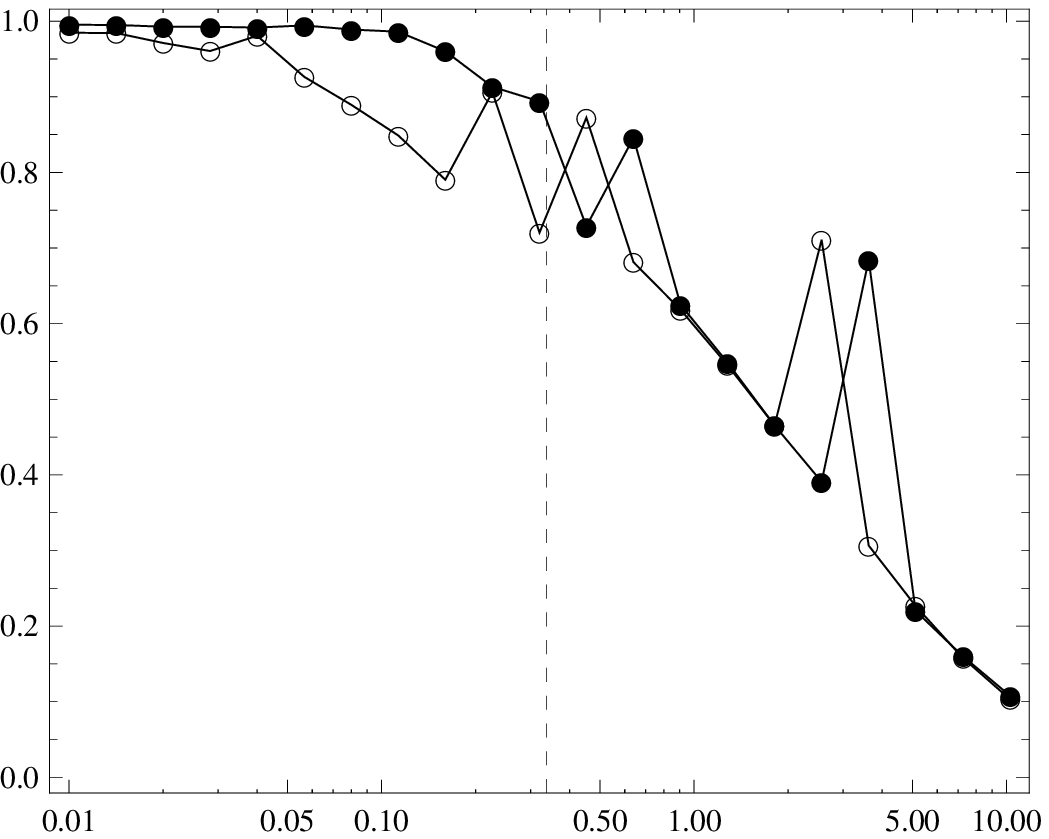}}
 \put(115,100){\epsfxsize=110\unitlength\epsfbox{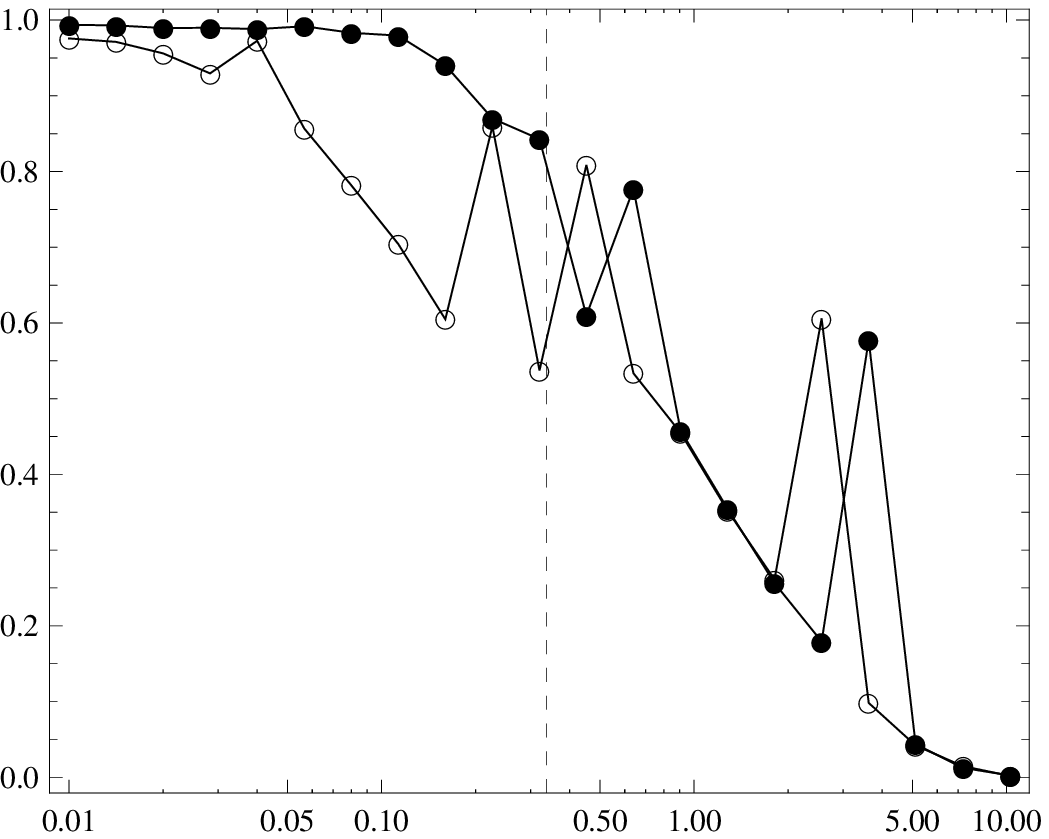}}
\put(230,100){\epsfxsize=108\unitlength\epsfbox{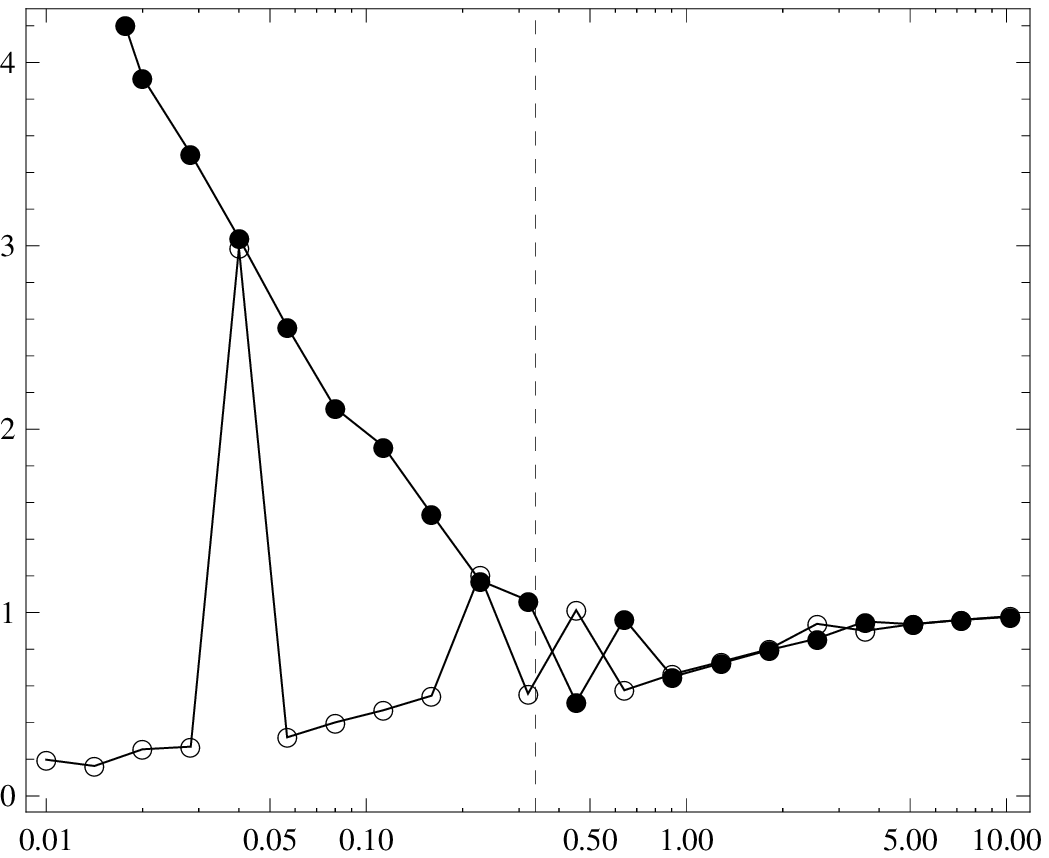}}

\put(0,200){\epsfxsize=110\unitlength\epsfbox{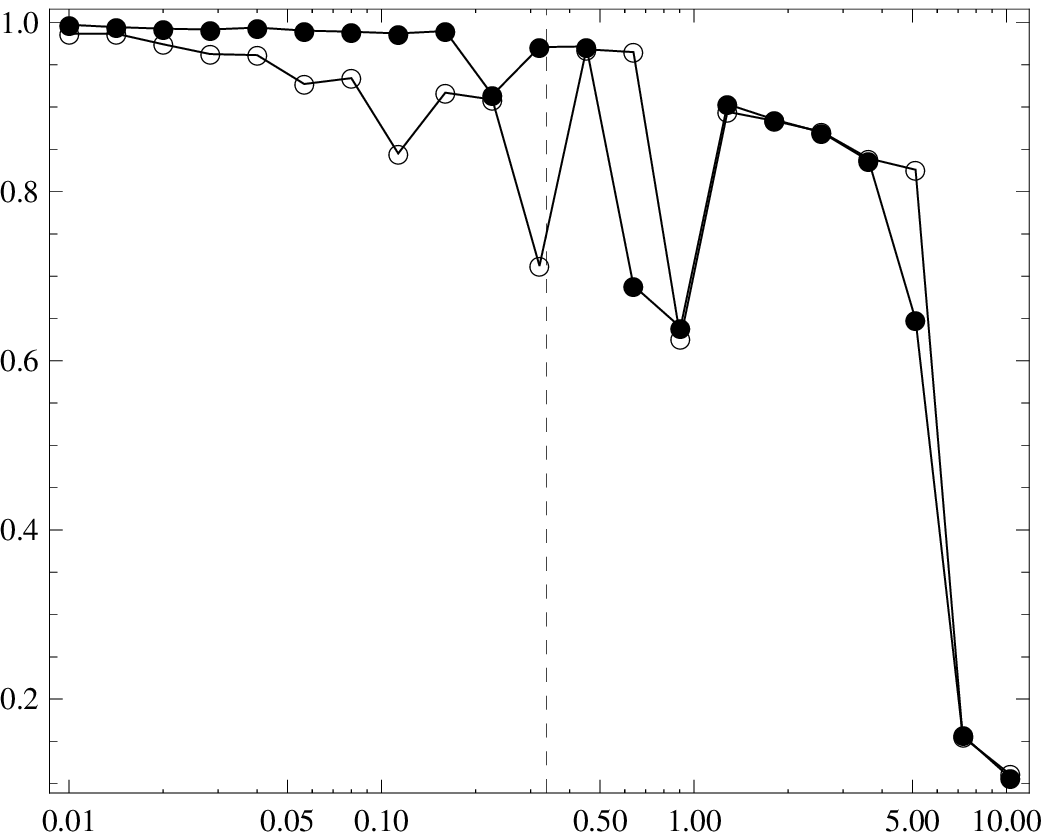}}
 \put(115,200){\epsfxsize=110\unitlength\epsfbox{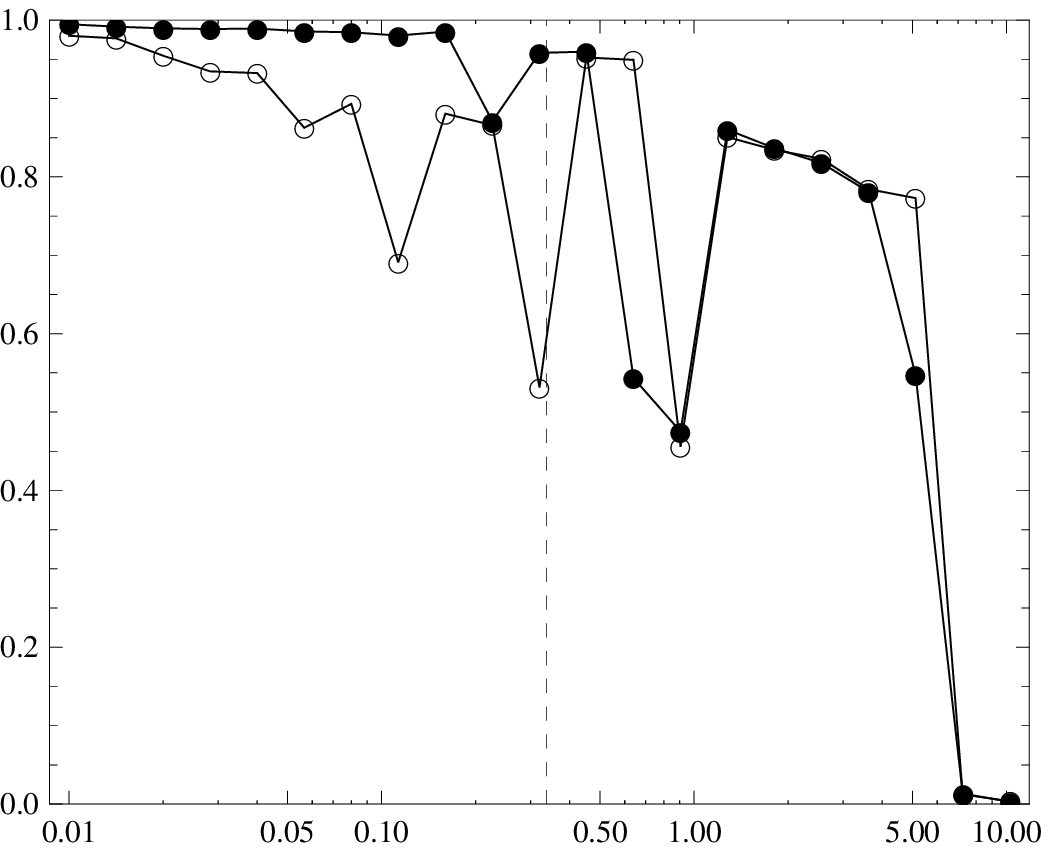}}
\put(230,200){\epsfxsize=108\unitlength\epsfbox{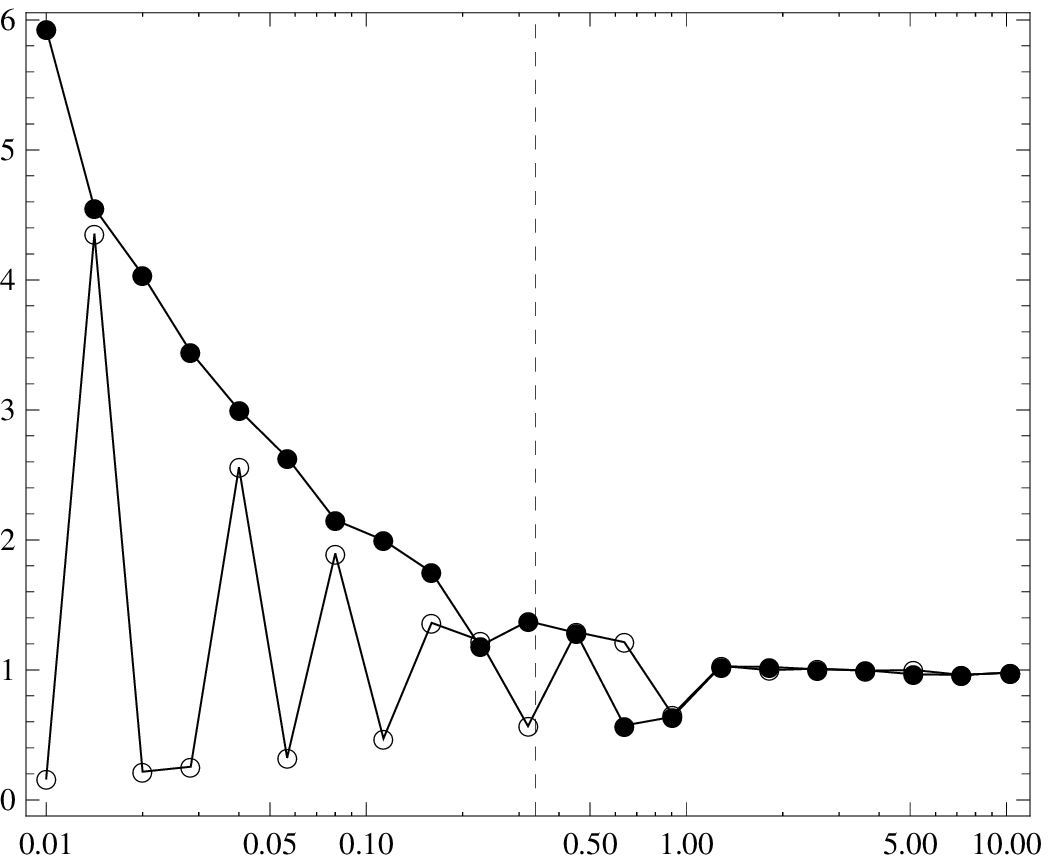}}

\put(0,300){\epsfxsize=110\unitlength\epsfbox{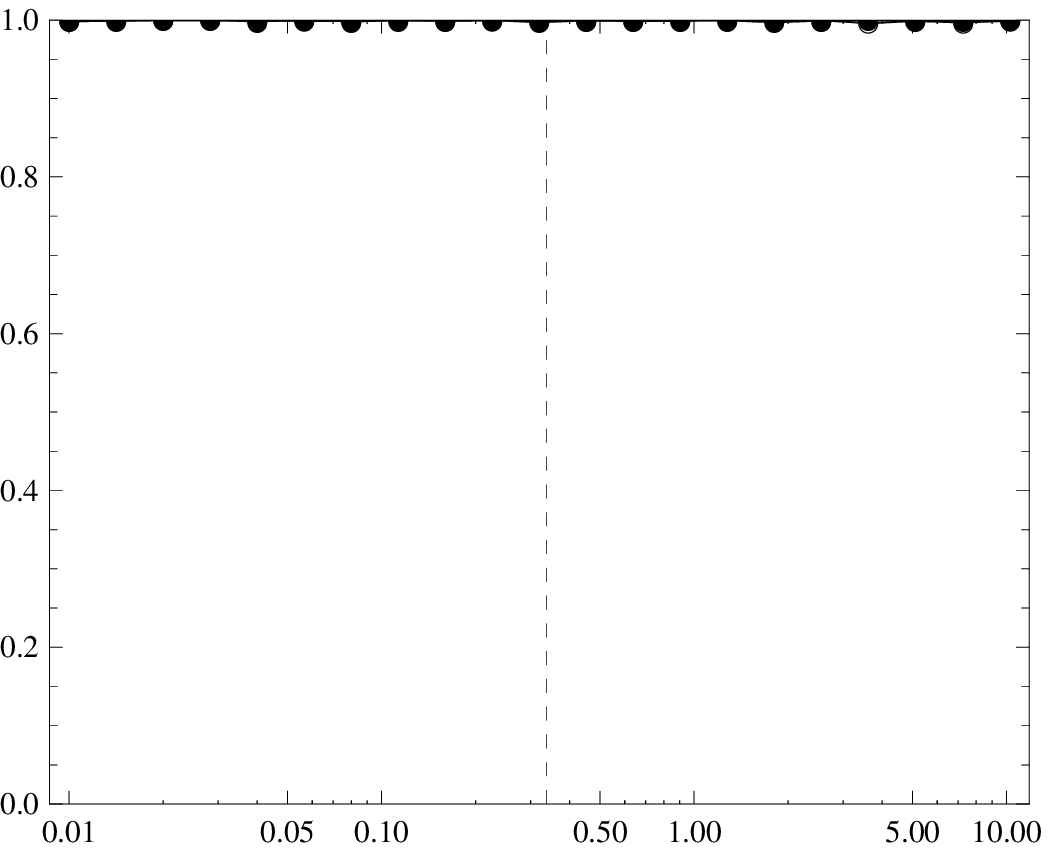}}
 \put(115,300){\epsfxsize=110\unitlength\epsfbox{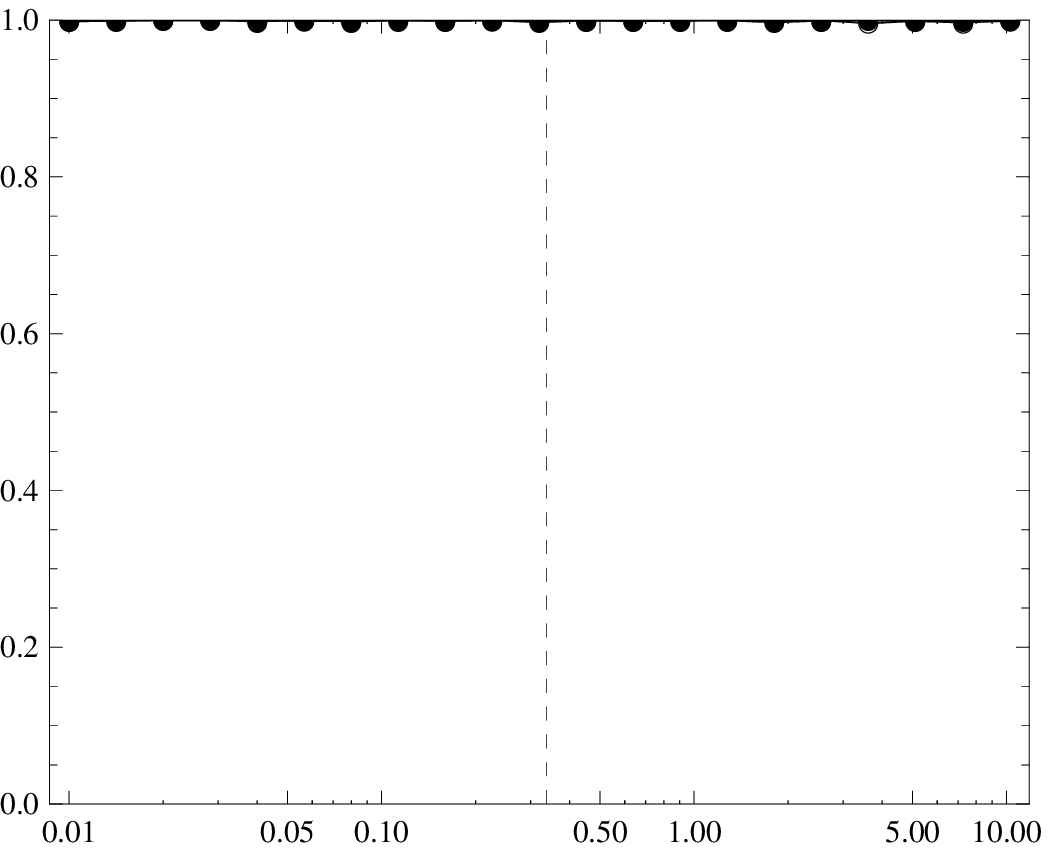}}
\put(230,300){\epsfxsize=108\unitlength\epsfbox{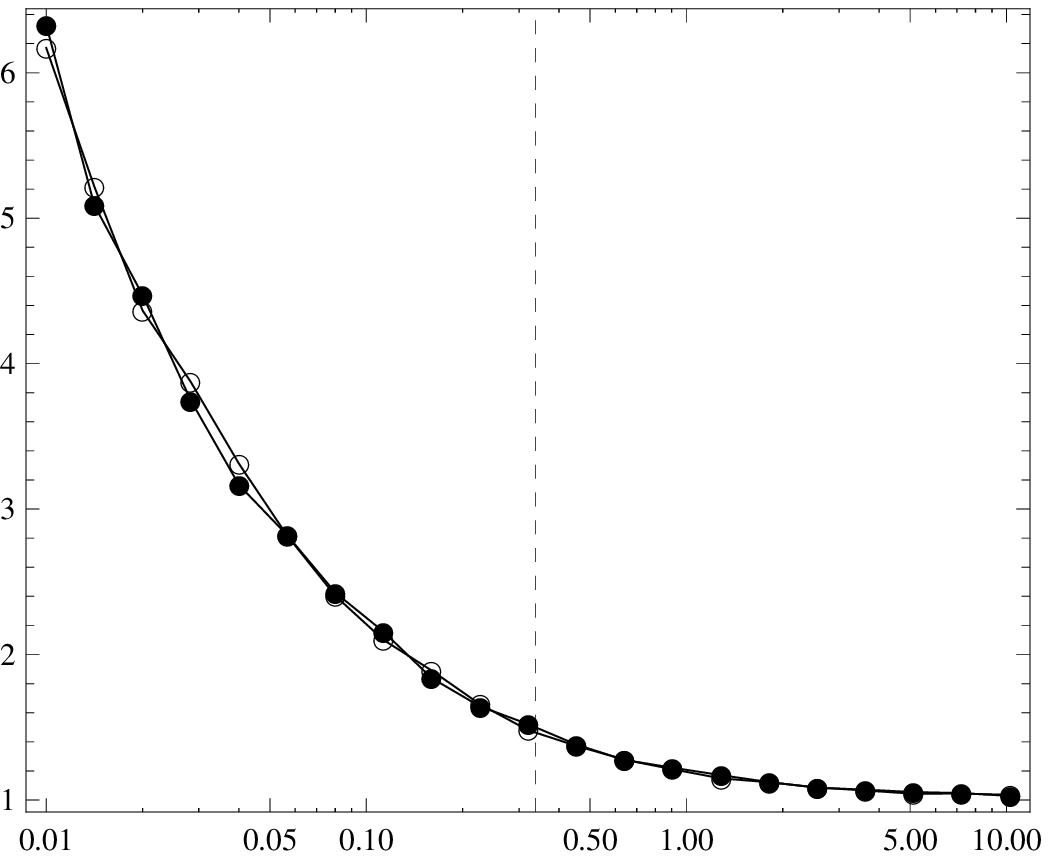}}

\put(55,-10){$\alpha$}\put(170,-10){$\alpha$}\put(288,-10){$\alpha$}

\put(50,395){$c$}
\put(170,395){$\phi$} \put(280,395){$\sigma$}
 \put(-55,50){$A_0^{-1}\!=0.25$}
 \put(-55,150){$A_0^{-1}\!\approx 0.29$}
 \put(-55,250){$A_0^{-1}\!\approx 0.32$}
 \put(-55,350){$A_0^{-1}\!=0.50$}

\end{picture}
\vspace*{1mm}
\caption{
Simulation results for the persistent correlations $c$, the
fraction of frozen agents $\phi$, and the volatility $\sigma$, for the troublesome case $\tau=1$ (i.e. $F[A]=A-A^3/A_0^2)$.
According to the phase diagram, one should for small $A_0^{-1}$
find a conventional $\chi=\infty$ phase separating an ergodic from a non-ergodic MinGame phase. For larger $A_0^{-1}$ we should at some point enter
a MajGame phase.
Full markers: \emph{tabula rasa} initial conditions; empty markers: biased initial conditions.
Although the observed behaviour agrees with the theory for small $A_0^{-1}$, and indeed a MajGame phase appears as $A_0^{-1}$ is increased,
the point where this happens does not agree with the phase diagram.  This underlines the problems in the identification of transition lines in the remanent region of the phase diagram, which are indeed much more profound  for $\tau=1$.
}\label{A0taupos}
\end{figure}

In figure \ref{A0taupos} we present simulation results for the most difficult case $\tau=1$, i.e. $F[A]=A-A^3/A_0^2$. These show that
the anticipated transition from a MinGame phase at small values of $A_0^{-1}$ to an MajGame phase for larger $A_0^{-1}$ indeed occurs, but  for some  value $0.25\leq A_0^{-1} \leq 0.5$, which differs from the prediction in the phase diagram. Furthermore, one notes instabilities of multiplicities of states in the remanent regime, to the right of the $\chi=\infty$ transition.
These deviations for $\tau=1$ are of course not unexpected, since the $\tau=1$
 phase diagram is plagued by remanence. They could be caused by many things, possibly in combination: incorrect selection of mathematical solutions in the remanent phase, insufficiently equilibrated simulations made worse by remanence, alternative transitions (e.g. onset of weak long-term memory),
etc.

\clearpage
\section{Discussion}

Minority-game type models have an amazing ability to describe new phenomena within an accessible mathematical framework, and to throw up new mathematical surprises and puzzles. In this paper we have studied generalized agent-based market models of the  MG-type
(in their so-called `fake history' batch version),
in which the agents' strategy valuation update
rule is allowed to depend on the overall market bid, via an impact function $F[A]$ (which would be $F[A]=\pm A$ for the standard minority and majority games, respectively). The function $F[A]$  allows us to model  the effect of agents' interpretation
of the state of the market (the magnitude of the fluctuations), assuming that whether price fluctuations
are perceived to be large or small should somehow influence how agents trade in financial markets.
Our motive is to incorporate into solvable market models such behavioral
elements. In this study we focus on two model classes: one in which agents are always contrarians, but where we can control their greed
and willingness to take risks, viz. $F[A]=\sgn(A)|A|^\gamma$, and one in which agents can switch between trend following and contrarian trading,  viz. $F[A]=\pm A[1-A^2/A_0^2]$, somewhat reminiscent of how they would adapt to booming and `bear' markets.

At a mathematical level, the generating functional analysis of the present class of models requires studying the overall bid evolution explicitly, similar to how one would study models with real  market history.
It turns out that a key issue in working towards closed self consistent equations for persistent order parameters is whether  the overall bid process
is remanent. In non-remanent cases, such as $F[A]=\sgn(A)|A|^\gamma$, solution is direct and relatively easy, and the agreement between theory and simulations is excellent (for this particular model: unless $\gamma$ becomes too large, where we lose the ergodic phase altogether).
In remanent cases, such as $F[A]=\tau A[1-A^2/A_0^2]$ with $\tau=\pm 1$,
we need to rely on ans\"{a}tze and Maxwell-type approximations to select a solution from the possible stationary states, especially for $\tau=1$, and the agreement between theory and experiment is consequently limited.

 Once more, what appear at first sight to be simple modifications of the standard
MG lead to highly nontrivial and unexpected behaviour (even in batch models with fake histories).
As soon as agents are allowed to adapt their trading style to the magnitude of the fluctuations, in the spirit of \cite{demartino2},
one introduces a non-trivial effective overall bid process, with new instabilities and new transitions.
We are now approaching
the point where theories based on persistent order parameter equations only, the ones that benefit most from
the simplifications induced by having `frozen agents', are no longer giving us the information we need.
In models dominated by remanence, which are the type one needs when including more realistic agent behaviour,
we can no longer avoid solving the full dynamics more explicitly.

\appendix

\section{Scaling of overall market bid covariances}
\label{app:messypart}

We defined the overall bid covariance matrix as $\Xi_{tt^\prime}(z)=\bra \tilde{A}(t,z)\tilde{A}(t^\prime,z)\ket$.  It must be time-translation
invariant, so we write $\Xi(t,z)=\Xi_{s+t,s}(z)$, i.e. $\Xi(t,z)=\bra \tilde{A}(s,z)\tilde{A}(t+s,z)\ket$. From (\ref{eq:bid_eqn_2})  we extract:
\begin{eqnarray}
\hspace*{-20mm}
\Xi(t,z)&=&
\frac{1}{2}\tilde{C}(t)
 - \sum_{s}G(s)\Big\{\bra F[\overline{A}(z)+\tilde{A}(t\!-\!s,z)]\tilde{A}(0,z)\ket
 +\bra F[\overline{A}(z)+\tilde{A}(\!-\!s,z)]\tilde{A}(t,z)\ket\Big\}
\nonumber
\\
\hspace*{-20mm}
&&- \sum_{ss^\prime}G(s)G(s^\prime)\bra \Big\{F[\overline{A}(z)+\tilde{A}(t-s,z)]F[\overline{A}(z)+\tilde{A}(-s^\prime,z)]-\overline{F}^2(z)\Big\}
\ket
\end{eqnarray}
Since the non-persistent bids are Gaussian this equation gives an explicit relation from which to solve their covariances; along the lines
of \cite{Hatchett} we apply the general relation
\begin{eqnarray}
\hspace*{-20mm}
\bra G[\tilde{A}(t,z),\tilde{A}(t^\prime,z)]\ket&=&\int\! DxDy
\\
\hspace*{-20mm}&&\hspace*{-15mm}\times
G\big[\frac{x}{\sqrt{2}}\sqrt{S_1(t-t')}+
\frac{y}{\sqrt{2}}\sqrt{S_2(t-t')},
\frac{x}{\sqrt{2}}\sqrt{S_1(t-t')}-\frac{y}{\sqrt{2}}\sqrt{S_2(t-t')}]
\big]\nonumber
\end{eqnarray}
where $S_1(t-t')=\Xi(0,z)\!+\!\Xi(t\!-\!t^\prime,z)$,$~S_2(t-t')=\Xi(0,z)\!-\!\Xi(t\!-\!t^\prime,z)$. In particular we need the two quantities
\begin{eqnarray}
\bra F[\overline{A}(z)+\tilde{A}(u,z)]\tilde{A}(v,z)\ket
&=&\Xi(u\!-\!v,z)\overline{F^\prime}(z)
\end{eqnarray}
and
\begin{eqnarray}
\hspace*{-25mm}
\bra F[\overline{A}(z)\!+\!\tilde{A}(u,z)]F[\overline{A}(z)\!+\!\tilde{A}(v,z)]\ket
&=& \int\!DxDy~F\big[\overline{A}(z)+\!\frac{x}{\sqrt{2}}\sqrt{S_1(u\!-\!v)}+
\!\frac{y}{\sqrt{2}}\sqrt{S_2(u\!-\!v)}\big]
\nonumber
\\
\hspace*{-15mm}
&&
\times F\big[\overline{A}(z)+\frac{x}{\sqrt{2}}\sqrt{S_1(u\!-\!v)}-\frac{y}{\sqrt{2}}\sqrt{S_2(u\!-\!v)}
\big]
\end{eqnarray}
For $u=v$ this gives simply $\bra F^2[\overline{A}(z)+\tilde{A}(u,z)]\ket=\overline{F^2}(z)$.
The first average above is of order $\Xi$. The second average can for $u\neq
v$ be expanded in powers of $\Xi(u-v,z)$ (which should decay to zero quickly), using
\begin{eqnarray}
\hspace*{-15mm}
I(\xi)
&=& \int\!DxDy~F\big[\overline{A}+\frac{x}{\sqrt{2}}\sqrt{\Xi(0)\!+\!\xi}+
\frac{y}{\sqrt{2}}\sqrt{\Xi(0)\!-\!\xi}\big]
\nonumber
\\
\hspace*{-15mm}&&\hspace*{10mm}\times
 F\big[\overline{A}+\frac{x}{\sqrt{2}}\sqrt{\Xi(0)\!+\!\xi}-\frac{y}{\sqrt{2}}\sqrt{\Xi(0)\!-\!\xi}
\big]
\nonumber\\
\hspace*{-15mm}
&=&\Big\{ \int\!Dx~F[\overline{A}+x\sqrt{\Xi(0)}]\Big\}^2+
\xi\Big\{\int\!Dx~F^\prime\big[\overline{A}+x\sqrt{\Xi(0)}\big]\Big\}^2
+\order(\xi^2)
\end{eqnarray}
Thus we conclude that
\begin{eqnarray}
\hspace*{-15mm}
\bra F[\overline{A}(z)+\tilde{A}(u,z)]F[\overline{A}(z)+\tilde{A}(v,z)]\ket-\overline{F}^2(z)
&=& \delta_{uv}\Big[\overline{F^2}(z)-\overline{F}^2(z)\Big]\nonumber
\\
\hspace*{-15mm}&&\hspace*{-30mm}+(1\!-\!\delta_{uv})\Xi(u\!-\!v,z) \Big[
\Big\{\overline{F^\prime}(z)\Big\}^2
+\order(\Xi(u\!-\!v,z))\Big]
\end{eqnarray}
which leads to
\begin{eqnarray}
\hspace*{-15mm}
\Xi(t,z)&=&
\frac{1}{2}\tilde{C}(t)
 - \overline{F^\prime}(z)\sum_{s>0}G(s)\Big\{\Xi(-s-t,z)+\Xi(-s+t,z)\Big\}
\nonumber
\\
\hspace*{-15mm}
&&- \Big[\overline{F^2}(z)-\overline{F}^2(z)\Big] \sum_{s>0}G(s)G(s-t)
\nonumber
\\
\hspace*{-15mm}
&&-\Big[\overline{F^\prime}(z)\Big]^2
\sum_{s>0}\sum_{v>t-s,v\neq 0}G(s)G(s-t+v)
\Xi(v,z)[1+\order(\Xi)]
\label{eq:Xiformula}
\end{eqnarray}
At this point we inspect how various terms scale. We know from earlier MG work (especially from approximate calculations of the volatility, see e.g. \cite{Book2})
that the relaxation of the response function $G$ is very slow. Upon making an ansatz
of the form $G(t)=\chi(\rme^\mu-1)\rme^{-\mu t}$
one can show that in the stationary state one must put $\mu\to 0$ (the relaxation time seems to diverge with $N$ in an
as yet undetermined way).
For such an exponential ansatz one would find $\sum_{s>0}G(s)G(s-t)=\chi^2 \rme^{-\mu |t|}(\rme^\mu-1)/(\rme^\mu+1)=\order(\mu\chi^2)$.
Thus  if $\mu\to 0$ the second term in
(\ref{eq:Xiformula}) vanishes in the ergodic regime. In the final term we use our earlier ansatz that the correlations $\Xi(s)$
decay on short times; formula (\ref{eq:Xiformula})
suggests that this relaxation time will be that of $\tilde{C}(t)$. If this relaxation time is $\tau_C$, we may estimate the scaling
of the last term in
(\ref{eq:Xiformula}) as
$\order(\mu\tau_C \chi^2)$.
So, upon writing the relaxation time of the response function as $\tau_G$, we may replace $\mu\to 1/\tau_G$
and write (\ref{eq:Xiformula}) as
\begin{eqnarray}
\Xi(t,z)&=&
\frac{1}{2}\tilde{C}(t)
 - \overline{F^\prime}(z)\order(\tau_C/\tau_G)
- \Big[\overline{F^2}(z)-\overline{F}^2(z)\Big] \order(1/\tau_G)
\nonumber
\\
&&
-\Big[\overline{F^\prime}(z)\Big]^2\order(\tau_C/\tau_G)
\label{eq:Xiformula_expanded}
\end{eqnarray}

\end{document}

%% file: Commands.tex
\newcommand{\bra}{\langle}
\newcommand{\ket}{\rangle}

\newcommand{\order}{{\mathcal O}}

\newcommand{\sgn}{\textrm{sgn}}

\newcommand{\one}{{\rm 1\!\!I}}

\newcommand{\bnull}{{\mbox{\boldmath $0$}}}
\newcommand{\be}{\begin{equation}}
\newcommand{\ee}{\end{equation}}
\newcommand{\bd}{\begin{displaymath}}
\newcommand{\ed}{\end{displaymath}}
\newcommand{\vsp}{\vspace*{3mm}}

\newcommand{\R}{{\rm I\!R}}

\newcommand{\D}{{\mathcal D}}

\newcommand{\bq}{\ensuremath{\mathbf{q}}}

\newcommand{\bz}{\ensuremath{\mathbf{z}}}

\newcommand{\bA}{\ensuremath{\mathbf{A}}}

\newcommand{\bD}{\ensuremath{\mathbf{D}}}

\newcommand{\bR}{\ensuremath{\mathbf{R}}}

\newcommand{\bxi}{{\mbox{\boldmath $\xi$}}}

\newcommand{\bpsi}{{\mbox{\boldmath $\psi$}}}
\newcommand{\bphi}{{\mbox{\boldmath $\phi$}}}

\newcommand{\hq}{\hat{q}}

\newcommand{\hC}{\hat{C}}

\newcommand{\hK}{\hat{K}}
\newcommand{\hL}{\hat{L}}

\newcommand{\here}{\makebox(0,0)}

\newcommand{\blacksquare}{\rule{1.3\unitlength}{1.3\unitlength}~}

\newcommand{\rate}{\tilde{\eta}}

\newcommand{\chiR}{\chi_{\!\!\!~_R}}